\documentclass[%
reprint,
superscriptaddress,
nofootinbib,
 amsmath,amssymb,
 aps, prd,
 showpacs,
 showkeys,
]{revtex4-1}

\usepackage{paralist}
\usepackage{graphicx}%
\usepackage{dcolumn}%
\usepackage{bm}%
\usepackage{color}%
\usepackage{units}  %
\usepackage{subfig} %
\usepackage{xspace} %
\usepackage{multirow}

\newcommand{\abu}
{\affiliation{Department of Physics, Boston University, Boston, MA
02215, USA}}

\newcommand{\auiuc}
{\affiliation{Department of Physics, University of Illinois at
Urbana-Champaign, Urbana, IL 61801, USA}}
\newcommand{\ajmu}
{\affiliation{Department of Physics, James Madison University,
Harrisonburg, VA 22807, USA}}
\newcommand{\aky}
{\affiliation{Department of Physics and Astronomy, University of
Kentucky, Lexington, KY 40506, USA}}
\newcommand{\akvi}
{\affiliation{KVI, University of Groningen, NL-9747AA Groningen, The
Netherlands}}
\newcommand{\akwu}
{\affiliation{Department of Mathematics and Physics, Kentucky
Wesleyan College, Owensboro, KY 42301, USA }}
\newcommand{\aregis}
{\affiliation{Department of Physics and Computational Science, Regis
University, Denver, CO 80221, USA}}
\newcommand{\apsi}
{\affiliation{Paul Scherrer Institute, CH-5232 Villigen PSI,
Switzerland}}
\newcommand{\ayork}
{\affiliation{Department of Earth and Physical Sciences, York College,
  City University of New York, Jamaica, NY 11451, USA}}
\newcommand{\auw}
{\affiliation{Department of Physics, University of Washington,
Seattle, WA 98195, USA}}

\newcommand{\GF}[0]{\ensuremath{G_F}\xspace}

\def\nuc#1#2{\relax\ifmmode{}^{#1}{\protect\text{#2}}\else${}^{#1}$#2\fi}

\makeatletter
\renewcommand\paragraph{\@startsection{paragraph}{4}{\z@}%
  {-3.25ex\@plus -1ex \@minus -.2ex}%
  {1.5ex \@plus .2ex}%
  {\normalfont\normalsize\itshape}}
\makeatother

\usepackage{ragged2e} 
\DeclareCaptionJustification{justified}{\justifying}
\begin{document}

\title{Detailed Report of the MuLan Measurement of the Positive Muon Lifetime and Determination of the Fermi Constant}%

\author{V.\,Tishchenko}
\aky
\author{S.\,Battu}
\aky
\author{R.M.~Carey}
\abu
\author{D.B.\,Chitwood}
\auiuc
\author{J.\,Crnkovic}
\auiuc
\author{P.T.\,Debevec}
\auiuc
\author{S.\,Dhamija}
\aky
\author{W.\,Earle}
\abu
\author{A.\,Gafarov}
\abu
\author{K.\,Giovanetti}
\ajmu
\author{T.P.\,Gorringe}
\aky
\author{F.E.\,Gray}
\aregis
\author{Z.\,Hartwig}
\abu
\author{D.W.\,Hertzog}
\auiuc \auw
\author{B.\,Johnson}
\akwu
\author{P.\,Kammel}
\auiuc \auw
\author{B.\,Kiburg}
\auiuc
\author{S.\,Kizilgul}
\auiuc
\author{J.\,Kunkle}
\auiuc
\author{B.\,Lauss}
\apsi
\author{I.\,Logashenko}
\abu
\author{K.R.\,Lynch}
\abu \ayork
\author{R.\,McNabb}
\auiuc
\author{J.P.\,Miller}
\abu
\author{F.\,Mulhauser}
\auiuc \apsi
\author{C.J.G.~Onderwater}
\auiuc \akvi
\author{Q.~Peng}
\abu
\author{J. Phillips}
\abu
\author{S.\,Rath}
\aky
\author{B.L.\,Roberts}
\abu
\author{D.M.\,Webber}
\auiuc
\author{P.\,Winter}
\auiuc
\author{B.\,Wolfe}
\auiuc

\collaboration{MuLan Collaboration}%

\date{\today}

\begin{abstract}

We present a detailed report of the method, setup, analysis and results of a precision measurement 
of the positive muon lifetime. The experiment was conducted at the Paul Scherrer Institute using 
a time-structured, nearly 100\%-polarized, surface muon beam and a segmented, fast-timing, plastic scintillator array. 
The measurement employed two target arrangements; a magnetized ferromagnetic target with a $\sim$4~kG internal magnetic field 
and a crystal quartz target in a 130~G external magnetic field. Approximately $1.6 \times 10^{12}$ positrons 
were accumulated and together the data yield a muon lifetime of 
$\tau_{\mu} ($MuLan$) = 2\, 196\, 980.3(2.2)$~ps (1.0~ppm), thirty times more precise than
previous generations of lifetime experiments. The lifetime measurement yields the most accurate value 
of the Fermi constant $G_F ($MuLan$) = 1.166\, 378\, 7(6) \times 10^{-5}$ GeV$^{-2}$ (0.5~ppm).
It also enables new precision studies of weak interactions 
via lifetime measurements of muonic atoms.

\end{abstract}

\maketitle

\section{Introduction}

Weak interactions are germane to topics
from big-bang nucleosynthesis and stellar hydrogen burning,
to archaeological dating and medical imaging,
as well as fundamental nuclear and particle physics.
Unlike electromagnetic and strong interactions, 
all particles participate 
in weak interactions, and every process
appears consistent with a universal strength for the weak force. 
This strength is governed by the Fermi constant \cite{Commins:1983}.

The value of the Fermi constant $G_F$ is best determined by
the measurement of the muon lifetime $\tau_{\mu}$ for two reasons; 
the muon lifetime is well suited to precision time measurements 
and the pure-leptonic character of muon decay permits
an unambiguous theoretical interpretation. 
Until recently, the uncertainty in $G_F$ 
was 17~parts-per-million (ppm); it originated from a 9~ppm contribution
from the existing measurements of the muon lifetime
and a 15~ppm contribution from the knowledge of the 
theoretical relation between the muon lifetime and the Fermi constant.

The theoretical uncertainty arose 
from a limited knowledge of the 2-loop QED corrections
to the muon decay rate. 
In 1999, van Ritbergen and Stuart (vRS) evaluated these corrections,
thus establishing the relation between $G_F$ and $\tau_{\mu}$ to 
sub-ppm level \cite{vanRitbergen:1999fi,vanRitbergen:1998yd,vanRitbergen:1998hn}. 
Their work was the motivation 
for our part-per-million measurement of the positive muon lifetime.

\subsection{Relation between \GF and the muon lifetime}

\begin{figure}
\begin{center}
\includegraphics[width=0.8\linewidth]{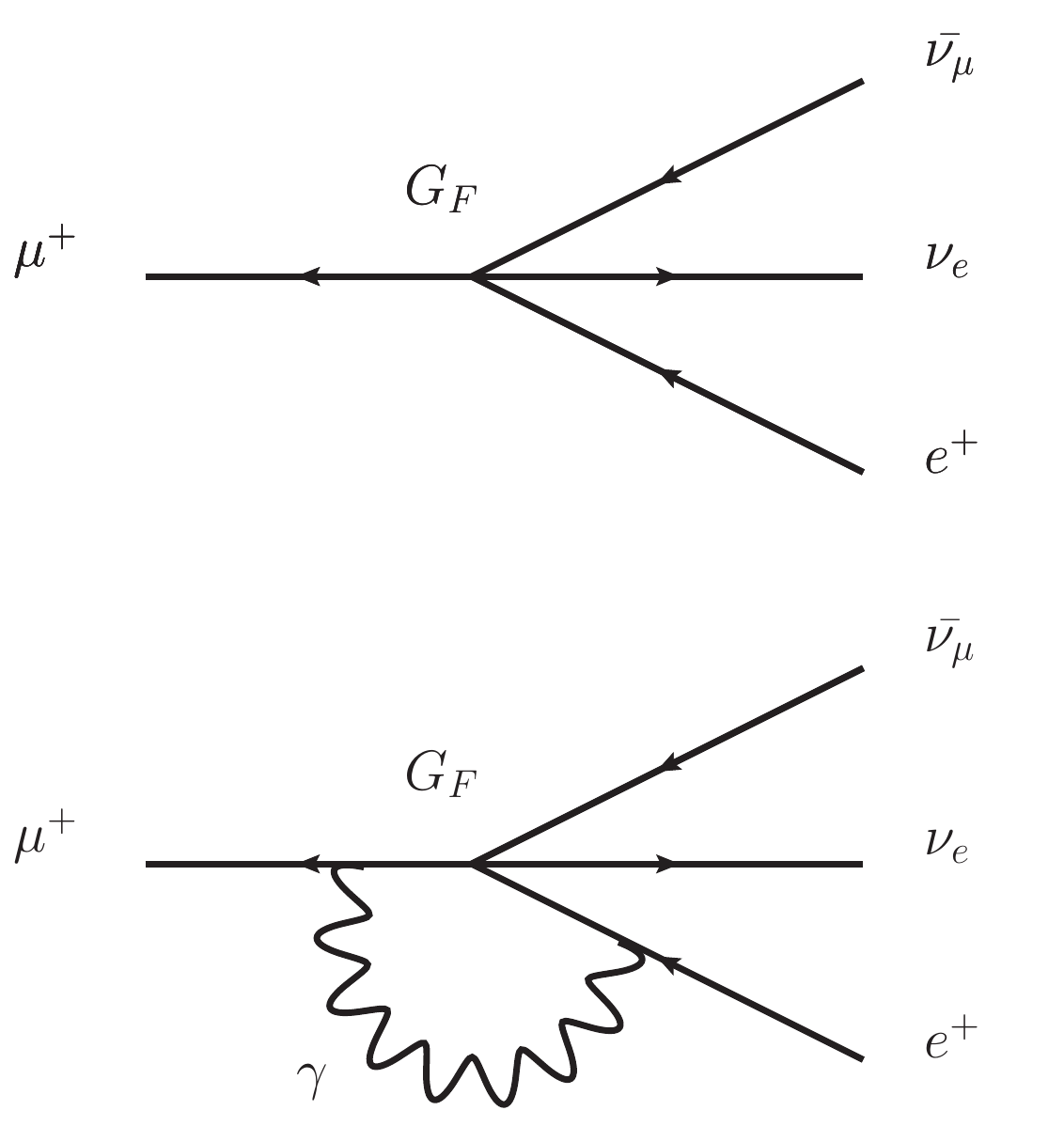}
\end{center}
\caption{Feynman diagrams for ordinary muon decay in Fermi theory.  
The upper graph shows the tree-level diagram for
the ordinary decay amplitude while
the lower graph shows the 1-loop QED correction
to the ordinary decay amplitude.}
\label{fig:fermi_theory}
\end{figure}

The V-A current-current Fermi interaction is a 
convenient effective theory 
for the description of the charged-current weak interaction
and the definition of the Fermi constant.
Although non-renormalizable,
it is well suited to the low-energy limit 
of the electroweak interaction
and neatly separates 
the parameterization of the weak interaction 
from corrections due to electromagnetic interactions.  
In particular, \GF\ encapsulates
all the weak interaction effects 
in the low-energy effective theory \cite{vanRitbergen:1999fi}.  

Feynman diagrams for ordinary decay 
at tree- and 1-loop level in Fermi theory 
are shown in Fig.\ \ref{fig:fermi_theory}. 
At the part-per-million level 
all appreciable corrections
to the tree-level diagram 
arise from QED radiative corrections.
Most importantly the hadronic effects 
are virtually non-existent, their 
leading terms arising as 2-loop QED corrections 
that contribute only 0.2~ppm to $\tau_{\mu}$ \cite{vanRitbergen:1998hn}.

Despite such simplifications the extraction of \GF\ from $\tau_{\mu}$
to sub-ppm level requires an exhaustive calculation of
first- and second-order QED radiative corrections.
In vRS, the authors expressed the 
relation between the Fermi constant and the muon lifetime
as a perturbative expansion 
\begin{equation}
\frac{1}{\tau_\mu} = \frac{\GF^2 m_\mu^5}{192\pi^3} \left(1 +
\sum_i \Delta q^{(i)}\right)\ 
\label{eq:vrs_lifetime}
\end{equation}
where  $m_{\mu}$ is the muon mass and  $\sum \Delta q^{(i)}$ describes the 
phase space ($\Delta q^{(0)}$), first-order QED
($\Delta q^{(1)}$), second-order QED ($\Delta q^{(2)}$), {\it etc}., 
theoretical corrections.

With an uncertainty arising from the theoretical corrections 
of 0.14~ppm \cite{vanRitbergen:1999fi,vanRitbergen:1998yd,vanRitbergen:1998hn,Pak:2008qt}\footnote{The original
derivation of vRS was conducted 
in the massless electron limit. 
The authors assessed a theoretical 
uncertainty of 0.3~ppm in $G_F$ arising from
estimated mass terms at 2-loop order
and neglected Feynman diagrams at 3-loop order.
In 2008, Pak and Czarnecki \cite{Pak:2008qt} extended 
the work of vRS to include the
2-loop order, electron mass terms.
Their work reduced the theoretical uncertainty in $G_F$ to 0.14~ppm.}
and an uncertainty arising from the muon mass 
of \unit[0.08]{ppm} \cite{Mohr:2012tt}, 
the overwhelming uncertainty in the Fermi constant 
became the experimental determination of the muon lifetime.

\subsection{Relation between $G_F$ and the Standard Model}

\begin{figure}
\begin{center}
\includegraphics[width=0.8\linewidth]{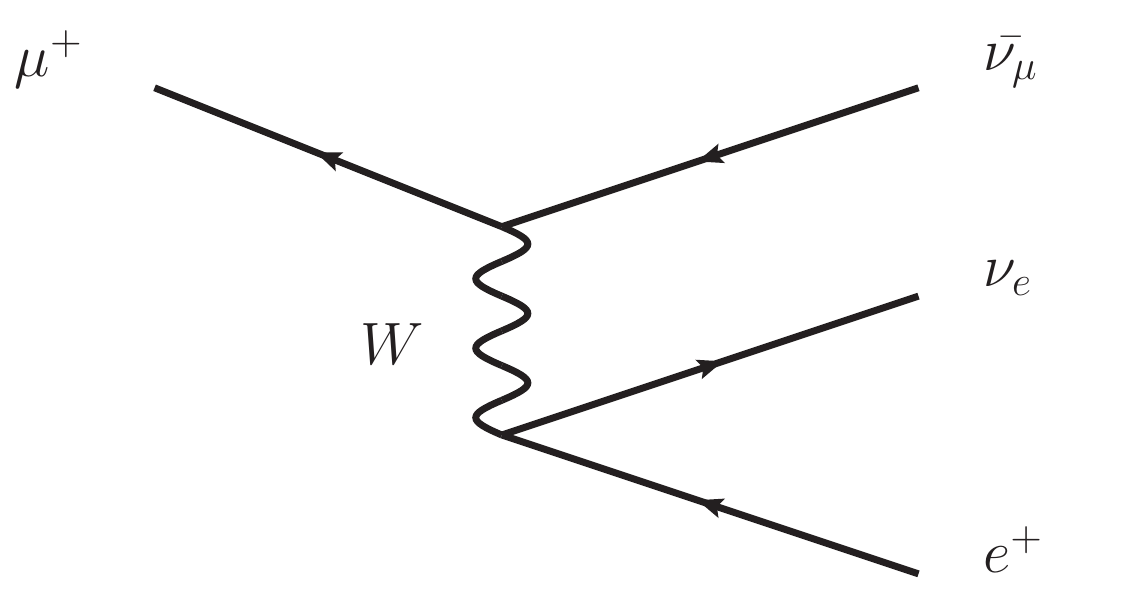} 
\end{center}
\caption{Standard model tree-level diagram of ordinary muon decay
showing the W-boson mediating the electroweak interaction between
the weak leptonic currents.}
\label{fig:W_GF}
\end{figure}
Once \GF\ is determined from $\tau_{\mu}$ using Eqn.\ \ref{eq:vrs_lifetime},
it can be related to the fundamental parameters of any high energy theory, 
{\it e.g.}\ the  electroweak 
interaction in the Standard Model.
The relation between the Fermi constant 
and the Standard Model can be written as a perturbative expansion 
of the electroweak interaction
\begin{equation}
\frac{\GF}{\sqrt{2}} = \frac{g^2}{8M_W^2}\left( 1 + \sum_i r_i\right)\ 
\label{eqn:GFSM}
\end{equation}
where $1 / M^2_W$ represents the low-energy limit
of the tree-level propagator corresponding to $W$-boson exchange 
(See Fig.\ \ref{fig:W_GF})
and $\sum r_i$ incorporates all higher-order electroweak interaction
corrections \cite{Awramik:2003rn}.
Note that Eqn.\ \ref{eqn:GFSM}  may be recast
in a form with only observable quantities and the higher-order corrections
\begin{equation}
M_W^2\left( 1- \frac{M_W^2}{M_Z^2} \right) = \frac{\pi
\alpha}{\sqrt{2} \GF} \left( 1 + \sum_i r_i\right)\ .
\label{eqn:SMtest}
\end{equation}
This relation between the Fermi constant $G_F$, 
fine structure constant $\alpha$, and weak boson masses 
$M_W$ and $M_Z$, is an important prediction of the Standard Model.

\subsection{Summary of previous measurements}

In 1999, at the time of the vRS publications, the
PDG world average of the muon lifetime 
was $\tau_{\mu} = 2.19703\pm0.00004$~$\mu$s \cite{Caso:1998tx}.
This average was largely determined 
by three measurements: Giovanetti {\it et al.}\ \cite{Giovanetti:1984yw}, 
Bardin {\it et al.}\ \cite{Bardin:1984ie} 
and Balandin {\it et al.}\ \cite{Balandin:1975fe}, 
all conducted more than twenty five years ago. 
The experiments all observed the decay positrons 
from stopped muons in various  targets.
 
The experiments of Giovanetti {\it et al.}\ and Balandin {\it et al.}\
both used a continuous beam. 
To avoid any mis-identification 
of parent muons with detected positrons, the beams 
were restricted to low rates,
thereby limiting the accumulated statistics 
of decay positrons.

The Bardin {\it et al.}\ experiment used 
a pulsed beam with the associated time structure
of the Saclay electron linac.
In principle, a pulsed beam technique can overcome the aforementioned rate limitations 
by first accumulating multiple muon stops during the beam-on period
and then detecting multiple decay positrons during the beam-off period.
In practice, the duty cycle at Saclay was not ideal and therefore limited 
the accumulated statistics of decay positrons.
 
Following the vRS publications, two new measurements
of the muon lifetime were initiated at the Paul Scherrer Institute.
The FAST experiment 
\cite{Barczyk:2007hp}
used a finely-segmented active target to identify
decay positrons with parent muons and thereby operate at
somewhat higher continuous beam rates. 
Herein we describe the MuLan experiment \cite{MuLan_proposal},
which made a part-per-million measurement of the 
positive muon lifetime using a customized time-structured muon beam.
Letters describing our 2004 commissioning measurement and 
2006-2007 production measurements were published
in Refs.\ \cite{Chitwood:2007pa,Webber:2010zf}.
\section{Relevant features of decays and interactions of muons}

\subsection{Muon decay.}

The known decay modes of positive muons are 
ordinary muon decay $\mu^+ \rightarrow e^+ \nu_e \bar{\nu}_\mu$,
radiative muon decay $\mu^+ \rightarrow e^+ \nu_e \bar{\nu}_\mu \gamma$ ($BR = 1.4\pm0.4 \times 10^{-2}$),
and the rare decay $\mu^+ \rightarrow e^+ e^+ e^- \nu_e \bar{\nu}_\mu $ ($BR = 3.4\pm0.4 \times 10^{-5}$).
In this experiment the neutrinos from muon decay were completely undetectable, 
the photons from radiative decay were barely detectable, and 
the detected positrons were overwhelmingly from ordinary decay.
Although the setup did not measure the energy of the positrons,
it did identify their direction.  

\begin{figure}
\includegraphics[width=0.9\linewidth]{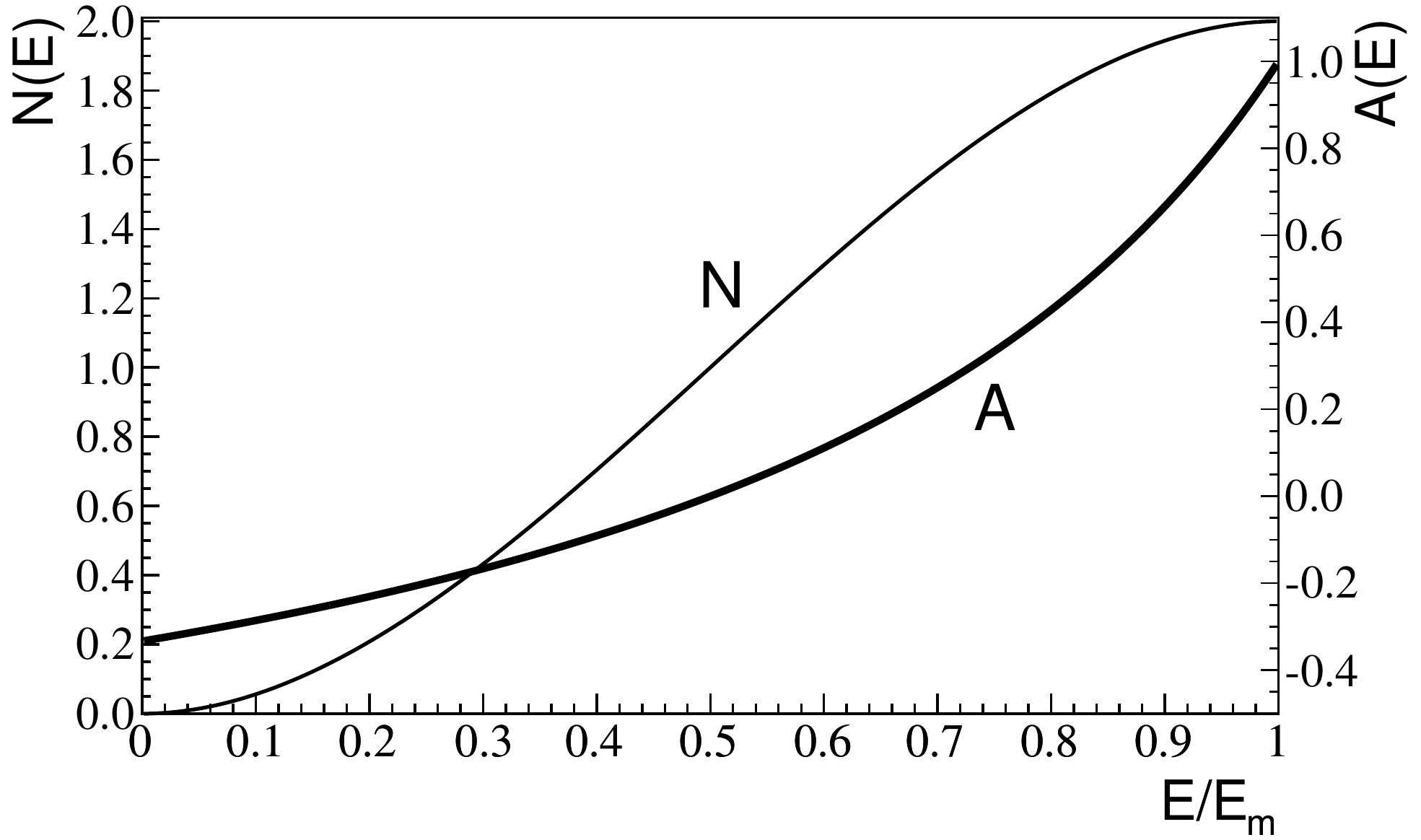}
\caption{The positron energy distribution $N(E)$
and energy-dependent positron asymmetry $A(E)$ in ordinary muon decay 
$\mu^+ \rightarrow e^+ \nu_e \bar{\nu}_\mu$. $E_m$ is the end-point energy 
of the ordinary decay spectrum.}
\label{fig:AsymmetryByEnergy}
\end{figure}

Due to parity violation in the weak interaction,
the positrons emitted in ordinary decay
are distributed asymmetrically relative to the muon spin axis.
The distribution is given by \cite{Commins:1983}
\begin{equation}
N ( \theta , E ) \propto 1 + A(E) ~ \cos{ \theta } ~,
\label{eq:differentialdecay}
\end{equation}
where $\theta$ is the angle between the muon spin
direction and the positron momentum direction and
$A(E)$ is the energy-dependent positron asymmetry.
As shown in Fig.~\ref{fig:AsymmetryByEnergy}, 
the asymmetry $A(E)$ varies with positron energy $E$ according to 
\begin{equation}
A ( E ) = ( 2 E - E_{m}) / ( 3 E_{m} - 2 E ) ~,
\label{eq:differentialdecay}
\end{equation}
\noindent
where $E_{m} = m_{\mu}/2$ is the end-point
of the positron energy spectrum. For high energy positrons
the asymmetry approaches $A = +1$ and positrons are preferentially emitted
in the muon spin direction. For low energy positrons
the asymmetry approaches $A = -1/3$ and positrons are preferentially emitted
opposite the muon spin direction.
When integrated over the theoretical energy 
distribution $N(E)$ of emitted positrons, the asymmetry is $A = +1/3$. 

\subsection{Muon slowing-down in condensed matter}
\label{sec:muslowing}

In the experiment a nearly 100\%-longitudinally 
polarized surface muon beam was stopped in various target materials.
The dynamics of a muon slowing-down in condensed matter 
has important series for the initial properties 
of the stopped muon ensemble in the target material.

On entering the target material the incident muons are slowed 
by ionization and excitation.
In the final  stages of slowing-down, 
the muons can undergo a series of electron pick-up 
and stripping reactions
and exist as either neutral muonium atoms or charged muon ``ions''.
Importantly, the formation of muonium can cause the rapid depolarization
of stopping muons via the large hyperfine interaction between the 
$\mu$/$e$ magnetic moments.

The initial depolarization and magnetic species\footnote{The initial depolarization
refers to depolarization during the slowing-down process.
The magnetic species refer to diamagnetic $\mu^+$ states 
and paramagnetic $\mu^+$$e^-$ states with comparatively small and  
large magnetic moments, respectively.}
of stopped muons are consequently dependent on the details 
of the muonium formation and its hydrogen-like chemical reactions
in the stopping material. 
Empirically, large populations of diamagnetic $\mu^+$ states
are generally associated with muon stops in metals.
By comparison muonium populations are less common,
although their existence is well established in such materials 
as ice, quartz, silicon, and germanium.
Any surviving polarization of stopped muon
or muonium populations is potentially worrisome for lifetime measurements.

\subsection{Muon spin rotation in condensed matter}\label{sec:muSR}

Following the slowing-down process the spin vectors 
of stopped muons will precess and relax in the local magnetic fields 
of the target material. This phenomenon is known as muon spin rotation ($\mu$SR).\footnote{Alternatively,
$\mu$SR can refer to muon spin rotation, relaxation or resonance.}
This time-dependent muon-ensemble polarization  yields, through the
angular correlation between the muon spin and the
positron direction, a time-dependent decay-positron
angular distribution.  In turn, this effect is 
seen---when detecting positrons in specific directions---as a 
geometry-dependent modulation 
of the exponential decay curve by the $\mu$SR signals.

In discussing muon spin rotation it is helpful to distinguish
the cases of $\mu$SR in a transverse magnetic field;
{\it i.e.}, a magnetic field perpendicular to the polarization axis, 
and  $\mu$SR  in a longitudinal magnetic field;
{\it i.e.}, a field parallel to the polarization axis.
Of course, in general, both effects will contribute
to the time distributions of the decay positrons.

\subsubsection{Transverse field (TF) $\mu$SR}\label{sec:muSR:TF}

A transverse field $\vec{B}$ causes the muon/muonium magnetic moment $\vec{\mu}$ to 
undergo Larmor precession in the plane perpendicular to the field axis.
The precession  frequency is $\omega = \mu B / 2 m c$ 
where $m$ represents the muon/muonium mass. It yields frequencies of 
13.6~kHz per Gauss for free diamagnetic muons and
1.39~MHz per Gauss for free paramagnetic muonium.\footnote{The muonium 
ground state is a two-state system consisting of the $F = 1$
triplet state and the $F = 0$ singlet state.
The precession frequency of triplet atoms is  1.39~MHz per Gauss.
Singlet atoms are rapidly depolarized by hyperfine oscillations 
and therefore unobservable by $\mu$SR techniques.}

In matter, the ensemble-averaged polarization will generally relax.
One source of relaxation is the spin dephasing 
of the individual muons in the differing local $B$-fields 
at their stopping locations in the target material. 
Another source is the spin interactions
between the stopped muons and the surrounding electrons 
and crystal lattice. 

\begin{figure}
\includegraphics[width=0.85\linewidth]{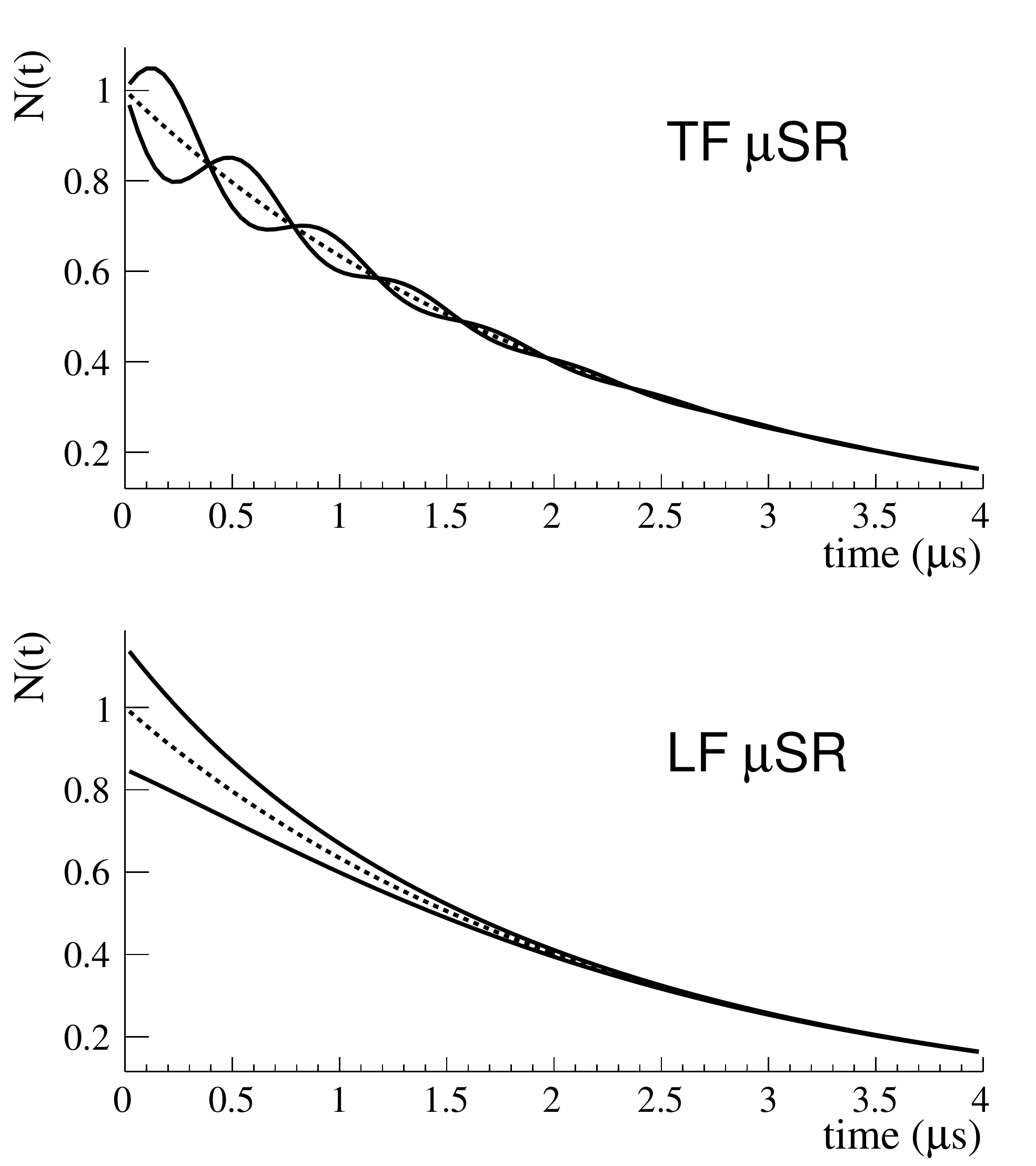}
\caption{Depiction of typical time spectra from TF $\mu$SR (upper panel) and 
LR $\mu$SR (lower panel). The paired-solid lines represent the time distributions 
of outgoing positrons in geometrically opposite detectors. The single-dashed line
represents a positron distribution without $\mu$SR effects. The TF $\mu$SR spectrum 
shows both spin precession and spin relaxation  whereas the LF $\mu$SR spectrum
shows a relaxation signal but no precession signal. Note, these figures greatly exaggerate
the observed effects of $\mu$SR in this experiment.}
\label{fig:muSRcartoon}
\end{figure}

For positrons detected at an angle $\theta_B$ relative
to the $B$-field axis, the effects of TF $\mu$SR yield 
the time distribution depicted in Fig.\ \ref{fig:muSRcartoon}.
A common, semi-empirical representation 
of the time dependence of the TF $\mu$SR signal is
\begin{equation}
N(t) \propto ( 1 + A_2 ( \theta_B )~ e^{-t/T_2} ~ \sin{( \omega t + \phi ) } ) ~
e^{ -t / \tau_{\mu} } ,
\label{eq:N_TF}
\end{equation}
where $A_2 ( \theta_B )$ is the amplitude 
of the $\mu$SR signal, $\omega$ and $\phi$ are the 
frequency and the phase of the precession, and $T_2$ 
is the time constant of the relaxation.
Note the geometry-dependent amplitude $A_2 ( \theta_B )$ 
is maximum for positrons emitted perpendicular to the $B$-field axis
and zero for positrons emitted parallel to the $B$-field axis.
Also---as shown in Fig.\ \ref{fig:muSRcartoon}---the  TF $\mu$SR signals of positrons emitted in opposite directions 
have opposite phases.

\subsubsection{Longitudinal field (LF) $\mu$SR}

Because the torque $\vec{\tau} = \vec{\mu} \times \vec{B}$ on the muon spin
vanishes for a longitudinal $B$-field, LF $\mu$SR exhibits 
no precession signal.
However, in matter the polarization will 
relax via the spin interactions with the surrounding electrons
and the crystal lattice. 
Because relaxation via spin dephasing in differing $B$-fields is absent, 
the time constants for LF-relaxation are 
generally longer than TF-relaxation.

For positrons detected at an angle $\theta_B$ relative
to the $B$-field axis, the effects of LF $\mu$SR 
yield the time distribution depicted in Fig.\ \ref{fig:muSRcartoon}.
A common, semi-empirical representation 
of the time dependence of the LF $\mu$SR signal is
\begin{equation}
N(t) \propto  ( 1 \pm A_1 ( \theta_B ) ~ e^{ -t/T_1 } ) ~ e^{ -t / \tau_{\mu} } ,
\label{e:LFmuSR}
\end{equation}
where $A_1 ( \theta_B )$ and $T_1$ are the amplitude and the time constant 
of the longitudinal relaxation.
Note the geometry-dependent amplitude $A_1 ( \theta_B )$
is maximum for positrons emitted along  the $B$-field axis
and zero for positrons emitted perpendicular to the $B$-field axis 
({\it i.e.}\ the reverse of the TF-case).
Again---as shown in Fig.\ \ref{fig:muSRcartoon}---the LF $\mu$SR signals of positrons 
emitted in opposite directions 
have opposite signs.

\section{Experimental setup}\label{sec:Setup}

The experiment was conducted at Paul Scherrer Institute 
(PSI) in Villigen, Switzerland, using a 
nearly 100\% longitudinally polarized, surface $\mu^+$ beam.
Incoming muons were stopped in a solid target 
and outgoing positrons were detected by a
finely-segmented, large-acceptance, scintillator array
instrumented with fast-sampling waveform digitizers.

One important feature of the experimental setup was the 
time structure of the muon beam.
This time structure or ``fill cycle'' consisted of a 5~$\mu$s-long, 
beam-on accumulation period followed by
22~$\mu$s-long, beam-off measurement period.
The time structure was important in avoiding 
the need to associate decay positron with parent muons, 
a requirement that limited earlier experiments 
using continuous muon beams. Rather, 
the time of the decay positrons are associated
with the beam-off transition.

Another important feature was a large reduction of the 
muon ensemble polarization via the spin precession during the accumulation period. 
Since muons arrive randomly during accumulation,
their spins precess through different angles
under the influence of the local $B$-field,
thus dephasing the individual spins 
and reducing the ensemble polarization.

Additionally, the positron detector comprised
a symmetric array of target-centered, geometrically-opposite, detector pairs.
As $\mu$SR signals have equal magnitudes but opposite signs 
in opposite detectors (See Sec.\ \ref{sec:muSR}), any remaining effects 
from $\mu$SR signals are largely canceled by detector symmetry.

\subsection{Muon beam}\label{sec:Beam}

The experiment used the $\pi$E3 beamline
at the laboratory's 590~MeV proton cyclotron. The
1.4-1.8~mA proton beam provided 
the intense secondary muon beam required for this high statistics experiment. The
proton beam has a continuous macroscopic time structure consisting
of roughly 1~ns duration proton pulses at a 50.63~MHz repetition rate. 

The $\pi$E3 secondary beamline viewed the  E-station production target
in the primary proton beamline. The station comprised
a 40~or 60~mm-long graphite target, which was rotated at roughly 1~Hz to prevent damage by beam heating.
The rotation resulted in a slight modulation of the muon intensity.

The surface muons originate from at-rest $\pi^+ \rightarrow \mu^+ \nu_{\mu}$ decays
in the outer layer of the production target. The decays yield
a back-to-back neutrino and positive muon with momenta 29.8~MeV/c.
As a consequence of parity violation in the weak interaction,
the surface muons were nearly 100\% longitudinally polarized, 
with their spins anti-aligned to their momenta.

\begin{figure*}[htbp]
\begin{center}
\includegraphics[width=0.8\linewidth]{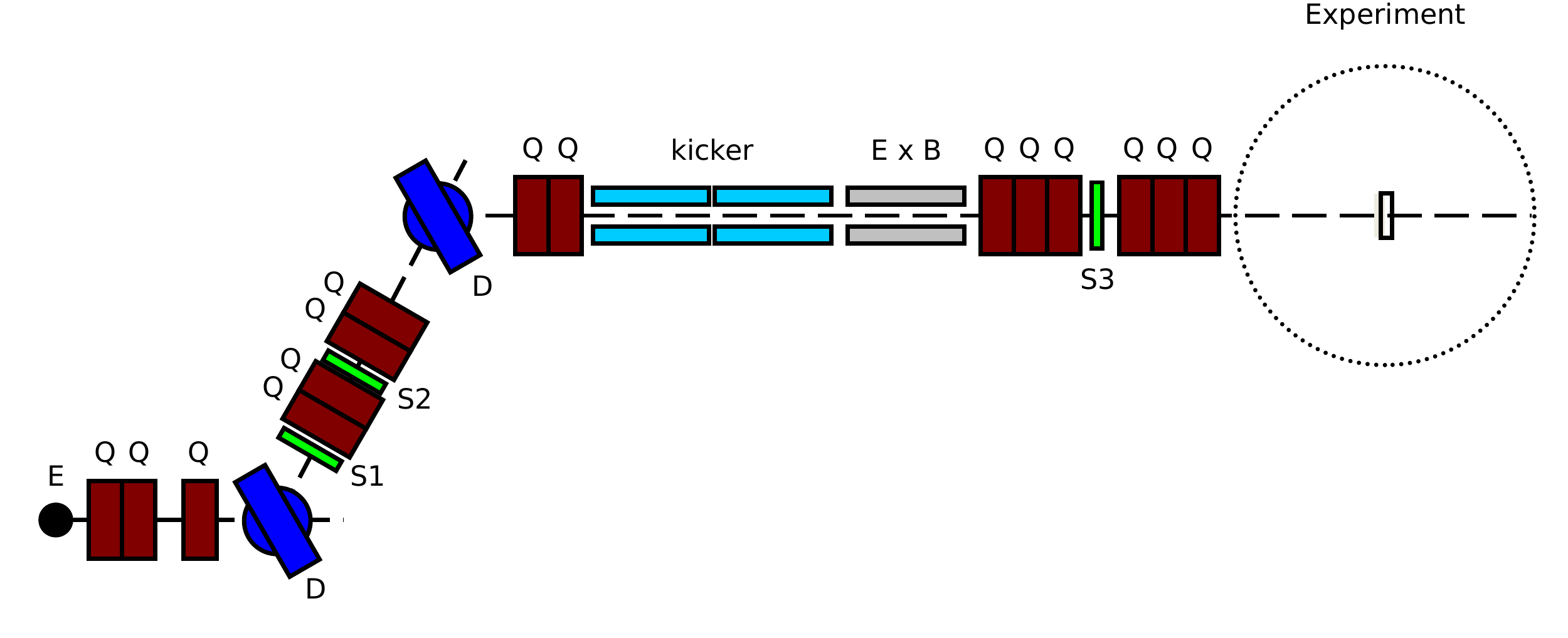}
\end{center}
\caption{Schematic view of the $\pi$E3 beamline including the
E target station, opposing vertical bending magnets (D),
quadrupole focusing magnets (denoted Q), slit systems (S), 
electrostatic kicker, $\vec{E} \times \vec{B}$ velocity separator,
and the location of the experiment. See text for details.}
\label{fig:beamline}
\end{figure*}

A schematic of the $\pi$E3 beamline elements from the E target station
to the muon stopping target is shown in Fig.~\ref{fig:beamline}.
The beamline includes a fixed-element section, which views the E target station at 90~degrees
to the through-going proton beam.  Surface muons produced by pion decays were collected 
by a magnetic lens system and momentum-selected through a 60-degree, vertically oriented, dipole.  
Additional quadrupoles and two horizontal and vertical slit systems provide momentum- 
and flux-limiting apertures.  A second opposing dipole and a quadrupole doublet were tuned 
to redirect the beam parallel to a raised experimental floor, which is roughly 6~m above 
the proton beam.  

The customized beamline elements that follow transport muons without deflection through 
the uncharged electrostatic kicker and next through an $\vec{E}\times\vec{B}$ velocity separator, 
which removes unwanted positrons.  The beam is then focused by a quadrupole triplet through 
an $x$-$y$ slit system and then refocused by a second triplet at the stopping target. 
The three slit systems were adjusted to optimize the combination of beam-on intensity, 
momentum acceptance, and beam-off extinction.  

\subsubsection{Electrostatic kicker}\label{sec:Kicker}

A time structure was imposed on the continuous beam using a
custom-built, fast switching, electrostatic kicker~\cite{Barnes20041932,Armenta:2005eu}.
With  the kicker high voltage off, the muons were transported straight
along the beamline axis.
With the kicker high voltage on, the muons were displaced vertically by 46~mm 
and deflected downward by 45~mrad at the kicker exit.
They were then focused onto the bottom edge of the third slit system.

The kicker system includes two pairs of 75~cm-long by 20~cm-wide aluminum plates
that were aligned in series with a spacing of 5~cm.
The plates were positioned symmetrically above and below the beamline axis
with a 12~cm gap.
Collimators with apertures of 12.5$\times$12~cm$^2$
were located both upstream and downstream to prevent particles
reaching regions either above or below the plate gap.
The plates were housed in a 60-cm diameter, 2.0-m long, cylindrical vacuum tank
with high-voltage, vacuum feed-throughs.

When energized, a potential difference of 25~kV was applied between
the top plates and the bottom plates ($+12.5$ kV was applied to the two upper plates
and $-12.5$~kV was applied to the two lower plates with
a virtual ground at the kicker mid-plane).
The switching of plate voltages between ground potential and $\pm$12.5 kV (and vice versa) 
was accomplished using series circuits (``stacks'') of seventeen fast-transition, high-power,
MOSFETs with each MOSFET switching 735~V.
Pairs of MOSFET stacks were operated together in push-pull mode
with one stack pushing the plate voltage to $\pm 12.5$~kV and the other stack
pulling the plate voltage to ground potential.

The kicker was typically operated at a cycle frequency of about 37~kHz
with a 5~$\mu$s beam-on period (plates grounded) and a
22~$\mu$s beam-off period (plates energized).
The transition time between ground and $\pm$12.5~kV
was 67~ns \cite{Barnes20041932,Armenta:2005eu}.
Measurements performed during 2006 running indicated 
the $\pm$12.5~kV plate high voltage was stable to better than 300~mV.
Improved measurements during 2007 running indicated a 
stability of better than 150~mV. For further details see Sec.\ \ref{sec:beam stability}.

\subsubsection{Beam monitor}\label{sec:EMC}

A planar, high-rate, multi-wire proportional chamber 
with associated readout electronics
was used for both extensive beam studies prior to
data taking and for periodic beam monitoring during the data taking.
For the beam studies, the chamber was positioned at the stopping target location
with the positron detector rolled to a downstream location.
For the beam monitoring, the chamber was positioned immediately after the
150~$\mu$m-mylar beampipe window at the downstream end of the positron detector.

The beam monitor was designed to measure the $x$-$y$ muon profile
during the high instantaneous rates of the accumulation period
and the low instantaneous rates of the measurement period.
The chamber was assembled from two perpendicular anode wire planes interleaved between
three high-voltage cathode planes. Each anode plane consisted of 96 tungsten anode wires
of 20~$\mu$m diameter and 1.0~mm wire spacing and each cathode plane consisted
of 12.5~$\mu$m aluminized Mylar foils held at 3.0~kV.
A ``fast'' chamber gas of 30\% isobutane and 70\% tetrafluoromethane was used to
handle the high rates. The beam monitor was about
98\% efficient for surface muons and about 8\% efficient for beam positrons.

Adjacent pairs of anode wires were connected together into 2~mm-spaced wire-pairs
and read out by six, 16-channel, amplifier-discriminator boards.
The resulting logic signals were recorded by a custom-built, field programmable gate array (FPGA)
that combined individual hits into $x-y$ positions and time stamps of detected particles
(during the accumulation period the data was pre-scaled).
The FPGA data was read out through a Struck Innovative Systeme GmbH SIS3600 32-bit VME event latch \cite{Struck3600}
by the data acquisition system.

\subsubsection{Beam characteristics}\label{sec:Beamline_summary}

\begin{figure}
\begin{center}
\includegraphics[width=\linewidth]{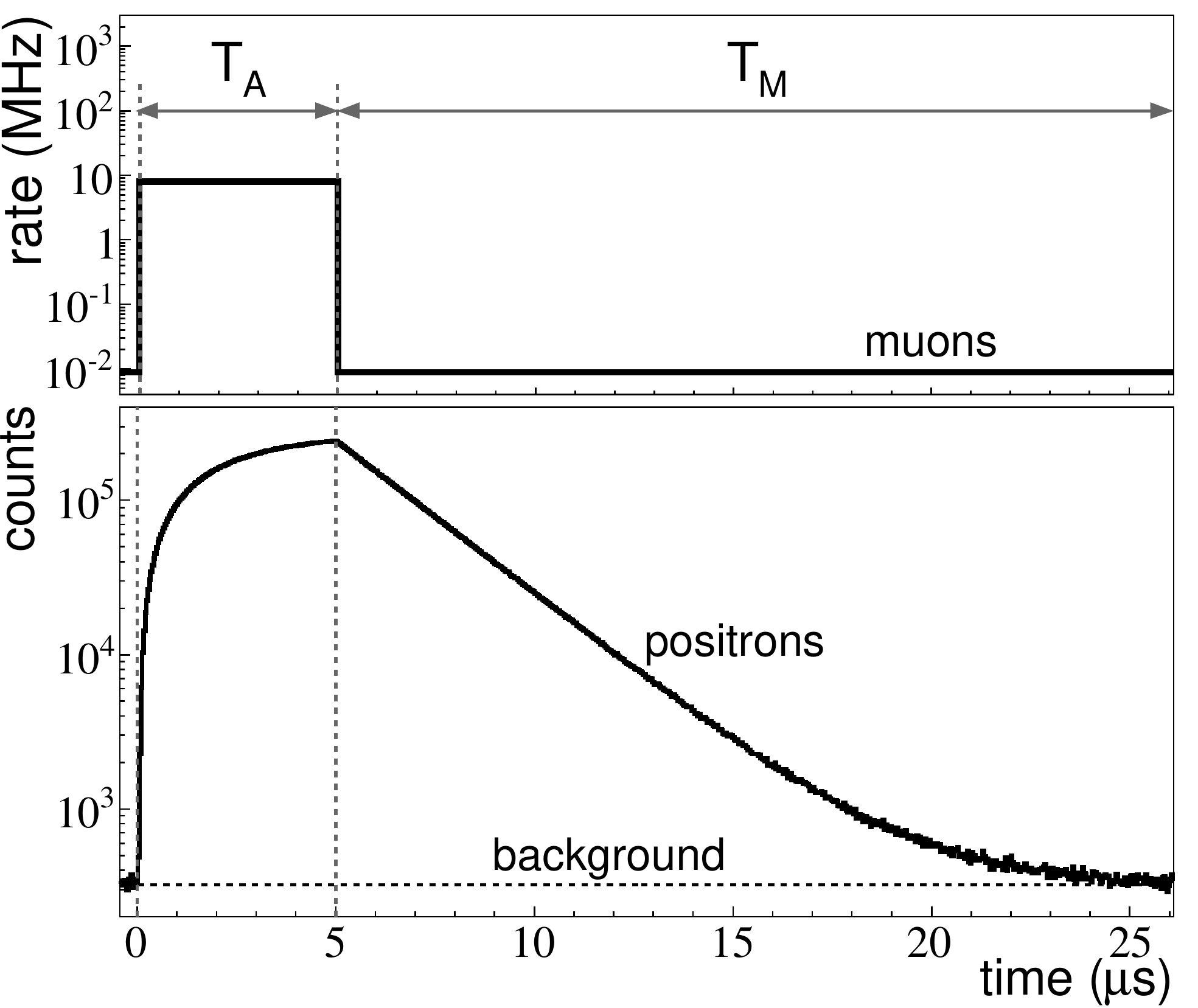}
\end{center}
\caption{Plot of the muon arrival times (upper panel) and decay positron times (lower panel)
that are produced by the pulsed beam technique. The fill cycle comprises
the $T_A = 5$~$\mu$s beam-on accumulation period 
and the $T_M = 22$~$\mu$s beam-off measurement period.}
\label{fig:fill_structure}
\end{figure}

The beam time structure is depicted in Fig.~\ref{fig:fill_structure}.
The upper panel shows the muon rate over the fill cycle and indicates 
the 5~$\mu$s~beam-on accumulation period, 22.0~$\mu$s beam-off measurement period,
and the fast transitions between the beam-on/off states.
The lower panel shows the corresponding positron rate over the fill cycle,
which increases monotonically during the accumulation period and 
decreases exponentially during the measurement period.

The muon beam was operated at a central momentum of 28.8~MeV/c
and a momentum bite of $\sim$2.6\% ($\sigma$).
During the beam-on accumulation period, the instantaneous flux of
surface muons at the stopping target was about 8~MHz %
in a spot size approximately 21$\times$10~mm$^2$.
This rate yielded about 40~stopped muons per accumulation period
and 15 detected positrons per measurement period.
The surface muons arrive at the stopping target with their polarization tilted
at a $6^\circ$-angle below the beamline axis. This tilt arises from 
spin precession in the
transverse $B$-field of the velocity separator.

During the beam-off measurement period the instantaneous rate of
surface muons was reduced by an ``extinction'' factor $\epsilon \sim 900$
and the spot size was increased to about 43$\times$32~mm$^2$.
   
\subsection{Target arrangement}\label{sec:Targets}

For the 2006/07 data taking the $\pi$E3 beamline 
was extended through the positron detector with the 
stopping target mounted in the vacuum pipe.
The in-vacuum target reduced the number of upstream muon stops
compared to our 2004 commissioning experiment \cite{Chitwood:2007pa}.
These stops were worrisome as their $\mu$SR signals 
are not canceled by the opposite detector sums. Such distortions were 
identified in Ref.\ \cite{Chitwood:2007pa}.

The beamline extension necessitated a constriction
of the vacuum pipe from 35.5-cm diameter to
20.2-cm diameter at a location 52~cm upstream of the stopping
target. A hinge enabled the target to be swung 
into the beam path for production running 
and out of the beam path for beam monitoring.
A collar enabled the stopping target to be rotated azimuthally 
about the beam axis.

Two combinations of stopping target and transverse $B$-fields 
were employed to maximize the spin dephasing of stopped muons. The first strategy used 
the large internal $B$-field of a magnetized ferromagnetic alloy 
target (Arnokrome-III) to rapidly  
dephase its population of diamagnetic muons.
The second strategy used a moderate external $B$-field
with a non-magnetic quartz crystal (\nuc{}{Si}\nuc{}{O}$_2$)
to rapidly dephase its population of paramagnetic muonium atoms.

To further reduce the dangers of upstream stops,
the beampipe wall was also lined with Arnokrome-III ferromagnetic foil.
The 0.3~mm-thick lining was arranged with its magnetization perpendicular to the beamline axis
and was extended 67~cm upstream of the stopping target.
The lining dephases the ensemble-averaged polarization of upstream stops.

\subsubsection{Arnokrome-III target setup}\label{sec:AK3}

The Arnokrome-III (AK-3) stopping target was used
in the 2006 production run.
AK-3 is a ferromagnetic alloy consisting of 
about $30\%$ \nuc{}{Cr}, $10\%$ \nuc{}{Co} and $60\%$ \nuc{}{Fe} 
and manufactured by Arnold Engineering Co.\ \cite{AK3}.
The disk-shaped AK-3 target had a diameter of 200~mm and thickness of 0.5~mm 
and was mounted on the hinged frame in the vacuum pipe at the geometrical
center of the positron detector. 
A SRIM~\cite{Ziegler20041027} calculation of the AK-3 target stopping power 
for the incident 28.8~MeV/c $\mu^+$ beam  
gave a mean range of 0.17~mm and a range straggling of 0.02~mm. 

The AK-3 target was magnetized in the plane of the disk. The internal 
$B$-field was approximately 0.4~T and oriented for production
running with the magnetization axis to the beam left or the beam right
({\it i.e.}\ at 90 degrees to the muon polarization).
In the $\sim$0.4~T transverse $B$-field the diamagnetic $\mu^+$'s 
have a $\sim$50~MHz precession frequency. The resulting dephasing 
during the accumulation period reduced the ensemble-averaged transverse polarization 
by roughly a factor of 700.

For further details on $\mu$SR effects in Arnokrome-III see Sec.\ \ref{supplemental muSR} 
and Ref.\ \cite{Morenzoni}.

\subsubsection{Quartz target setup}\label{sec:Quartz}

The quartz (SiO$_2$) stopping target was used
in the 2007 production run.
The target was a disk-shaped, artificially grown, single crystal 
with a 130-mm diameter, 2.0-mm thickness that was oriented with its
$Z$-crystallographic axis perpendicular to the disk face.
The crystal was purchased 
from Boston Piezo-Optics Inc.\ \cite{Quartz} and grown by 
so-called seed-free technology to minimize crystal imperfections
that possibly lead to reduced formation of muonium atoms.
A SRIM~\cite{Ziegler20041027} calculation of the quartz target stopping power
for the 28.8~MeV/c muon beam gave a mean range of 0.53~mm and a range straggling of 0.03~mm. 

The 130-mm diameter quartz disk was the largest diameter single crystal that was commercially available. 
The remaining region between the 130-mm quartz disk outer diameter and the 200-mm beam pipe inner diameter 
was therefore covered by a magnetized AK-3 annulus to guarantee depolarization of off-axis muon stops.

A 130-G transverse $B$ field was used for the spin precession of  
the muon stops in the quartz target.  
The field was produced by a Halbach arrangement \cite{Halbach:1981}
of twenty four, 2.5$\times$2.5$\times$2.5~cm$^3$, neodymium magnets
mounted in an aluminum ring of inner radius 14~cm, outer radius 19~cm
and thickness 7.5~mm. 

Note the magnet was mounted to the detector assembly rather than the beam pipe. The mounting scheme
allowed for both its rotation about the beam axis and its inclination relative to
the vertical axis. The rotation degree-of-freedom enabled the direction
of the transverse $B$ field to be varied and the inclination degree-of-freedom 
enabled the introduction of a longitudinal $B$ field at the target. 
This setup required a careful alignment to ensure the muon polarization axis 
and magnetic field axis were mutually perpendicular during production running. 

In the 130-G transverse magnetic field the precession frequency of muonium atoms
was roughly 180~MHz and that of positive muons was roughly 1.8~MHz. The resulting dephasing during the
accumulation period reduced the ensemble-averaged transverse polarization by 
roughly a factor of 1000 for the large paramagnetic muonium population
and 25 for the diamagnetic $\mu^+$ population.

For further details on $\mu$SR effects in crystal quartz 
see Sec.\ \ref{supplemental muSR}
and Refs.\ \cite{Dawson:1994,Brewer2000425}.

\subsection{Positron Detector}\label{sec:MuLan_ball}

\begin{figure}
\begin{center}
\includegraphics[width=\linewidth]{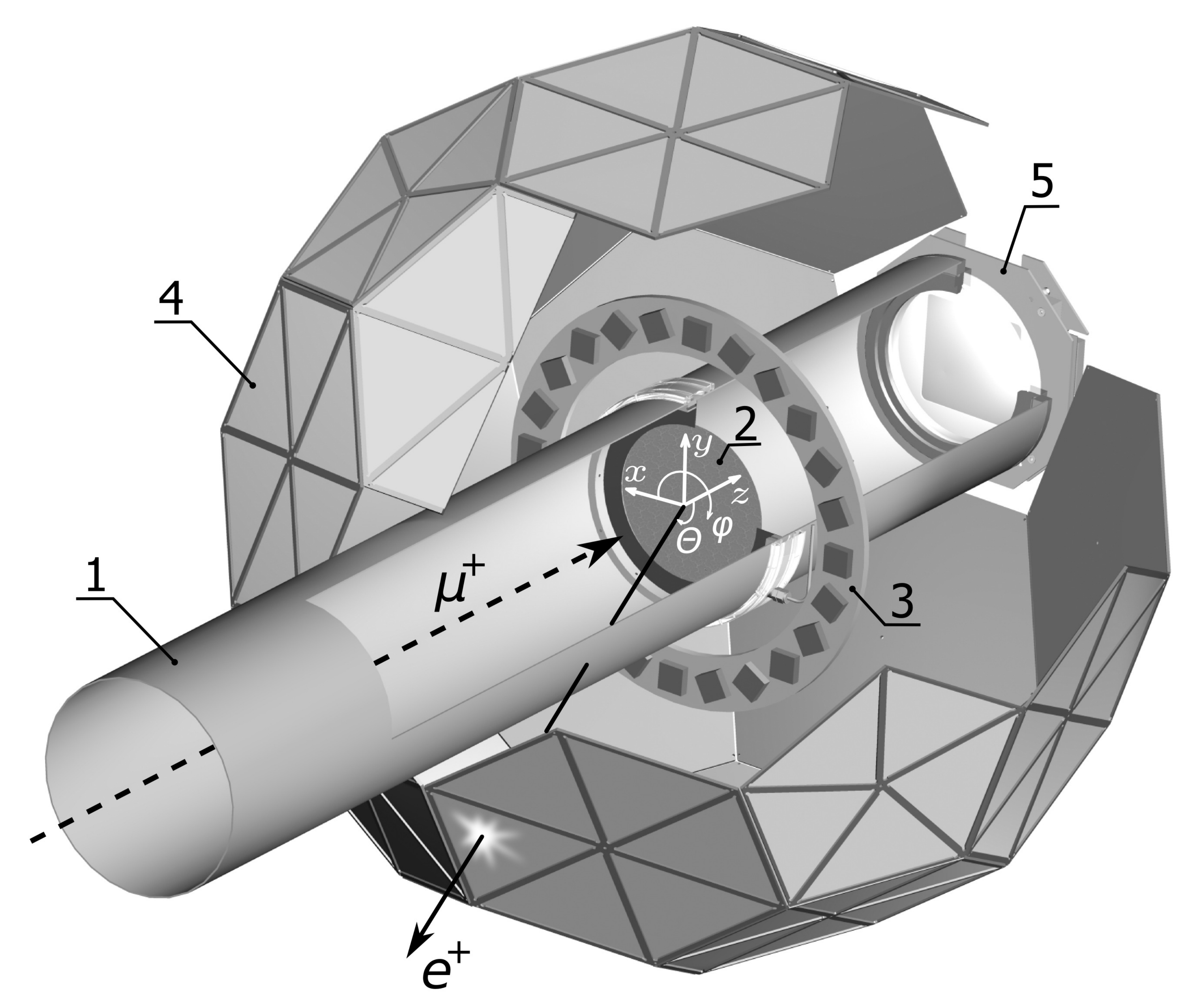}
\end{center}
\caption{Cutaway drawing of the experimental setup showing the vacuum beampipe (1), 
stopping target (2), Halbach magnet (3), scintillator array (4)
and the beam monitor (5) at the downstream window of the beam line.
Note that the Halbach magnet was not installed for the AK-3 data taking
and the stopping target was rotated out of the
beam path for the beam monitoring.}
\label{fig:MuLan_ball}
\end{figure}

The positron detector is depicted in Fig.~\ref{fig:MuLan_ball}. The detector was 
a fast-timing, finely-segmented, large-acceptance, plastic scintillator array that was
geometrically centered on the stopping target. The detector enabled an inclusive measurement
of positron time and angular distributions from the muon decays in the stopping target.
The detector segmentation was important in minimizing positron pileup
and the detector geometry was important in handling $\mu$SR distortions.

The detector was constructed of 170 triangle-shaped scintillator pairs  
arranged in a truncated icosahedral (soccer ball) geometry.
Each pair comprised an inner plastic scintillator tile 
and outer plastic scintillator tile. The pairs were grouped into ten
pentagonal  enclosures (pent-houses) containing
five tile-pairs and twenty hexagonal enclosures (hex-houses) containing 
six tile-pairs; the ten pent-houses and twenty hex-houses together forming the detector geometry. 
The $\pi$E3 beamline traversed the positron detector 
through one upstream pentagonal hole and  
one downstream pentagonal hole in the detector array. 

For later reference a coordinate 
system with the z-axis parallel to the beam axis
and the x(y)-axes in the horizontal (vertical) directions
perpendicular to the beam axis is introduced.
The location of detectors are specified by
their polar angle $\theta$ measured from the z-axis
and their azimuthal angle $\phi$ measured in the xy-plane
(See Fig.\ \ref{fig:MuLan_ball}).
The detector arrangement yielded pairs of
geometrically-opposite tiles that viewed the target
with coordinates $( \theta , \phi )$ and 
$( 180^{\circ} - \theta , \phi + 180^{\circ} )$.

The triangular tiles were 3-mm thick BC-404 scintillator
that produced light signals with 0.7~ns rise times 
and 1.8~ns fall times \cite{BC404}.
The tiles in hex-houses were equilateral triangles
with base lengths of 15.0~cm for inner tiles and 14.0~cm for outer tiles.
The tiles in pent-houses were isosceles triangles
with base lengths of 15.0~cm (13.8~cm) and heights of 10.0~cm (9.5~cm) for inner (outer) tiles.
The scintillator array had an inner radius of approximately 40.5~cm
and subtended a solid angle of approximately 70\% of 4$\pi$ steradians.

Two edges of each tile were covered with reflective tape
while the remaining edge was glued to an adiabatic 
lightguide followed by a 29-mm diameter photomultiplier tube (PMT).
The light guides and PMTs---not shown 
in Fig.~\ref{fig:MuLan_ball}---emerged radially from the geometrical center 
of the detector array.   
Two types of photomultipiers were used, 
with 296 tiles (eighteen hex-houses and eight pent-houses) 
instrumented by Photonis XP 2982 PMTs \cite{Photonis}
and 44 tiles (two hex-houses and two pent-houses) instrumented 
by Electron Tube 9143 PMTs \cite{Electron}.  
The same brands of 29-mm PMTs were used to read out geometrically-opposite detector pairs.
A clip-line was used to shorten the durations of the pulses from the photomulipliers
to full-widths at 20\% maximum of 10~ns.   

An important requirement was the accurate centering of the detector array on the stopping distribution.
The detector was supported by a steel platform with
an arrangement of vertical posts and horizontal rails 
enabling the fine adjustment of the detector horizontal 
and vertical coordinates.
Using this arrangement, the detector array and stopping distribution
were centered to better than  \unit[2]{mm} in each direction.

\subsubsection{Detector operation}

On average each minimally-ionizing positron (MIP) yielded a photomultiplier signal of $\sim$70 photo-electrons.
The initial settings of the high voltages for each photomultiplier
were performed by adjusting the MIP peak 
to correspond to one half of the 1.0~V dynamic range of the readout electronics.
During data taking the gain stability was continuously monitored
via the amplitude of the MIP peak and showed typical drifts 
of $\sim$2~mV per day. 
To compensate for long-timescale gain drifts, the PMT high voltages were adjusted 
on several occasions during each running period.\footnote{Slow gain changes
over the data taking period do not interfere with the positron time distribution
during the measurement period.}

The detector was designed 
to minimize vulnerability to gain changes.
The vast majority of detected positrons 
were minimally ionizing particles with well-defined energy loss
in scintillator tiles.
By constructing the detector with
two layers of scintillator tiles, and designing the detector
for efficient collection of scintillation light, 
the resulting coincident pulses from through-going positrons 
were well separated from background noise.

\begin{figure}
\begin{center}
\includegraphics[width=0.7\linewidth]{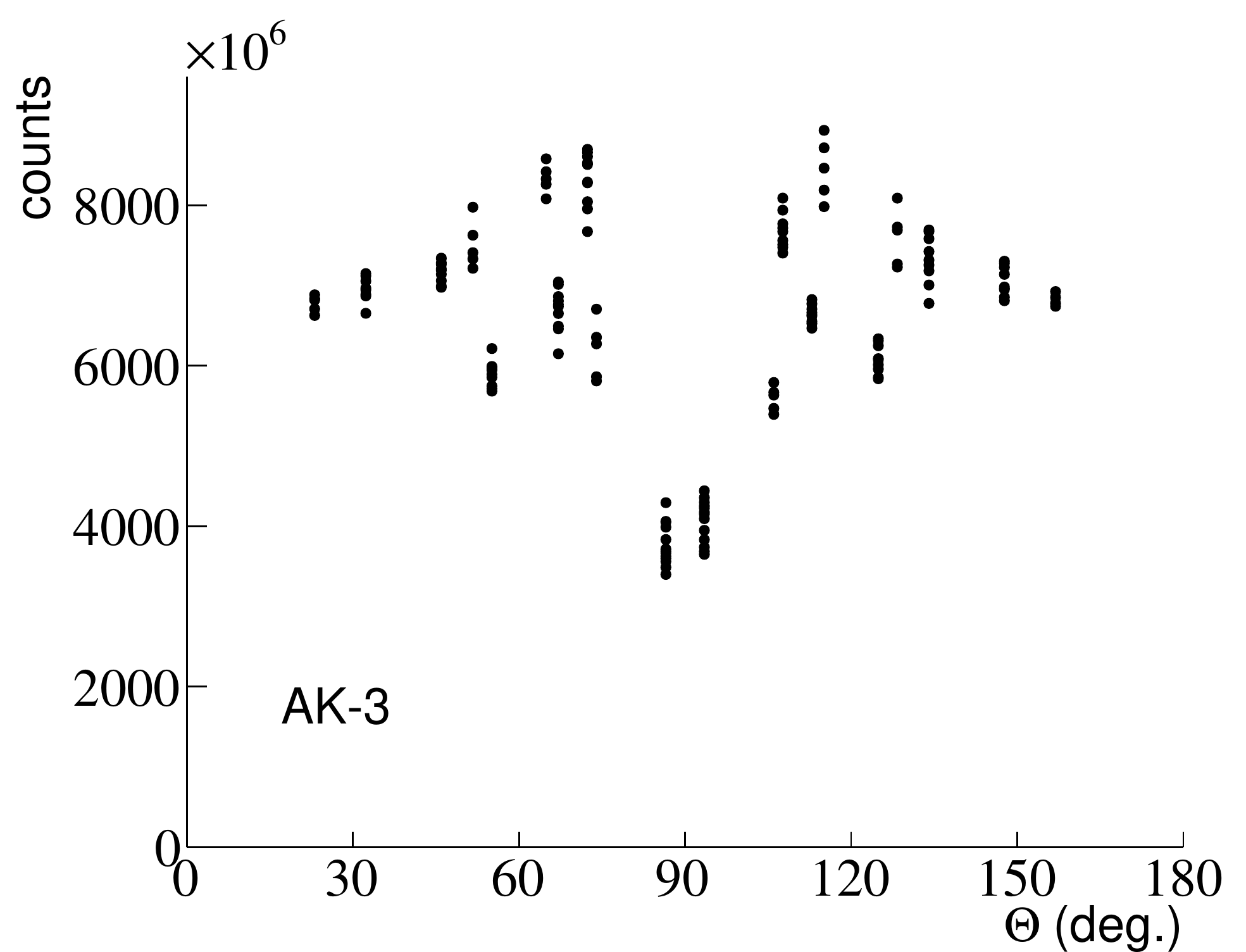}\\
\includegraphics[width=0.7\linewidth]{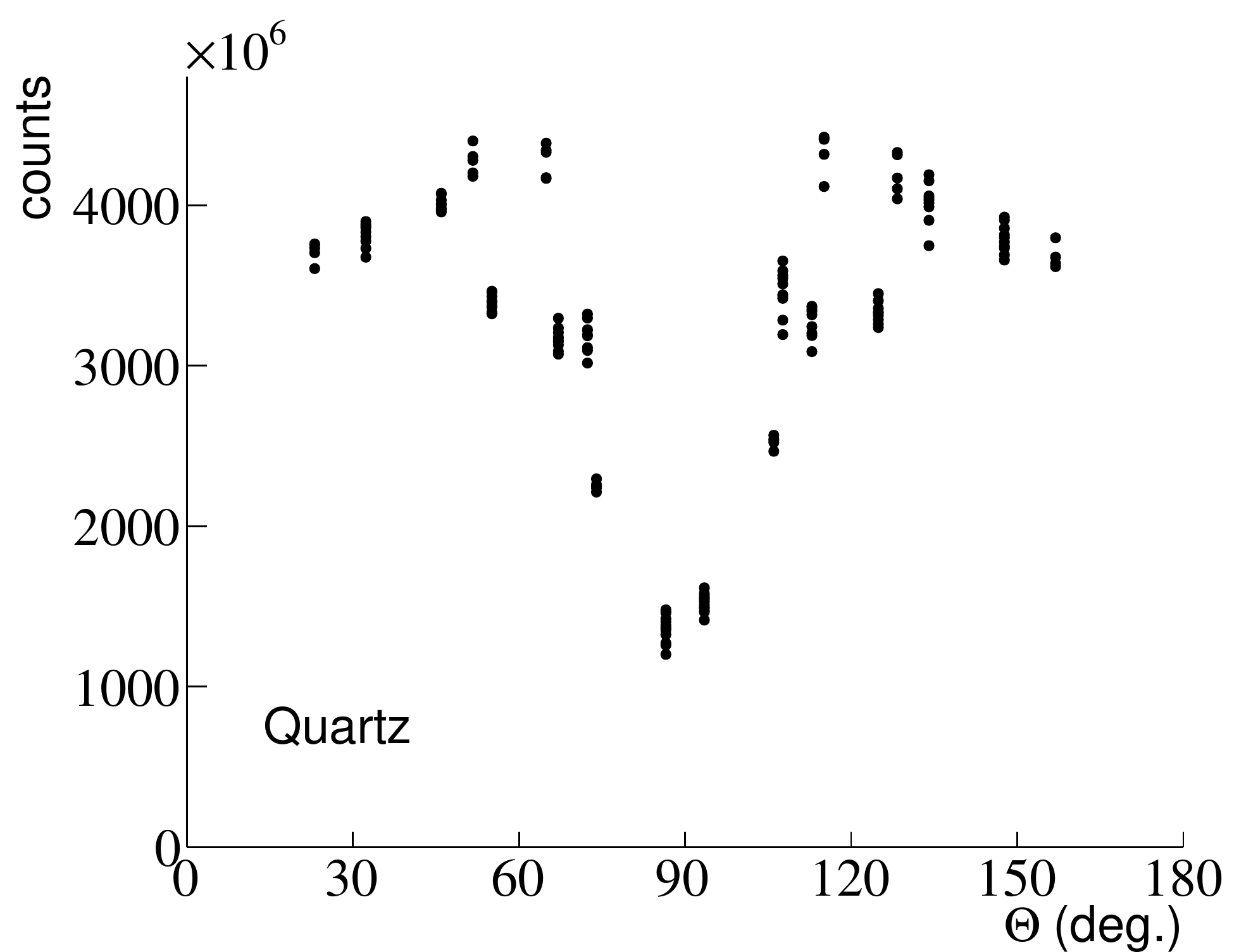}
\end{center}
\caption{Hit distribution for the 170 tile pairs versus 
the angle $\theta$ relative 
to the beamline axis
for the entire AK-3/quartz datasets. 
The upper panel is the 
hit distribution for the AK-3 setup without the Halbach magnet and
the lower panel is the hit distribution for the quartz setup 
with the Halbach magnet.
The vertical groupings with common $\theta$-values  
are tile pairs with different values of angle $\phi$.}
\label{fig:hit distribution}
\end{figure}

Fig.\ \ref{fig:hit distribution} shows the distribution of hits versus the angle $\theta$ 
relative to the beamline axis. The dip at 90 degrees
is a consequence of the shadow of the target disk, target flange
and Halbach magnet (in quartz data taking). The increase
in counts either side of 90 degrees is attributed to scattering
in the same components.

\subsection{Laser system}\label{sec:Laser}

A laser system was used to monitor the time stability of
the positron detector during the measurement period.
The LN203 nitrogen laser \cite{Laser_Photonics} generated  337~nm wavelength, 
600-ps duration, UV pulses that were split and then 
distributed via a fiberoptic cable to 24 scintillator tiles 
and one reference photomultiplier.
The reference PMT was situated outside the experimental area
in a location that was well-shielded from the beam-related radiation
and the kicker-related electrical noise.
The PMT pulse shapes corresponding to decay positrons 
and laser signals were similar.
 
In normal operations the laser system
was run asynchronously with the fill cycle 
at a 33~Hz repetition rate. Typically the system 
was operated for one hour each 8-hour shift.

\subsection{Waveform digitizers}\label{sec:WFD}

The PMT signals from scintillator tiles were read out via
340 channels of 450~MHz sampling-rate, 8-bit resolution, 
VME-based waveform digitizers (WFDs).
Data from the custom-built WFDs was used to determine the times and amplitudes 
of the tile pulses. The latter were also used in the detailed analysis of pulse pileup 
and gain changes.

The WFD modules were single-slot VME-64 boards. Each module
had four analog channels with 0-1~V analog input ranges,
one  enable input, and one clock input. The four channels
of each module were used to digitize the signals
from one inner-outer tile pair and its geometrically-opposite tile pair.
Each enable input received a logic signal synchronized 
to the measurement period and each clock input received 
a 450~MHz sinusoidal signal 
distributed by the clock system.

The complete setup involved 85 WFD modules in six VME crates. The crates
were located in racks on the detector platform in the experimental area.
Approximately 12~meter-long, 50~$\Omega$ cables connected 
the PMT outputs to the WFD inputs.

Each analog input was continuously digitized by 
independent analog-to-digital converters (ADCs) 
that sample at the external clock frequency.
Two field programmable gate arrays (FPGAs) controlled 
the data flow from the four sampling ADCs to their
associated 512~kB, FIFO memories.
When the enable input was present
and the analog input exceeded a threshold,
a time stamp, fill cycle number, and  contiguous block of 24~ADC 
samples were written to memory.
The digitized data in the WFD memories were then read out 
via the VME backplane.
The FIFO-design permitted simultaneous read and write access 
to digitizer memory.

The threshold trigger that initiated storage of ADC samples  
was derived from a high-bandwidth voltage comparator at each analog input.
Note, while the ADCs digitized 
at the external clock frequency $f$,
the FPGA control electronics 
operated at a divided frequency $f$/4.
Consequently, the recorded blocks of 24~ADC sample had ranges 
of 4-7 pre-samples preceding the ``trigger'' sample.

Interesting waveforms occasionally exceeded 24~ADC samples, 
for example when two pulses were close in time or a single pulse
with a large amplitude was over threshold 
for a long time.
Therefore, if the WFD input was 
over threshold during the last ADC sample of the 24-sample block, 
the WFD was re-triggered and recorded a continuation block of  
24 additional samples.

The thresholds and offsets of each WFD channel
were individually programmable via the VME interface.
The WFD analog thresholds were set as low as tolerable 
for data taking rates. The WFD analog offsets were set 
to permit the recording of 
under-shoots following large pulses.

Due to variations in clock signal cable lengths 
and WFD board propagation times,
the relative timing of the enable signal and the ($f$/4)-FPGA clock
was not identical on all modules. This not only results in a time offset 
between the different modules, but also a time offset that steps back and forth
by 4 clock ticks depending on the relative timing of the enable signal 
and the ($f$/4)-FPGA clock. Because the relative timing of
the two signals changes for every measurement period, 
the displacements between WFD modules also change.
Consequently, accumulated time histograms 
were actually superpositions of physical time distributions 
with varying proportions of the 4 clock tick displacement. 
Note that because no ambiguity exists between the four inputs 
of a single digitizer, the definition of inner-outer
tile coincidences and $\mu$SR cancellation by 
opposite tile sums were unaffected. Most importantly,
the determination of the lifetime was not affected by the  
displacements.

\subsection{Clock system}\label{sec:MasterClock}

An Agilent E4400 function generator \cite{AgilentClock} was used for the
external clock inputs of the waveform digitizer modules.
The function generator derives its clock signal from a temperature-controlled
crystal oscillator with a quoted short-term and long-term
stability better than 0.1~ppm.

The clock was located on the detector platform in the experimental area. 
Its signal was amplified and split six ways to generate one copy 
for each WFD crate. The six copies were then amplified and passively split sixteen ways 
to produce one copy of the clock signal for each WFD module.

A variation of the clock timebase
that was correlated with the measurement period  
could cause distortions of the lifetime determination. 
Such effects could arise from beating between the clocks defining the
digitizer sampling and kicker transitions.
To eliminate this possibility the 
digitizer sampling frequencies and kicker transition frequencies
were derived from completely independent, free running, 
clock systems.

The collaboration was blinded 
to the exact frequency of the clock until the completion
of the data analysis. During data taking and data analysis
the frequency was known by the collaboration 
to be $\unit[451.0 \pm 0.2]{MHz}$.  
The unblinding involved first a relative unblinding of the two (2006 and 2007) 
datasets into commonly-blinded clock ticks and then an absolute unblinding 
of the entire dataset into real time units.

\subsection{Data acquisition}\label{sec:DAQ}

The data acquisition \cite{DAQ} provided the read out and data logging
of the waveform digitizers and the beam monitor
as well as run control, data monitoring, {\it etc}.
The acquisition was developed using the \textsc{midas} acquisition package \cite{midas} and the
\textsc{root} analysis package \cite{Brun199781} on a 
parallel, layered array of networked Intel Xeon 2.6~GHz processors running
Fedora Core 3. 

To handle the high data rates and avoid any deadtime-related distortions,
the readout was organized into repeating cycles of deadtime-free intervals
called ``data segments.''
Each data segment comprised a complete, deadtime-free record
of every over-threshold digitized pulse during 5000 consecutive fill cycles.
Between each data segment a short deadtime of 
roughly 2-4~ms was imposed to complete the transactions 
that packaged the data and synchronized the readout.
The readout scheme relied on the ability to read out and write to 
the WFD FIFOs simultaneously.

An important component of the acquisition system was a programmable pulser unit (PPU)
that synchronized the data segments and measurement periods.
The PPU was constructed from a 25~MHz Xilinx FPGA.
It generated a continuous stream of fill cycles for the electrostatic kicker and 
a gated stream of 5000 fill cycles for the waveform digitizers. 
The data acquisition system initiated the beginning of each 5000 fill cycle sequence 
via the PPU and the PPU reported the completion of
each 5000 fill cycle sequence to the data acquisition system.

The data acquisition design involved 
a frontend layer for data readout,
a backend layer for event building and data storage,
and a slow control layer for control and diagnostics.
The frontend layer consisted of seven dual-core processors for the parallel readout 
of the six VME crates containing the waveform digitizers
and a single VME crate for the beam monitor readout.
Each VME crate was read out using a Struck Innovative Systeme GmbH SIS3100/1100 PCI-to-VME interface \cite{Struck}.
The frontend programs incorporated lossless compression 
of raw data using Huffman coding \cite{Huffman52,rfc1952}
as implemented in ZLIB \cite{rfc1950}.

The backend layer received data fragments 
asynchronously from the frontend layer via a 1~Gb Ethernet network.
The fragments were assembled into complete events  
that comprised all the recorded pulses 
during the  data segment.
These events were then written 
to a RAID-10 disk array and finally migrated 
to 400~GB LTO3 tapes.

The slow control layer was responsible for operation and monitoring of various 
instruments such as high voltage supplies, beamline magnet supplies,
temperature monitors, and $B$-field monitors. Slow control data
was read out periodically with one copy written to the main datastream 
and another copy written to an online database.

Lastly, an online analysis layer was responsible for 
ensuring data quality.
The online analyzer received 
data ``as available'' to avoid introducing any
unnecessary deadtimes into data readout.
As well as recording basic quantities such as run start/stop times and event statistics,
the online analysis layer determined quantities such as detector gains and fitted lifetimes.
These run-by-run quantities were stored in an online database 
that provided a comprehensive record of data taking
and enabled a straightforward monitoring of data integrity.

\section{Data preparation}

The procedure to obtain the positron time histograms 
from digitized scintillator pulses
involves first determining the various parameters 
of individual tile pulses and then
constructing the coincidences between 
inner-outer tile hits. Finally, the
accumulated histograms of coincidence times are corrected for
systematic effects including pulse pileup and gain changes.

The data obtained from the 2006 production run (R06)
using the AK-3 target has  $1.1\times10^{12}$ decay positrons
and the data obtained from the 2007 production run (R07) 
using the quartz target has $5.4\times10^{11}$ decay positrons.
The R06(R07) data were recorded in runs of 
duration 2(10) minutes and size 2(10)~GBytes, 
each containing about $4\times10^{7}$($2\times10^{8}$) positrons.
In total they yielded about 130~terabytes of raw data.

As listed in Table \ref{tbl:R07:DS}, the R06/R07 production data
are divided into subsets with different orientations 
of the magnetic field relative to the beam axis
and different positions of the geometrical center 
of the detector relative to the
surveyed intersection of the beam axis with the target disk.
These subsets are important for consistency checks
and systematic studies.

Shorter systematics datasets
were also collected. Examples of systematics datasets omitted 
in the final statistics are (i) magnet orientations that 
introduced large longitudinal $B$ fields at the stopping target, 
(ii) detector alignments that introduced 
large geometrical asymmetries between opposite tile pairs, 
and (iii) shorter accumulation periods that decreased the spin dephasing.
Short measurements taken every shift with low rates of 
laser pulses illuminating a sub-set of scintillator tiles are included 
in the final statistics.

The analysis was conducted using the 88-teraflop ABE cluster and 
mass storage system (MSS) at the National Center for Supercomputing 
Applications (NCSA) \cite{NCSA}. The processing of digitized waveforms 
into time histograms took roughly 60,000 CPU-hours for each dataset.

\begin{table}[htbp]
\centering
\caption{Summary of R06/R07 production datasets indicating the subsets with different
magnetic field orientations and detector alignments. The magnetic field orientation
is denoted L for beam-left and R for beam-right. The detector position represents 
the different coordinates of the geometrical center of the detector relative to the
surveyed intersection of the beam axis with the target disk.}
\label{tbl:R07:DS}
\renewcommand{\arraystretch}{1.4}
  \begin{tabular}{ l >{\centering}p{1.0cm} >{\centering}p{0.8cm} >{\centering}p{0.8cm} c r@{$\times$}l }
    \hline
    \hline
      data        &  \multicolumn{3}{c}{detector position [cm] }&    magnet    & \multicolumn{2}{c}{detected} \\
      subset             & $x$ &    $y$  & $z$ & orientation                   & \multicolumn{2}{c}{positrons} \\
    \hline
      R06-A         &   0     &    0     & 0 &      L      & $6.1$ & $10^{11}$ \\  %
      R06-B         &   0     &    0     & 0 &      R      & $5.0$ & $10^{11}$ \\  %
      R07-A         &   0     &    0     & 0 &      L      & $2.1$ & $10^{11}$ \\  %
      R07-B         &   0     &    0     & 0 &      R      & $2.4$ & $10^{11}$ \\  %
      R07-C         &   1.0   &   -1     & 0 &      L      & $3.2$ & $10^{10}$ \\  %
      R07-D         &   0.5   &   -1     & 0 &      L      & $1.4$ & $10^{10}$ \\  %
      R07-E         &   0.5   &    0     & 0 &      L      & $3.0$ & $10^{10}$ \\  %
      R07-F         &   0.5   &  -0.5    & 0 &      R      & $1.8$ & $10^{10}$ \\  %
    \hline
    \hline
  \end{tabular} 
\end{table}

Prior to analysis, a number of checks of data integrity were 
performed for occasional hardware or software failures.
The checks ensured no loss or corruption of data 
between the digitizer electronics and the analysis code.  
About 1.2\% of runs were rejected by integrity checks.

\subsection{Pulse fitting}

The time and amplitude of pulses 
are determined by least square fits 
to the individual digitized waveforms of each scintillator pulse. 
The procedure involves fitting a 
relatively high-resolution standard waveform (0.022~ns sampling-interval) 
to a relatively low-resolution individual waveform (2.2~ns sampling-interval) 
and it assumes the pulse shape is independent of
pulse amplitude.

\subsubsection{Pulse templates}

The high-resolution standard waveforms (pulse templates) 
are constructed by combining a 
large number of 450~MHz sampling-rate, single-pulse,
digitized waveforms. One template is derived for 
each scintillator tile using single-pulse waveforms with
peak amplitudes of 120-220 ADC counts. 

The template construction relies on the independence of the
positron arrival times and the digitizer clock phase, which implies
a uniform time distribution of pulse maxima across the sub-sample time 
intervals between consecutive clock ticks (c.t.). 
The waveforms of individual pulses thus correspond 
to displaced sets of samples from the same underlying pulse shape.
This makes it possible to reconstruct a high-resolution pulse template 
from many time-adjusted low-resolution waveforms 
via the computation of the aforementioned sub-sample time interval
for each individual waveform.

The procedure for computing the sub-sample time intervals of individual waveforms 
follows Ref.\ \cite{Bennett:2006fi}.
It involves first calculating a sub-sample pseudo-time $t^{\prime}$ for each digitized waveform
\begin{equation}
  {\rm t^{\prime}} = \frac{2}{\pi}{\tan^{-1}}\left( \frac{S_m-S_{m-1}}{S_m-S_{m+1}} \right) 
\end{equation}
from the three consecutive ADC samples $S_{m-1}$, $S_m$ and $S_{m+1}$ centered on the
maximum sample $S_m$. Then, using the measured distribution of pseudo-times $t^{\prime}$
and the uniform distribution of true times $t$, the pseudo-time is mapped to true time.

Finally, using the true times the individual waveforms are time-aligned and superimposed
in the construction of the pulse template for each scintillator tile. 
Fig.\ \ref{fig:pulses} (upper panel) shows a representative pulse template
for an individual scintillator tile.

\subsubsection{Fitting algorithm}

The pulse fitting algorithm fits pulse templates to individual 
waveforms so as to minimize
\begin{equation}
D=\sum^{N~\mathrm{samples}} \left[ S_i -  P - \sum^{n~\mathrm{templates}} A_j f_i(t_j) \right]^2
\label{e:pulsefitting}
\end{equation}
where $S_i$ are the $N$ samples of the digitized waveform and 
$f_i ( t_j )$ are the $n$ templates of the scintillator pulses.
The free parameters in Eqn.\ \ref{e:pulsefitting} comprise the time $t_j$ and amplitude $A_j$ of each scintillator pulse 
and an assumed constant pedestal, $P$. In minimizing the quantity $D$, all samples are 
equally weighted, except for underflows or overflows which were excluded 
from the computation. The equal weighting of ADC samples 
is discussed further in Sec.\ \ref{sec:pile-on}.

Given that the datasets contain $\sim$10$^{12}$ pulses, 
the algorithm for fitting is optimized for speed and robustness.
As the probability of two pulses on one waveform is roughly 1\%, 
the algorithm first tries to fit a single template to each waveform.
Because the  fit-function $P + A_1 f_i (t_1)$  is linear in $A_1$ and $P$,
the minimum of $D$ is found  by a one-dimensional search on the non-linear parameter $t_1$,
with the other parameters calculated from their partial derivatives at the fit minimum.
This one-dimensional search 
is performed by parabolic minimization using Brent's Method \cite{NumericalRecipes}. 

If the quality of this fit is not acceptable (as indicated by a search for further pulses in
the residuals) the algorithm then tries to improve
the fit by adding pulses, removing pulses and using more sophisticated, time-consuming,
$\chi^2$-minimization procedures. 
In fitting a waveform to $n$$>$$1$ pulses, the fitter first tries an
$n$-dimensional search on the pulse times $t_i$ using the Minuit minimization code \cite{James:1975dr} 
with the remaining parameters calculated from the aforementioned partial derivatives. 
If the $\chi^2$ exceeds a maximum value, the fitter attempts a 2$n$+1 dimensional search 
over all fit parameters. The fitter continues to add pulses or remove pulses until either (i)
the residuals indicate no evidence for further pulses
or (ii) ten attempts to fit $n$$>$$1$ pulses are completed.

During fitting, the algorithm will reject any pulses with amplitudes $A_j$ less than 35 ADC counts,
times $t_j$ outside the waveform boundaries, or times $t_j$ within three samples of other pulses. 
When no pulses are found, the pulse fitter calculates an amplitude-weighted time of all ADC samples.

\begin{figure}[h!]
\includegraphics[width=0.85\linewidth]{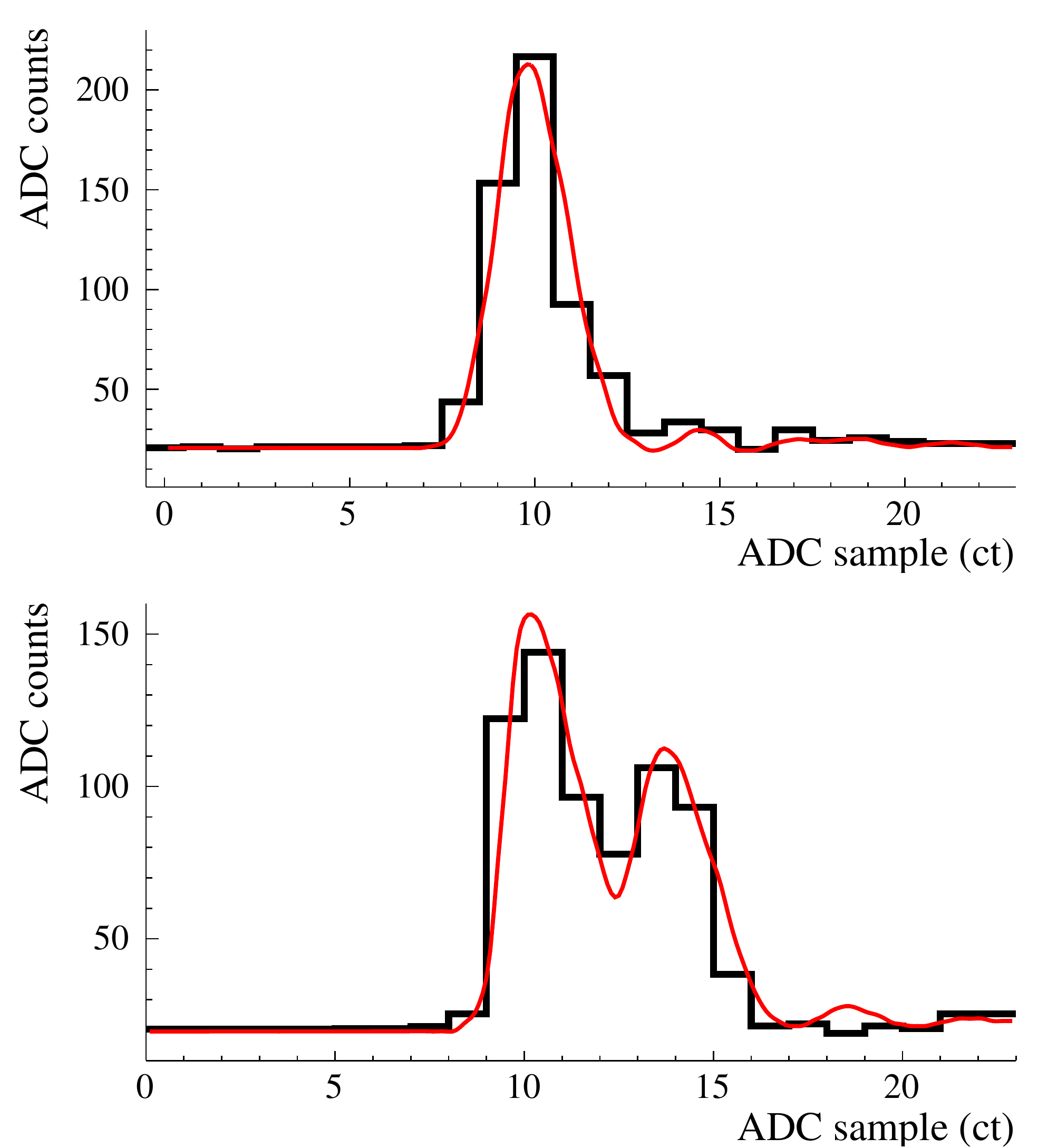}
\caption{Examples of the digitized waveforms and the corresponding fit results 
for a single pulse (upper panel) and two overlapping pulses (lower panel).
The waveforms are indicated by the  histogram and the 
fit function by the curve.}
\label{fig:pulses}
\end{figure}

Sample fits to digitized ``islands'' containing a single pulse and two overlapping pulses 
are shown in Fig.\ \ref{fig:pulses}. The digitized waveforms were 
overwhelmingly single pulse islands  with fractional contributions of roughly $1\times10^{-2}$ 
from two pulse islands and roughly $7\times10^{-4}$ from zero pulse islands.

\subsection{Hit definition}

Two cuts are applied before constructing the coincidences; one defining  
an unambiguous software threshold for pulses and another defining 
an unambiguous software deadtime between hits.
The construction of coincidence datasets with different 
threshold and deadtime settings is important for studying such effects as pulse pileup and gain variations.

\begin{figure}[ht]
\includegraphics[width=0.8\linewidth]{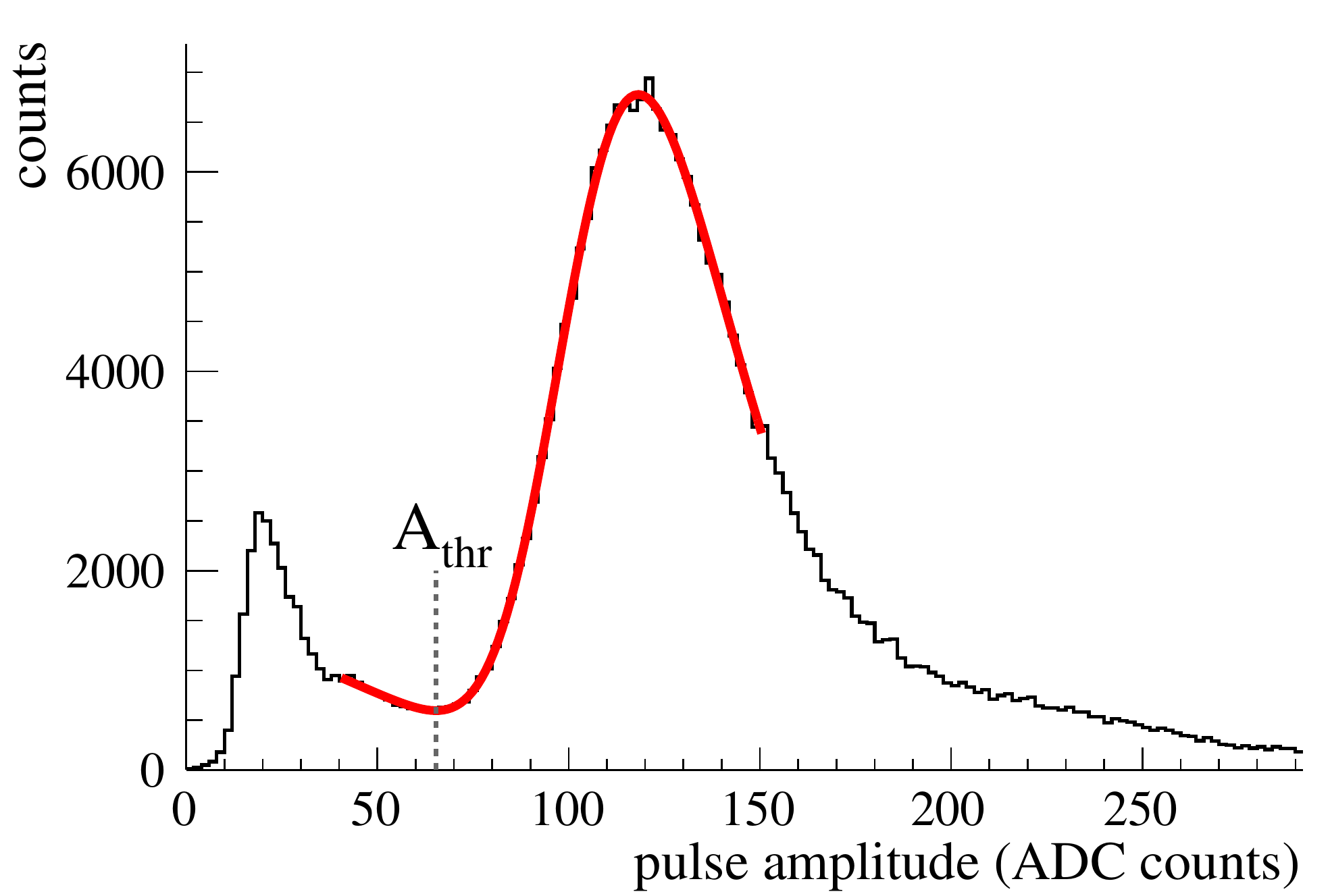}
\caption{Representative pulse amplitude distribution for an individual scintillator tile.
The histogram represents the measured data and the smooth curve the fit function (see text for details). 
The fit-function minimum defines the normal amplitude threshold $A_{thr}$ for each scintillator tile.}
\label{fig:fit_amp}%
\end{figure}

A typical pulse amplitude spectrum for an individual scintillator tile
is plotted in Fig.\ \ref{fig:fit_amp}.
It shows a broad peak due to decay positrons at $\sim$120~ADC units
and a rising background due to noise pulses below $\sim$50~ADC units.
As shown in Fig.\ \ref{fig:fit_amp}, these distributions are fit to the convolution of a
Landau function and a Gaussian peak plus a linear background term.
The fit-function minimum defines the normal amplitude threshold $A_{thr}$,
corresponding to the valley  between the low-energy noise and the positron MIP peak.

This threshold is re-calculated every 100(10) runs in R06(R07) data taking 
in order to account for the observed, long-timescale drifts 
in the detector gains over the data-taking period.
Additional datasets, with higher thresholds of
$A_{thr} + 20$, $A_{thr} + 30$ and $A_{thr} + 40$, are processed for studying the effects 
of gain variations during the measurement period.

The pileup of positrons within the instrumental deadtime distorts
the measured time distribution.
Specifically, when two positrons are separated by $>$$3$~c.t.\ (6.7~ns),
the corresponding pulses are cleanly resolved by the fitting algorithm. 
However, when two positrons are separated by $<$$3$ c.t.\ (6.7~ns),
the algorithm cannot reliably discriminate the overlapping pulses.\footnote{Subsequently, 
an ADT of $\geq 6$ c.t.\ (13.3~ns) is used to
avoid any re-triggering on the trailing-edges of the scintillator pulses.}

To explicitly define the minimum resolving-time between neighboring pulses,
a software deadtime following above-threshold hits is introduced.
It rejects any additional hits within an artificial deadtime (ADT).
A detailed study of pileup distortions with different ADTs 
of 5-68 c.t.\ (11$-$151~ns) was made (see Sec.\ \ref{sec:pileupprocedure}).

\begin{figure}[ht]
\centering
\includegraphics[width=0.8\linewidth]{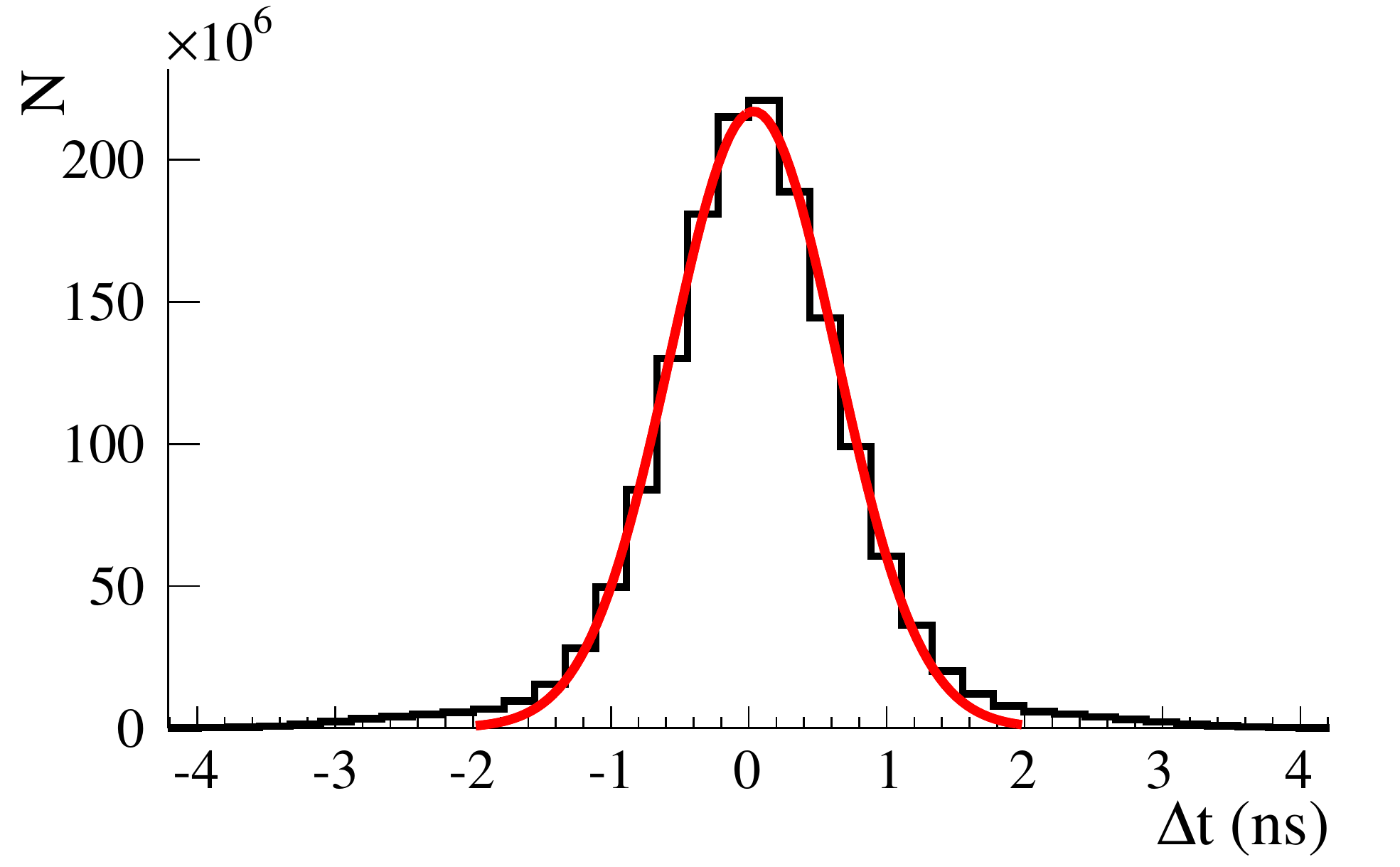}
\caption{Representative time difference distribution for a single inner-outer pair. 
The histogram is the measured distribution and the curve is the Gaussian fit. The observed timing jitter 
between inner-outer pairs is typically 0.55~ns (1~$\sigma$).}
\label{fig:coincidence_C}
\end{figure}

A typical inner-outer pair time difference spectrum 
is plotted in Fig.\ \ref{fig:coincidence_C}. It shows 
a sharp coincidence peak 
from through-going positrons with 
a time resolution of $0.55$~ns (1~$\sigma$). 
The  time offsets between inner-outer pairs, originating
from differences in  PMT voltages, cable lengths, and light guide configurations,
are determined by a Gaussian fit to each inner-outer time difference histogram. 
A run-by-run table of timing offsets
is used to account for changes of the offsets during the data-taking.

The individual tile hits are then
sorted into datasets of inner-outer coincidences,
inner singles, and outer singles. 
The inner-outer coincidences for each tile pair are identified using 
a coincidence window of $\pm1$~ADT.\footnote{Using the same coincidence window 
and artificial deadtime simplifies 
the pileup correction procedure.
It, however, does increase the contribution
from accidental coincidences between
unrelated singles hits in inner-outer tile pairs.}
The coincidence time is defined
as the time corresponding to the later hit of the inner-outer pair. 

In the unlikely case that two combinations of coincidence pairs 
are found within $\pm$1~ADT---for example when one outer hit falls  
between two inner hits with mutual separations 
of 1-2~ADT---the earlier combination of hits defines the inner-outer coincidence
(this reduces the sensitivity to coincidences originating from 
photomultiplier ``ringing'').  
Such circumstances yield one coincidence hit 
and one single hit.

Finally, the sorted datasets of coincidence hits and singles hits are used 
to fill per-run, per tile-pair, histograms of inner singles, outer singles
and inner-outer coincidence times.

\subsection{Histogram corrections}

Before fitting the time histograms to extract the muon lifetime, several corrections are applied. 
The largest corrections account for time-dependent distortions arising from   
pileup of the decay positrons and variations of the detector gains.
In addition, the statistical uncertainties on histogram bin contents
are corrected for multiple tile-pair hits originating
from positron scattering, positron annihilation and cosmic-rays.

\subsubsection{Pulse-pileup correction}
\label{sec:pileupprocedure}

In filling the histograms, a 
well-defined, fixed-length, artificial deadtime 
is applied so that subsequent detector hits
within the ADT of an earlier detector hit are not used.
Consequently, the measured time distribution corresponding to a
parent time distribution $N(t) \propto e^{-t/\tau_{\mu}}$ will be
\begin{equation}
N^{\prime}(t) \propto e^{-t/\tau_{\mu}} - r e^{-2t/\tau} + ...  
  \label{eq:muprime}
\end{equation}
where the second term represents the leading-order pileup losses
having a magnitude $r$ that depends on the beam rate and the ADT.

The procedure for correcting pileup takes advantage of the time structure of the incident beam.
The pileup losses are statistically recovered using so-called ``shadow windows'',
by replacing the lost hits in each measurement period with
measured hits at equivalent times in neighboring measurement periods.

To illustrate the method, consider the case of leading-order pileup, 
where a second hit within the artificial deadtime is lost.
To correct this loss, 
if a hit is observed at a time $t_i$ in a fill $j$ (denoted the ``trigger'' hit),
a hit is searched for in an interval $t_i$~$\rightarrow$~$t_i + \mathrm{ADT}$ 
in the fill $j+1$ (denoted the ``shadow'' hit). 
Adding the resulting histogram of shadow hit times to the original histogram of trigger hit times
thereby statistically restores the hits lost to leading-order pileup.

\begin{figure}[ht]
 \centering
\includegraphics[width=0.90\linewidth]{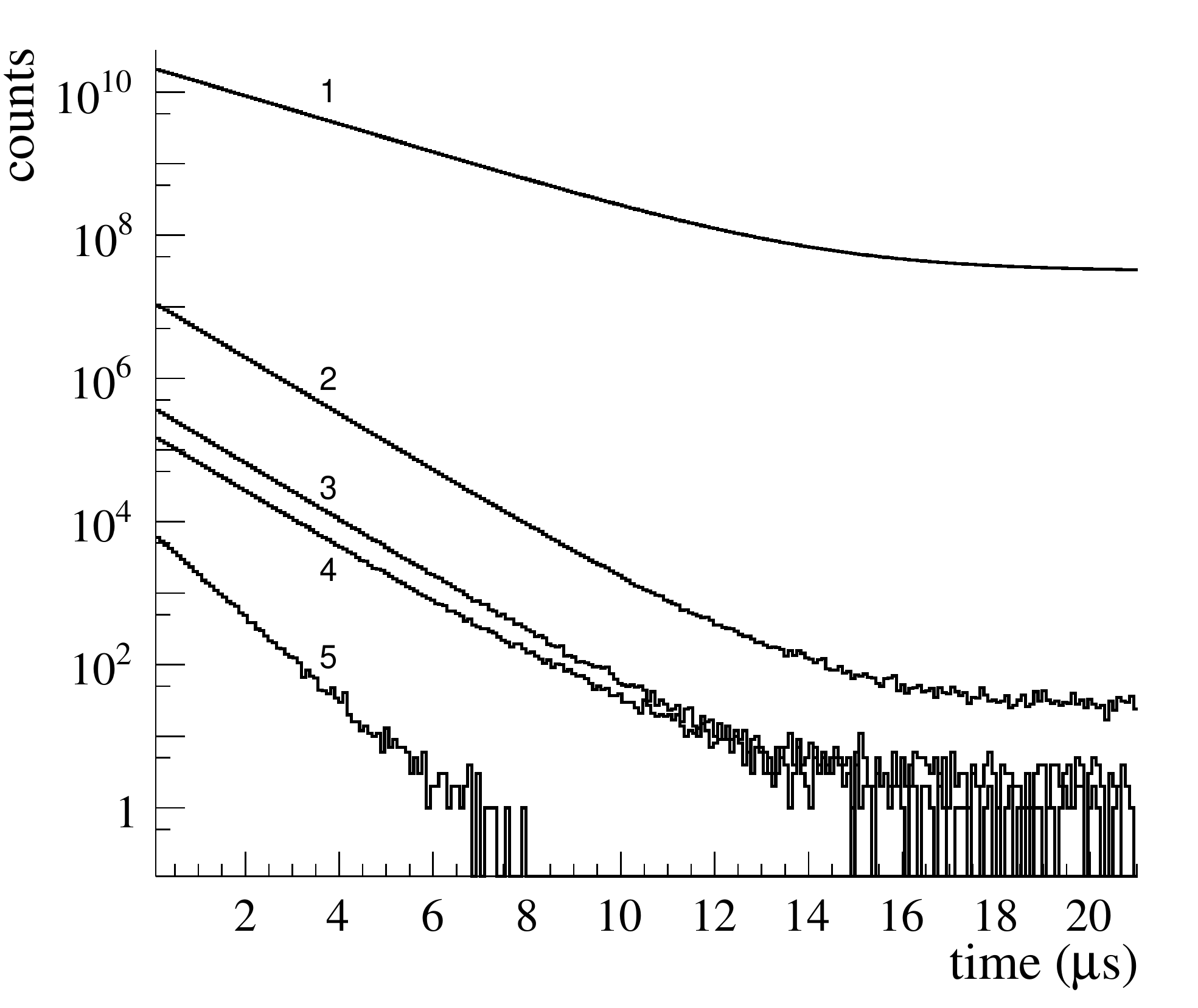}
\caption{Representative time distributions for the trigger hits before the pileup correction (curve 1)
and corrections accounting for leading-order pileup (curve 2), extended pileup (curve 3),
accidental coincidences (curve 4) and second-order pileup (curve 5). 
These histograms correspond to an artificial deadtime $\mathrm{ADT} = 6$~c.t.\ (13.3~ns).}
\label{fig:lo_pileup}
\end{figure}

The leading-order pileup correction for $\mathrm{ADT} = 6$~c.t.\ (13.3~ns)
is depicted in Fig.\ \ref{fig:lo_pileup} (curve 2). The correction ranges from roughly 
10$^{-3}$ at the start of the measurement period 
to roughly 10$^{-7}$ at the end of the measurement period.

The leading-order correction is insufficient to entirely account for pileup effects.
As the shadow windows have a width equal to the ADT, they contain either zero hits or one hit.
Consequently---at the level of the higher-order pileup---the leading order pile-up
is over-corrected and the higher-order pile-up is under-corrected by the shadow window 
(as $>1$ hits in the shadow window are treated as 1 hit in the shadow window).

To correct second-order pileup, 
if a shadow hit is found in the shadow window applied to fill $j+1$, 
another shadow hit is searched for in a second shadow window applied to fill $j+2$.\footnote{In practice, 
the shadow windows are applied using a five-fill ring buffer, with  fill 1 being
corrected by fills 2 and 4, fill 2 being corrected by fills 3 and 5, and finally fill 5 
being corrected by fills 1 and 3.}

The second-order pileup contribution for $\mathrm{ADT} = 6$~c.t.\
is also depicted in Fig.\ \ref{fig:lo_pileup} (curve 5). Its contribution 
is roughly 10$^{-6}$ at the start of the measurement period 
and is completely negligible by the end of the measurement period.
For a parent time distribution $e^{-t/\tau}$, 
the second-order pileup time distribution 
has a time dependence $e^{-3t/\tau}$
and an amplitude that increases quadratically with artificial deadtime.

The pileup corrections described above
assume the absence of the $\sigma \sim 0.5$~ns timing jitter between the inner and outer tile hits.
This timing jitter causes additional pileup losses of inner-outer coincidences
when one hit is lost but the other hit is not (recall the ADT is applied before defining
the coincidence). This 
``extended'' pileup is measured by searching for single tile hits within
the jitter of the trailing edge of each shadow window.

The extended pileup correction is plotted in Fig. \ref{fig:lo_pileup} (curve 3).
Its contribution is roughly 10$^{-5}$ at the start of the measurement period 
and is completely negligible by the end of the measurement period.
For a parent time distribution $e^{-t/\tau}$,
the extended pileup time distribution has the time dependence $e^{-2t/\tau}$ 
and an amplitude that is ADT-independent 
(at least for $\mathrm{ADT} \ll \tau$).
 
Random coincidences between unrelated singles hits
can cause accidental coincidence hits. 
They can originate 
from muon decays, environmental backgrounds, and electronic noise.
A significant number of inner tile hits without outer tile hits occur 
because inner tiles are slightly larger than outer tiles.
Overall about 27\% of inner hits are singles and 
about 3\% of outer hits are singles.

A shadow window technique is also used to correct for
accidental inner-outer coincidences. 
An inner singles hit at time $t_i$ in the trigger fill initiates a search for
an outer singles hit at time $t_i$~$\rightarrow$~$t_i + $ADT in the shadow fill.

The measured contribution of accidental coincidences 
is depicted in Fig.\ \ref{fig:lo_pileup} (curve 4). It ranges from 
roughly 10$^{-5}$ at the start of the measurement period 
and is negligible by the end of the measurement period.
For a parent singles time distribution $e^{-t/\tau}$,
the accidental coincidence time distribution has a 
time dependence $e^{-2t/\tau}$ and an amplitude that increases linearly with 
artificial deadtime.

A Monte Carlo (MC) simulation is performed
to verify the procedure for correcting pileup.
The MC generates a sequence of beam cycles with tile hits
from parent time distributions that reproduce the observed rates 
of coincidence hits and singles hits 
from muon decays and background sources. 
The simulation also incorporates the measured timing jitter 
between inner-outer pairs.

\begin{figure}[h!]
  \centering
  \includegraphics[width=0.9\linewidth]{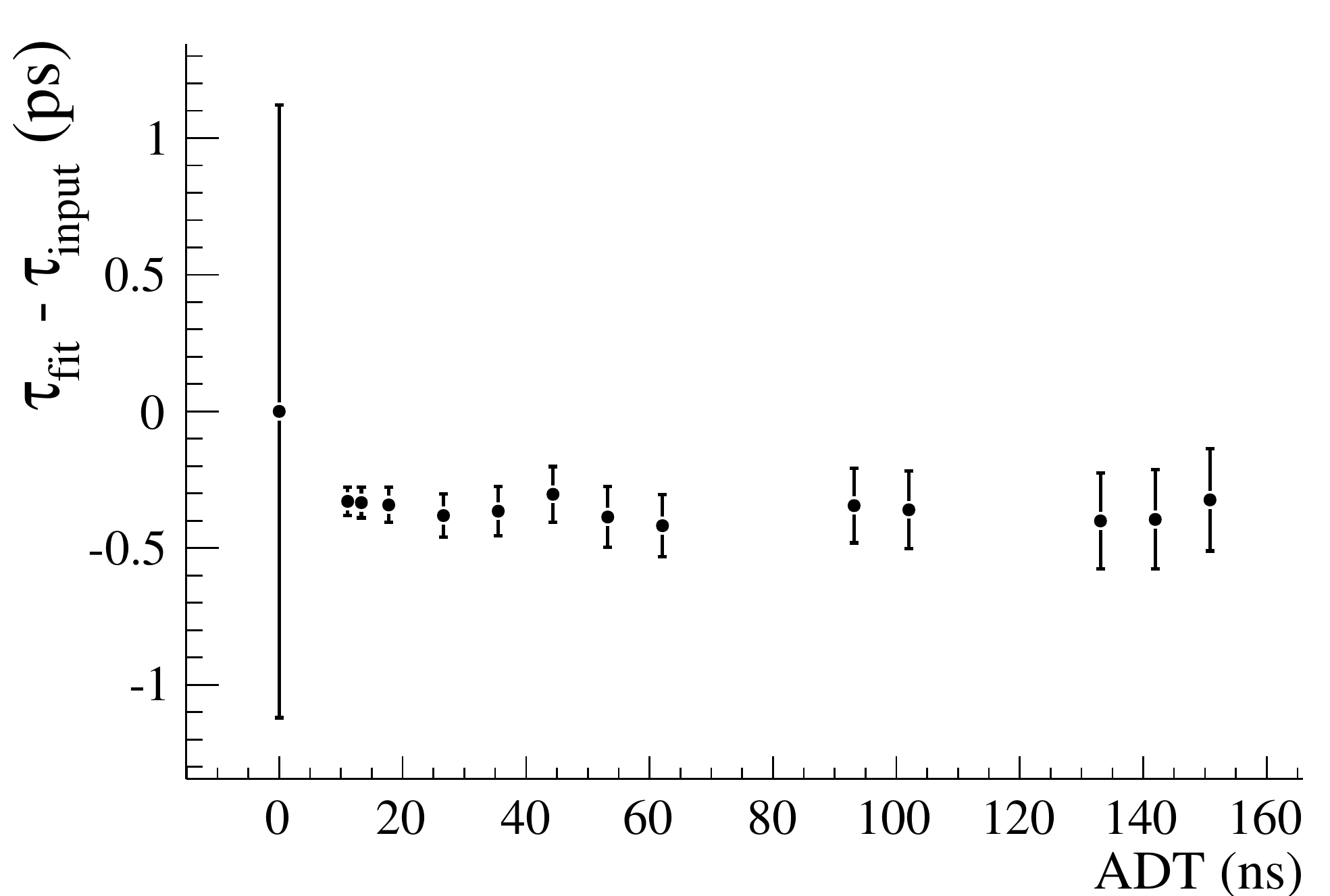}
\caption{Fitted muon lifetime versus artificial deadtime obtained from an analysis of a Monte-Carlo 
simulation of pileup effects, accidental coincidences and timing jitter. The simulated data are analyzed
using the same software and methods as measured data.
The fitted lifetime is ADT-independent to better than 0.1~ppm and in 
statistical agreement with the ``true'' lifetime depicted at the location $\mathrm{ADT} = 0$.
The error bars on the $\mathrm{ADT} > 0$ points correspond to the
statistical variations allowed by the ADT correction.
The error bar on the $\mathrm{ADT} = 0$ point corresponds to the
statistical variation allowed between the 
``true'' lifetime that omits the pulse pileup, accidental coincidences and timing jitter,
and the $\mathrm{ADT} > 0$ lifetimes 
that correct for pulse pileup, accidental coincidences and timing jitter.}
\label{fig:MC_reconstruction}
\end{figure}

The simulated data are analyzed using the same procedures 
as the measured data, {\it i.e.}, applying the artificial deadtime, forming the inner-outer coincidences,
and accumulating the histograms of trigger hits and shadow hits. 
The simulated trigger hit time distributions are corrected 
with the simulated shadow hit time distributions
and are found to reproduce the parent hit time distribution.
In particular, 
the ``true'' lifetime in the absence of pileup effects, accidental coincidences and timing jitter,
and corrected lifetime after the accounting for pileup effects, accidental coincidences and timing jitter,
are in good agreement.
The lifetimes derived for ADT ranges
of 5$-$68~c.t.\ (11$-$151~ns) are in agreement to a precision $\pm$0.1~ppm 
(see Fig.\ \ref{fig:MC_reconstruction}).

\subsubsection{Gain variation correction}
\label{sec:gaincorrection}

When applying deadtimes, constructing coincidences and filling histograms,
only hits with amplitudes that exceed the threshold $A_{thr}$ are used.
If the detector gain changes over the measurement period, 
then the time histogram will be distorted,
either by additional hits rising above the amplitude cut or  
by additional hits falling below the amplitude cut.\footnote{Because the threshold 
$A_{thr}$ is applied to pedestal-subtracted amplitudes the time histogram 
is not distorted by pedestal variations.} 

The detector gain versus measurement time 
is obtained by accumulating
a sequence of pulse amplitude spectra corresponding to 
consecutive, 220~ns-wide, time windows over 
the entire measurement period.
Because the amplitude distributions for decay positrons and cosmic rays 
are different, and the relative proportion of decay positrons and cosmic rays
is time dependent, the positron distributions are obtained
from measured distributions by subtracting the cosmic-ray distributions.
The subtraction uses the measured cosmic-ray distribution
in the last 220~ns time window of the measurement period.

\begin{figure}
\includegraphics[width=0.95\linewidth]{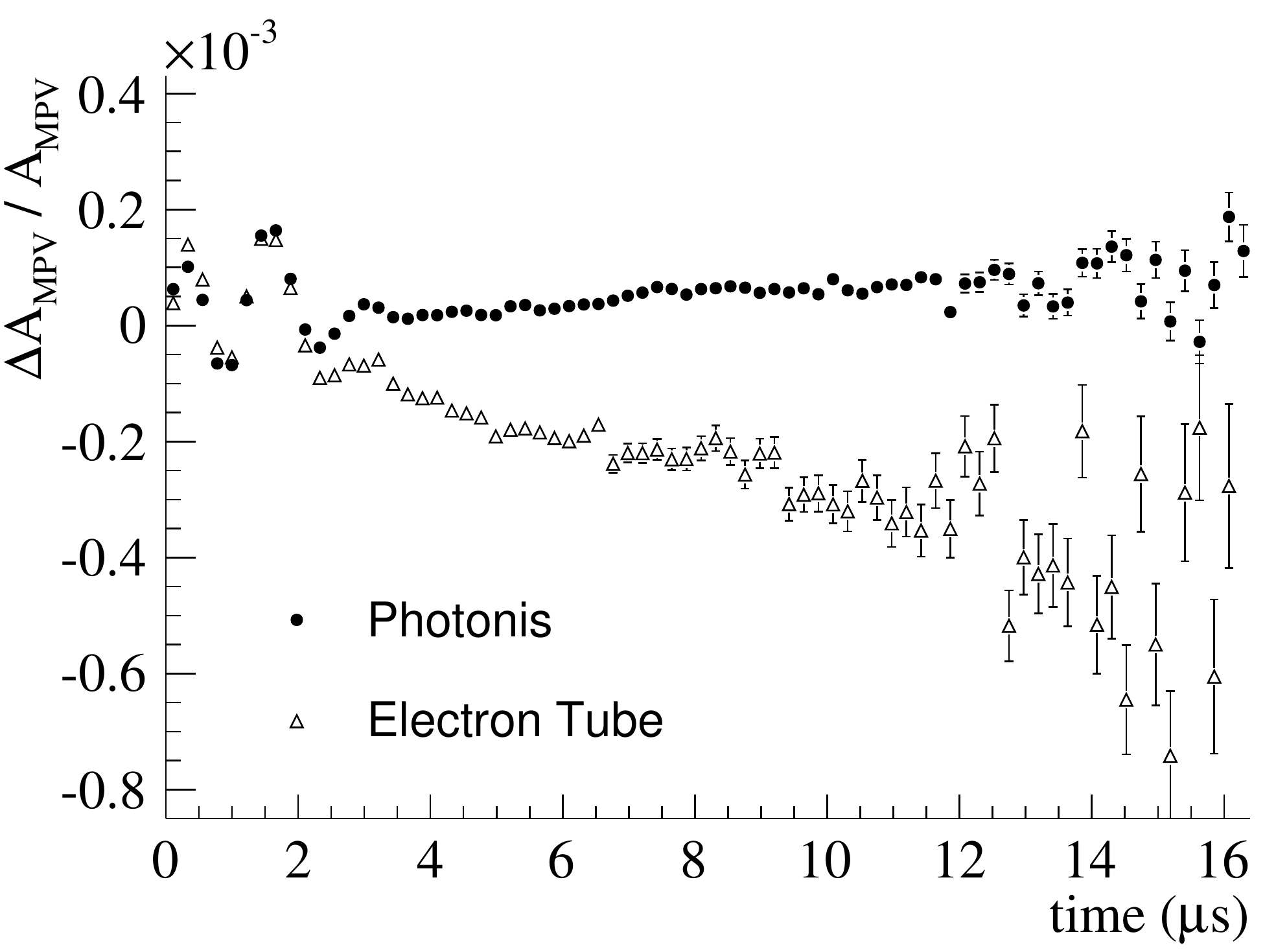}
\caption{Fractional MPV change versus the hit time 
in the measurement period. The solid circles correspond to the amplitude variation
of tiles with Photonis PMTs
and the open triangles correspond to the amplitude variation of 
tiles with Electron Tube PMTs.}
\label{fig:MPVversusFillTime}
\end{figure}

The positron amplitude distributions are then fit to a convolution of
a Landau energy loss distribution and a Gaussian instrumental resolution
to derive the Landau most-probable-value (MPV) versus measurement time.
In the absence of instabilities the MPV
would be independent of the measurement time. 

Fig.\ \ref{fig:MPVversusFillTime}---which plots MPV versus measurement time 
for scintillator tiles sub-divided by photomultiplier 
type---indicates a gain variation of roughly 5$\times$10$^{-4}$ over the measurement period.
One feature, which appears independent of PMT type, 
is an oscillation of the gain over the first microseconds of the measurement period. 
Another feature, which appears different for PMT types, 
is a drift of the gain over the entire duration of the measurement period. 

A gain variation during the measurement period will distort the extracted lifetime.
The change in the counts exceeding the amplitude cut, $\Delta N$,  
and the change in the MPV of the amplitude distribution, $\Delta A_{MPV}$,
are related by
\begin{equation}
\Delta N =  2 ~  N_{thr}   ~ \Delta A_{MPV} ~ ( { A_{thr} / A_{MPV} } )
\label{eqn:gainchange}
\end{equation}
where $N_{thr}$ is the counts in the histogram bin corresponding to the amplitude cut,
and $ \Delta A_{MPV} ~ ( A_{thr} / A_{MPV} ) $ is the inferred change in the 
amplitude cut via a scaling  
of the measured change in the Landau MPV. 
The factor of 2 in Eqn.\ \ref{eqn:gainchange} arises as gain changes 
of both inner and outer tiles can 
alter the above-threshold coincidences.

Using the measured MPV versus measurement time and Eqn.\ \ref{eqn:gainchange},
the time histograms are then corrected for gain variations. 
In one approach, the measured $\Delta A_{MPV}$ distribution is converted 
to a hit variation $\Delta N$ and
used to subtract (add) the extra (lost) hits from gain changes.
In another approach, the MPV distribution is first fit
to an empirical function describing the
gain changes. This function is then converted from
MPV versus measurement time to hits versus measurement time
and used to make the gain correction
(this approach is helpful in understanding
the sensitivity of $\tau_{\mu}$ to oscillations and drifts). 
The gain corrections are applied separately to Electron Tube
and Photonis PMTs.

Gain instabilities during the measurement period are identified
with the photomultipliers and the readout electronics.

\begin{figure}
\includegraphics[width=0.95\linewidth]{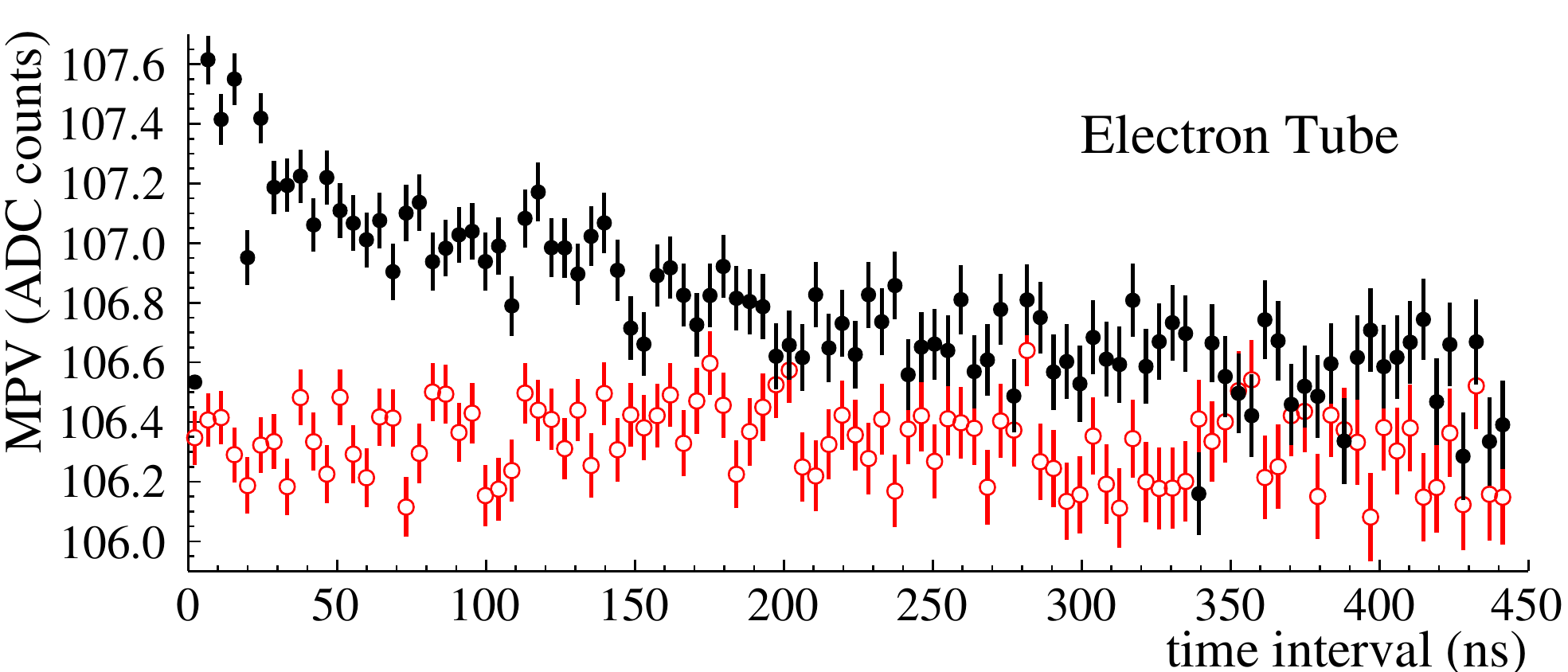}
\includegraphics[width=0.95\linewidth]{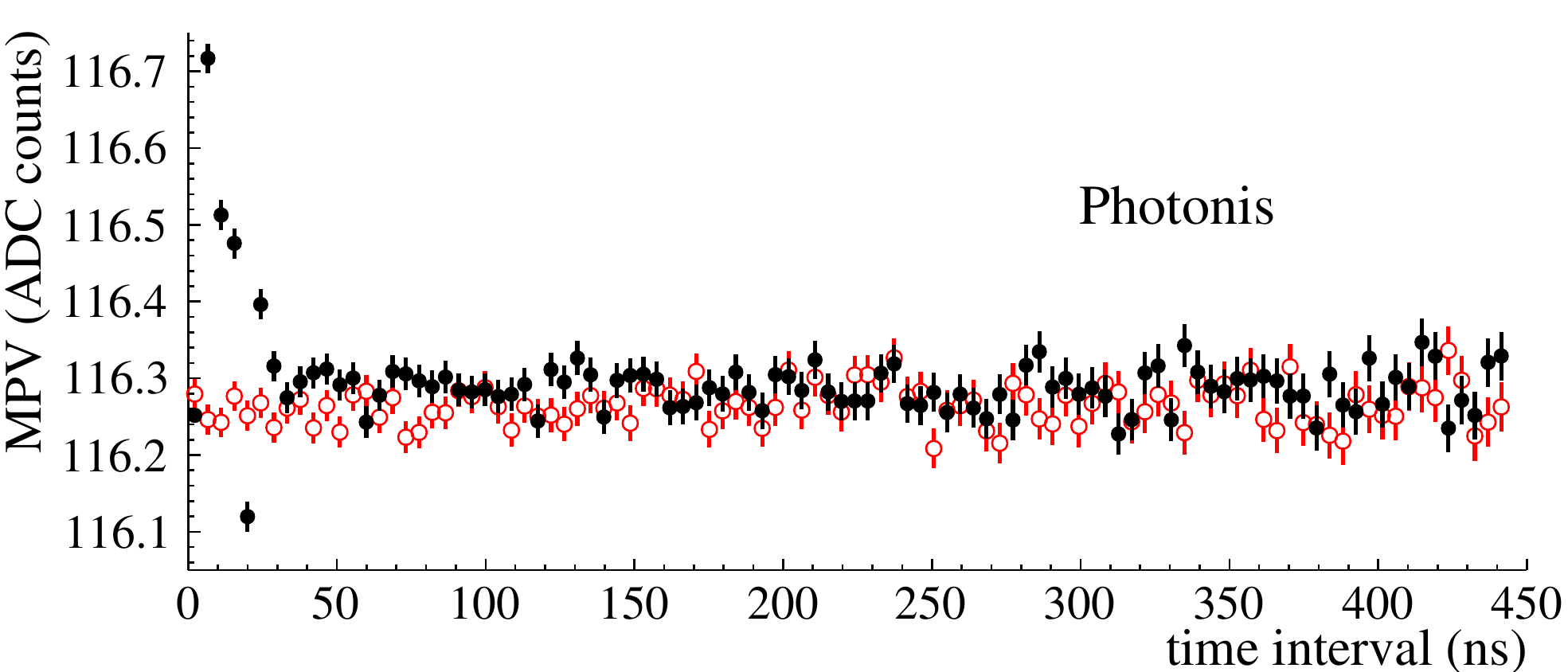}
\caption{Time dependencies of the Landau MPV versus the time interval after a preceding pulse
for representative groups of Electron Tube PMTs (upper panel) and Photonis PMTs (lower panel).
The black data points correspond to time intervals between two hits in the same tile 
pair and the same fill cycle. The red data points correspond to time intervals between two hits 
in the same tile pair but two neighboring fill cycles. The red points show no evidence 
of longer term effects from gain variations between neighboring fills.}
\label{fig:MPVversusDT}
\end{figure}

\begin{enumerate}
\item[(i)]
The gain oscillations in Fig.\ \ref{fig:MPVversusFillTime} originate from electronics instabilities 
due to fan-out modules that distribute the measurement period start signals to all waveform digitizer modules.
The oscillations are reproducible in bench tests of waveform digitizers and are observed in all channels 
of all waveform digitizer modules in all VME crates.
\item[(ii)]
The gain drift in Fig.\ \ref{fig:MPVversusFillTime} 
is dependent on PMT type. To understand this effect,
amplitude distributions corresponding
to different time intervals between consecutive detector hits are examined.
The time dependencies of the Landau MPV versus the 
elapsed time after the preceding pulse  
are plotted in Fig.\ \ref{fig:MPVversusDT} for each PMT-type.
The Photonis PMTs show a relatively short-timescale effect of duration $\sim$25~ns and 
fractional gain change $\sim$1\%.
The Electron Tube PMTs show a relatively long-timescale effect of duration $\sim$250~ns and 
fractional gain change $\sim$1\%.
Because rates change with measurement time, 
these analog pileup effects contribute 
to the gain drift during the measurement period.
\end{enumerate}

The pileup effect in item (ii) is distinct from
the ``digital'' pileup associated with the artificial deadtime 
that was discussed earlier in Sec.\ \ref{sec:pileupprocedure}. Effect (ii) is not corrected
by the shadow window technique.

\subsubsection{Hit multiplicity correction}

The final correction accounts for multiple coincidence hits 
arising from single parent events. For example, a 
through-going cosmic ray
or scattered decay positron may yield $>$1 coincidence hits. Consequently,
the number of coincidence hits in the time histograms exceeds 
the number of independent events populating the time histograms. 

To correctly account for statistical uncertainties,
the separate time histograms of multiplicity-one coincidences 
and multiplicity-two coincidences are accumulated.\footnote{Multiplicity-one 
coincidences are defined as one inner-outer hit 
in an 8~c.t.\ (17.7~ns) interval and multiplicity-two coincidences are 
defined as two inner-outer hits in an 8~c.t.\ (17.7~ns) interval.
The effects of higher multiplicities are negligible.}
These histograms are then fit to obtain the 
relative contribution of multiplicity-one and -two coincidences
to the decay curve and the time-independent background 
(the two contributions have 
different origins and different multiplicities). Denoting
the multiplicity one and two contributions as N$_1$ and N$_2$
(decay term) and  C$_1$ and C$_2$ (background term),
the ratio of independent events to coincident hits versus measurement time is
\begin{equation}
R = \frac{(N_1 + N_2) e^{-t/\tau_\mu} + (C_1 + C_2)}{(N_1 + 2 N_2) e^{-t/\tau_\mu} + (C_1 + 2 C_2)} ~~.
\end{equation}

The ratio $R$ 
decreases from $0.97$ at early times ({\it i.e.}, a $\sim$3\% contribution of multiplicity-two hits)
to $0.96$ at late times ({\it i.e.}, a $\sim$4\% contribution of multplicity-two hits).
To account for events with multiplicities $>$1 the
Poisson uncertainties on bin contents
are inflated by $1/\sqrt{R}$.
Note that this correction only affects the uncertainty 
on the lifetime and the $\chi^2$ of the fit---not the value of the lifetime.

\section{Lifetime Analysis}

The extraction of the positive muon lifetime 
from fits to the positron time histograms
is described next. 
Supplementary studies of transverse field and longitudinal field 
$\mu$SR in AK-3 and quartz and various checks
on data integrity are presented.
Systematic corrections and systematic uncertainties
that arise from $\mu$SR effects, pulse pileup, gain changes,
and time-dependent backgrounds are described.

\subsection{Fitting procedures in AK-3 and quartz}

The muon lifetime is obtained by fitting the time histograms of inner-outer coincidences.
In principle, each pileup-corrected, gain-corrected tile-pair time distribution can be fit to the
elementary function,
\begin{equation}
N(t) = N e^{-t/\tau_\mu} + C
\label{eq:fitfunction}
\end{equation}
where $\tau_\mu$ is the muon lifetime, $N$ is a normalization constant 
and $C$ represents the flat background.
In practice, distortions arising from residual $\mu$SR effects must be considered.
Separate fitting strategies are employed for AK-3 and quartz 
because the specific characteristics of their $\mu$SR signals are different.
Most important is the absence (presence) in AK-3 (quartz) of discernible
$\mu$SR effects in the individual tile-pair time histograms.

In all cases the benchmark fits use histograms prepared with an
artificial deadtime 6~c.t.\ (13.3~ns) and the aforementioned correction procedures
for pulse-pileup, gain changes and 
hit-multiplicities. The nominal 
fit start and stop times are 
$t_{start} = 0.22$~$\mu$s and $t_{stop} = 21.12$~$\mu$s, 
a range which begins and ends at a safe interval
from the beam (kicker) transitions.
However, the results do not include
the extrapolation to $\mathrm{ADT} = 0$ discussed 
in Sec.\ \ref{sec:pileup systematic} below. 

\subsubsection{Fit to the AK-3 data}

A simple fit method is used for AK-3 production data.
The summed tile-pair time histogram is fit 
to Eqn.~\ref{eq:fitfunction}, the lifetime result  
thereby relying on sufficient cancellation 
of $\mu$SR effects by spin dephasing, target choice 
and detector geometry.

\begin{figure}
\includegraphics[width=0.9\linewidth]{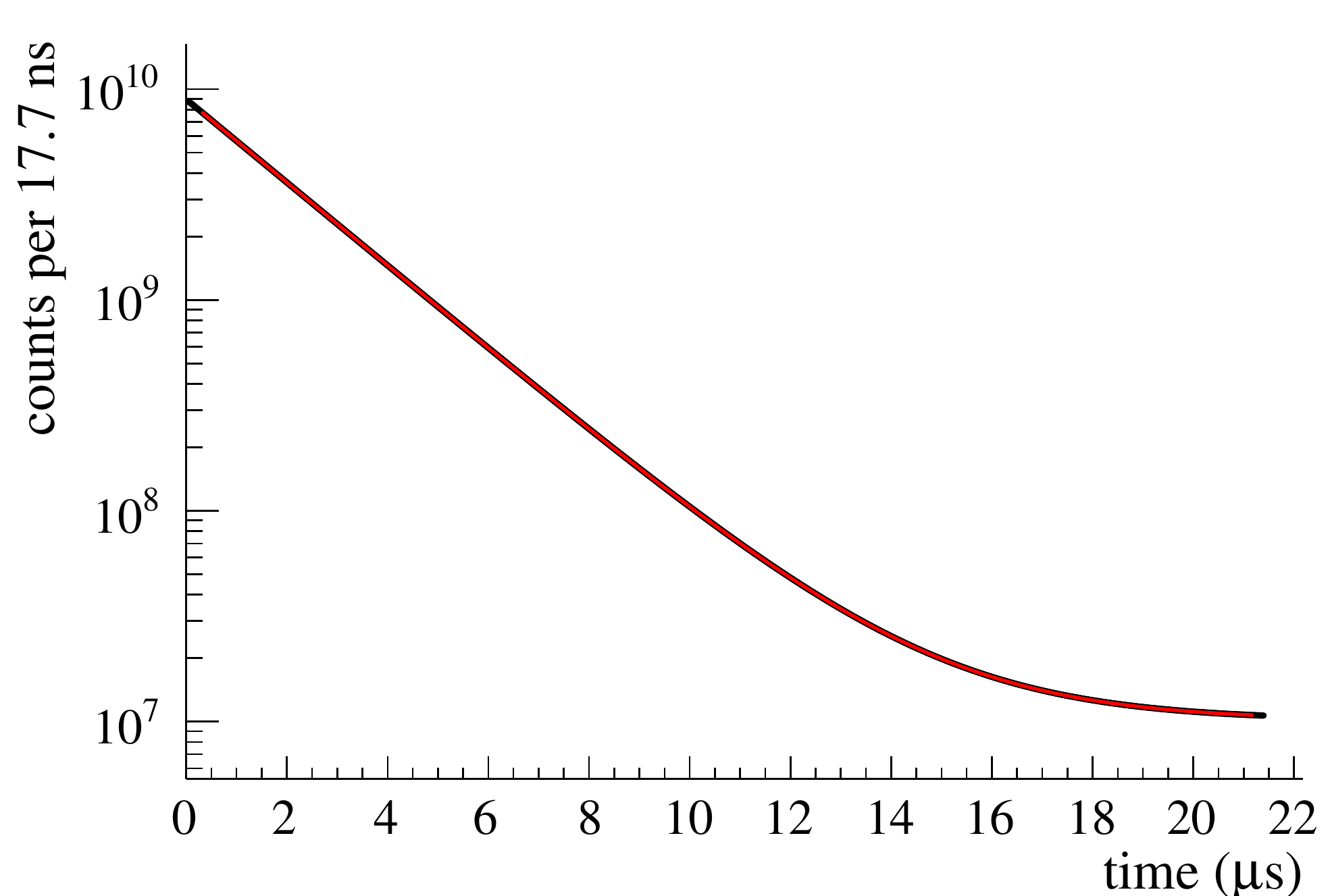}
\includegraphics[width=0.9\linewidth]{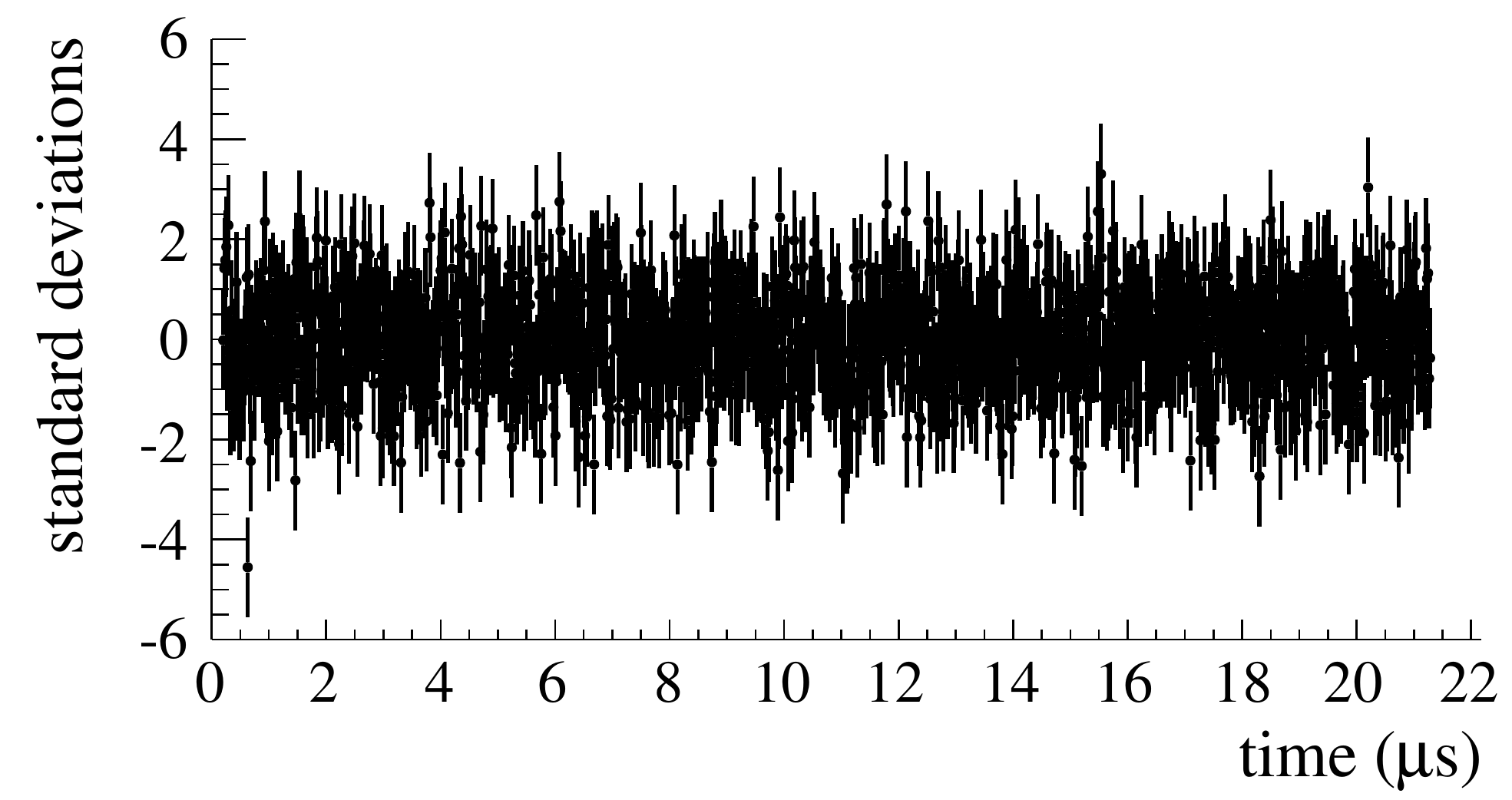}
\caption{Three-parameter fit to the entire AK-3 dataset.
The upper panel shows both the measured data and the fit function 
and the lower panel shows the normalized residuals between
the data points and the fit function.}
\label{fig:AK3fit}
\end{figure}

The benchmark fit to the entire AK-3 dataset (R06-A + R06-B)
with Eqn.~\ref{eq:fitfunction} is plotted in Fig.\ \ref{fig:AK3fit}.
The extracted lifetime is $\tau_{\mu} = 2~196~980.1 \pm 2.5$~ps.
Both the fit $\chi^2/\mathrm{dof}$ of $1224/1185$
and the distribution of the residuals between the 
data points and the fit function indicate
a reasonable fit and no evidence of any distortions
such as uncanceled $\mu$SR effects,
mis-corrected pulse pileup or time-dependent backgrounds.

The lifetime results for the benchmark fits
to the separate R06-A and R06-B datasets 
of $\tau_{\mu} =  2~196~979.2 \pm 3.4$~ps 
and $\tau_{\mu} = 2~196~981.3  \pm 3.8$~ps
are in good agreement (the two datasets have reversed $B$-fields).
The $\chi^2/\mathrm{dof}$ of the fits 
and the distributions of the residuals are also acceptable.

\subsubsection{Fit to the quartz data}
\label{sec:quartzfit}

A more detailed approach to LF/TF $\mu$SR effects is used 
to extract the lifetime from quartz data.
First, geometry-dependent effective lifetimes are extracted
for each tile-pair from fits using a modified version of Eqn.~\ref{eq:fitfunction}
accounting for transverse-field precession.
Then, $\tau_{\mu}$ is extracted from the effective lifetimes via a fit function 
that accounts for geometry-dependent longitudinal-field relaxation.
Together these two steps account for all observed features 
of TF/LF $\mu$SR in quartz.

In step one the 170 time histograms of all tile-pairs are fit to
\begin{equation}
N(t) = N e^{-t/\tau_\mathrm{eff}} [ 1 + f(t) ] + C
\label{eq:Lambda:d}
\end{equation}
where $\tau_\mathrm{eff}$, $N$ and $C$ are the effective lifetime, normalization constant
and background amplitude, respectively.
The additional term $f(t)$ accounts for the TF $\mu$SR signal
according to
\begin{equation}
f(t) =  A \sin{\theta_B} ~ P_2 e^{-t/T_2} \sin(\omega t+\phi) 
\label{eq:Lambda:d2}
\end{equation}
where $P_2$, $T_2$, $\omega$ and $\phi$ are the initial polarization, 
relaxation constant, angular frequency and phase
of the TF $\mu$SR signal. 
The quantity $A$ is the asymmetry parameter of the $e^+$-angular distribution
and the angle  $\theta_B$ is the tile coordinate relative to the $B$-field axis.
The product $A \sin{\theta_B}$ determines 
the geometry-dependent amplitude of the TF $\mu$SR signal 
in each tile pair (see Sec.\ \ref{sec:muSR}).

A 5-parameter fit using Eqn.\ \ref{eq:Lambda:d} is employed
to extract the 170 tile-pair effective lifetimes.
The parameters $\tau_\mathrm{eff}$, $N$, $P_2$, $\phi$
and $C$ are varied.
The parameters $T_2$ and $\omega$ are fixed to the weighted averages 
of their best-fit values from an initial 
7-parameter fit to a sub-set of tile-pairs having large TF signals.
The asymmetry parameter is always set at $A = +1/3$.
Any difference between this theoretical energy-integrated asymmetry
and the detector efficiency-weighted asymmetry is subsumed
into $P_2$.

In step two the distribution of effective lifetimes
versus tile-pair coordinates $( \theta_B , \phi_B )$ about the $B$-field axis 
is fit to
\begin{equation}
\tau_\mathrm{eff} ( \theta_B , \phi_B  ) = \tau_{\mu} ( 1 +  \delta ( \theta_B , \phi_B ) )  
\label{eq:fitfunction:sophisticated}
\end{equation}
where $\tau_\mu$ is the true muon lifetime and  $ \delta ( \theta_B, \phi_B  )$
is the tile-coordinate-dependent lifetime shift due to LF $\mu$SR effects.
The scheme for incorporating LF $\mu$SR, which holds
when $T_1 / P_1  \gg \tau_\mu $,\footnote{In the quartz production data
the longitudinal polarization is $P_1 \leq 0.01$ and longitudinal 
relaxation  constant is $T_1 = 28 \pm 8$~$\mu$s and therefore
the condition $T_1 / P_1  \gg \tau_\mu $ is fulfilled. For further details
see Sec.\ \ref{supplemental muSR}.} yields
\begin{equation}
\delta ( \theta_B , \phi_B ) = \tau_{\mu} \left( { P_1 } \over { T_1 } \right) A \cos{ \theta_B}   ,
\label{eq:fitfunction:sophisticated2}
\end{equation}
where $P_1$ is the initial longitudinal polarization
and $T_1$ is the longitudinal relaxation constant.
The factor $A \cos \theta_B$ incorporates the asymmetry parameter $A$ of 
the $e^+$-angular distribution and the tile coordinate $\theta_B$ relative 
to the $B$-field axis. It determines the amplitude of the LF $\mu$SR signal 
in each tile pair (see Sec.\ \ref{sec:muSR}).

To extract $\tau_{\mu}$,
a 2-parameter fit using 
Eqn.\ \ref{eq:fitfunction:sophisticated}
is performed varying both the muon lifetime $\tau_\mu$ 
and the initial polarization $P_1$.
The relaxation constant $T_1$ is fixed at $T_1 = 28$~$\mu$s 
(see Sec.\ \ref{sec:LFQuartzmuSR}), a choice 
that influences the ``best-fit'' value for the 
polarization $P_1$ but not $\tau_{\mu}$.
The asymmetry parameter $A$ in Eqn.\ \ref{eq:fitfunction:sophisticated2} depends
on the coordinates $( \theta_B , \phi_B )$ of the tile pair.
This occurs because positron absorption and scattering processes
by intervening materials yield a 
geometry-dependent angular and energy positron distribution in tile pairs.
Therefore tile-dependent values of $A ( \theta_B , \phi_B )$ obtained from a GEANT simulation 
of the experimental setup are used. 
The asymmetries range from $+0.30$ to $+0.40$.\footnote{In practice
the difference in $\tau_\mu$ between using the GEANT calculated asymmetries $A( \theta_B , \phi_B )$
and theoretical energy-integrated $A = 1/3$ is 0.11~ppm.}

Note the tile coordinates $( \theta_B , \phi_B )$ are calculated relative to the center of the
muon stopping distribution not the detector array.
The position offsets $( \delta x, \delta y, \delta z )$ of the stopping distribution
from the detector center are obtained from the measured distribution of the
outgoing positrons in the tile array. The offsets are determined
to a precision of $\pm$2~mm.

\begin{figure}
\includegraphics[width=0.9\linewidth]{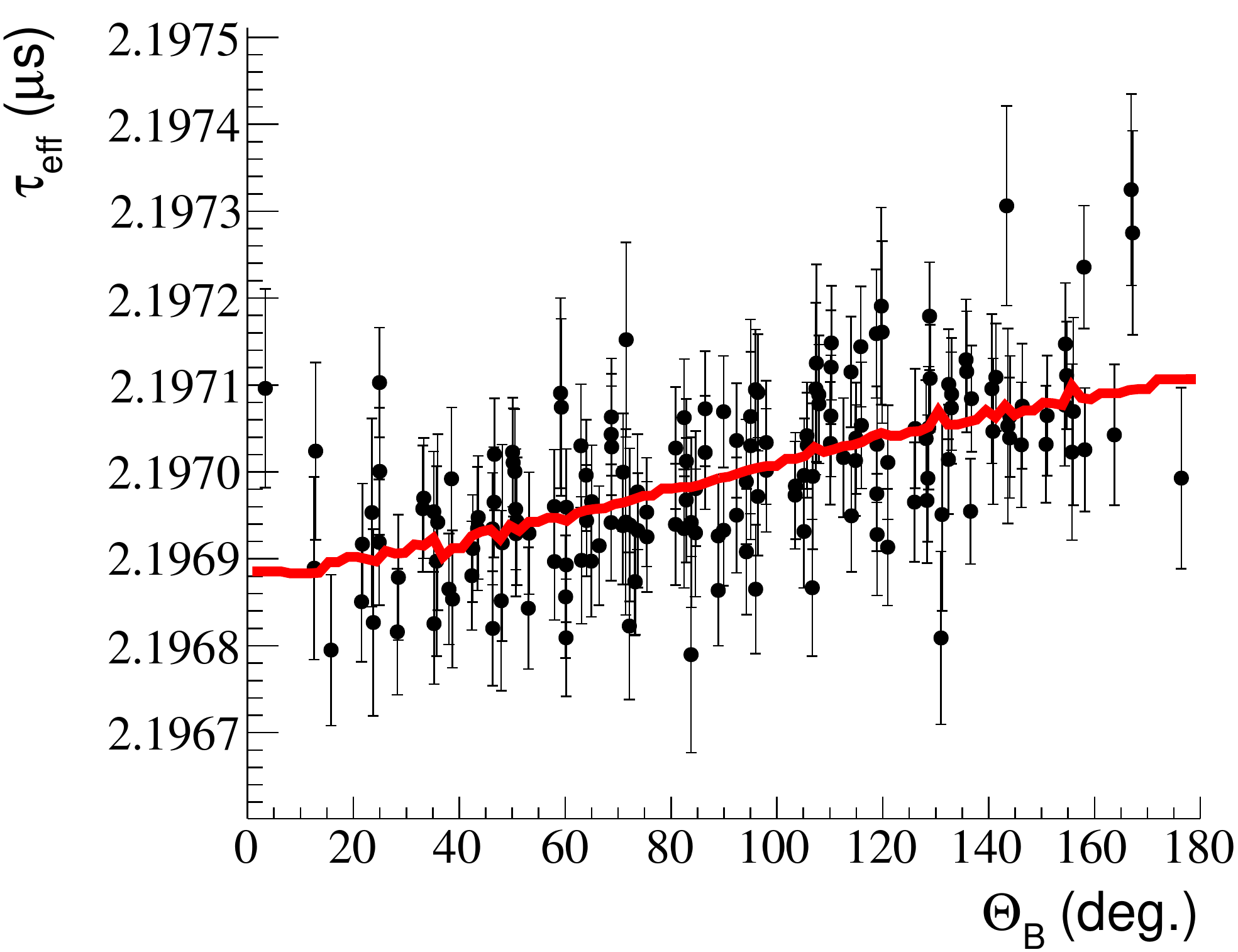}
\caption{
Effective muon lifetime $\tau_\mathrm{eff}$ versus tile coordinate $\theta_B$
relative to the $B$-field axis. The data points are the 170 individual effective lifetime measurements and the
solid curve is the effective lifetime fit from Eqn.\ \ref{eq:Lambda:d}.
The small irregularities in the solid curve arise from
the $\phi_B$-dependence of the asymmetry parameter $A$.}
\label{fig:tau_eff:sophisticated:ds201}
\end{figure}

\begin{figure}
\includegraphics[width=0.7\linewidth]{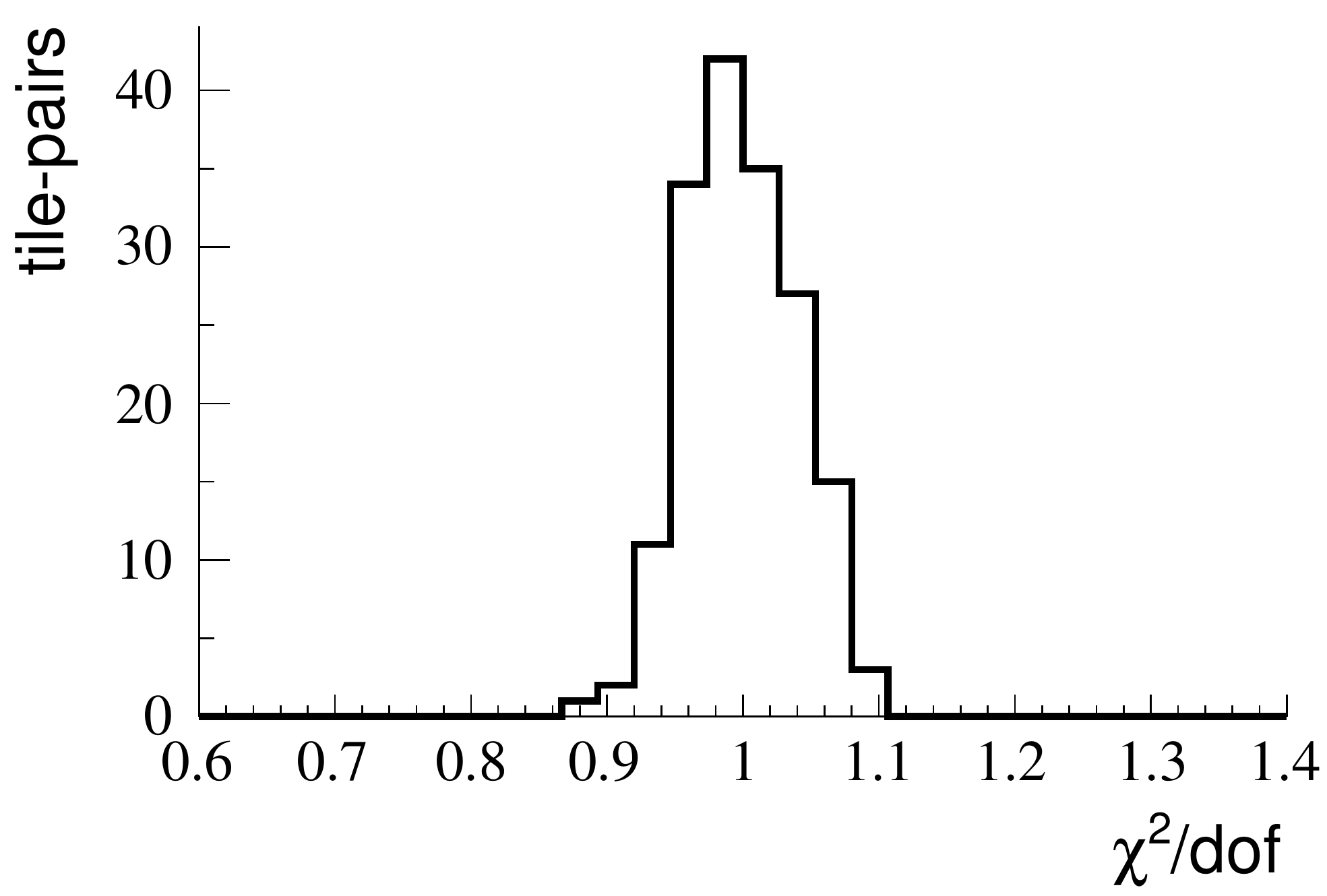}
\caption{The distribution of $\chi^2/$dof values from the 170 fits
to the individual time histograms for the R07-A dataset.
The fit function (Eqn.\ \ref{eq:Lambda:d}) incorporates
the TF $\mu$SR signal.}
\label{fig:chi2_sophisticated:ds_201}
\end{figure}

Figs.\ \ref{fig:tau_eff:sophisticated:ds201} and \ref{fig:chi2_sophisticated:ds_201}
show for dataset R07-A the fit to the angular distribution of the 170 effective lifetimes $\tau_\mathrm{eff} ( \theta_B , \phi_B )$
and the results for the $\chi^2$/dof values from the 170 tile-pair time fits.
The distribution of the $\chi^2$/dof values from the time fits
has an average $1.0$, which demonstrates that TF $\mu$SR effects are correctly handled.
The fit of the $\tau_\mathrm{eff}$-distribution 
has a $\chi^2/\mathrm{dof}$ of $201/168$, 
which demonstrates the LF $\mu$SR effects are resonably handled.
The $\chi^2$ of fits to decay time and $\tau_\mathrm{eff}$ distributions
for other quartz datasets are of similar quality (see Table \ref{tbl:R07:fitsummary}.

Table \ref{tbl:R07:fitsummary} summarizes the parameters derived 
from the application of this method to the quartz datasets.
There is good agreement between the $\tau_{\mu}$-values extracted from the six datasets.

\begin{table*}[htbp]
\centering
\caption{Results of the fitting method for the six quartz datasets.
The entries for the transverse polarization $P_2$ are weighted averages obtained 
from tile-pairs having large TF $\mu$SR signals. 
The entries for the longitudinal polarization $P_1$ are individual values obtained from  
effective lifetime fits.
The parameter $T_2 = 4.7$~$\mu$s is fixed
from fits to time histograms with large TF $\mu$SR signals
and the parameter $T_1 = 28$~$\mu$s is fixed from fits to supplemental
data with amplified LF $\mu$SR signals.
The offsets $( \delta x , \delta y, \delta z )$ are fixed 
from the geometrical distribution of hits over the positron detector.}
  \footnotesize
  \label{tbl:R07:fitsummary}
  \begin{tabular}{lcccccc}
    \hline
    \hline
     parameter            &  \multicolumn{6}{c}{data set}    \\
                          & R07-A                      
                          & R07-B                       
                          & R07-C                      
                          & R07-D                     
                          & R07-E                     
                          & R07-F \\
    \hline
    $P_2 \times 1000$     & $2.54\pm0.02 $      
                          & $2.45\pm0.02$        
                          & $2.35\pm0.05$ 
                          & $2.51\pm0.07$ 
                          & $2.28\pm0.05$ 
                          & $2.54\pm0.06$ \\
    $\omega$ (Rad/$\mu$s) & $11.4440 \pm 0.0061 $ 
                          & $11.5342 \pm 0.0066 $ 
                          & $11.570  \pm 0.016  $           
                          & $11.489  \pm 0.019  $           
                          & $11.579  \pm 0.016  $   
                          & $11.498  \pm 0.018  $ 
                          \\
    $P_1 \times 1000$     & $1.3\pm0.2$         
                          & $1.6\pm0.2$ 
                          & $1.3\pm0.4$ 
                          & $1.2\pm0.7$ 
                          & $0.7\pm0.5$ 
                          & $1.5\pm0.6$ 
                          \\
    $\tau_\mu$ (ps)       & $ 2196981.93 \pm  5.86 $
                          & $ 2196980.94 \pm  5.51 $
                          & $ 2196975.43 \pm 15.11 $
                          & $ 2196967.24 \pm 22.79 $
                          & $ 2196985.33 \pm 15.66 $
                          & $ 2196991.11 \pm 20.40 $ 
                          \\
    $\chi^2$/dof          & 200.9/168 %
                          & 174.2/168 %
                          & 154.9/168 %
                          & 153.0/168 %
                          & 167.1/168 %
                          & 172.1/168 %
                          \\
    beam $x$ (cm)         &  -0.3 &  -0.4 & -1.4  &  -0.7  &  -0.9 & -1.0  \\
    beam $y$ (cm)         &  -0.3 &   0.1 &  0.9  &   0.75 &  -0.1 &  0.6  \\
    beam $z$ (cm)         &  0.07 &  0.06 &  0.06 &   0.06 &  0.06 & -0.1  \\ 
    \hline
    \hline
  \end{tabular} 
\end{table*}

Because the six datasets involve different magnet orientations 
and different detector positions, the fitted values of
the initial longitudinal and transverse polarizations may differ.\footnote{In particular, 
$P_1$ is very sensitive to the exact angle 
between the Halbach magnet field 
and the muon spin axis.}
The small values of $P_1$ result from
the perpendicular arrangement of the muon spin and the $B$-field
while the small values of $P_2$ results 
from the spin dephasing of the muon stops.

\subsection{Supplemental information on  $\mu$SR effects}
\label{supplemental muSR}

In AK-3 the 3-parameter fit is successful 
because $\mu$SR signals are weak and cancel in sums of opposite tile pairs.
In quartz the 5-parameter time distribution fits 
and 2-parameter $\tau_\mathrm{eff}$ distribution fit
accounts for all observed TF/LF $\mu$SR distortions.  
We describe here supplemental information
that further establishes that the $\mu$SR effects 
in AK-3 and quartz were handled correctly.

\subsubsection{$\mu$SR studies in AK-3}
For decay times $t > 100$~ns, no evidence for TF $\mu$SR is observed in AK-3.
The absence of TF signals is expected as a consequence   
of the different axes
and the varying frequencies
of the spin rotation in the 
individual AK-3 magnetic domains.

A tiny TF $\mu$SR distortion, compatible with the fast relaxation
of the transverse polarization of the last muon arrivals
during the accumulation period, is seen at times $t < 100$~ns.
The signal is consistent with results from
$\mu$SR studies on AK-3 by Morenzoni {\it et al.}\ \cite{Morenzoni} and 
Scheuermann {\it et al.}\ \cite{Scheuermann}.
Both studies found evidence of spin relaxation
with time constants of $\sim$14~ns and $\sim$50~ns,
but no indications of longer timescales. 

\begin{figure}
\includegraphics[width=0.9\linewidth]{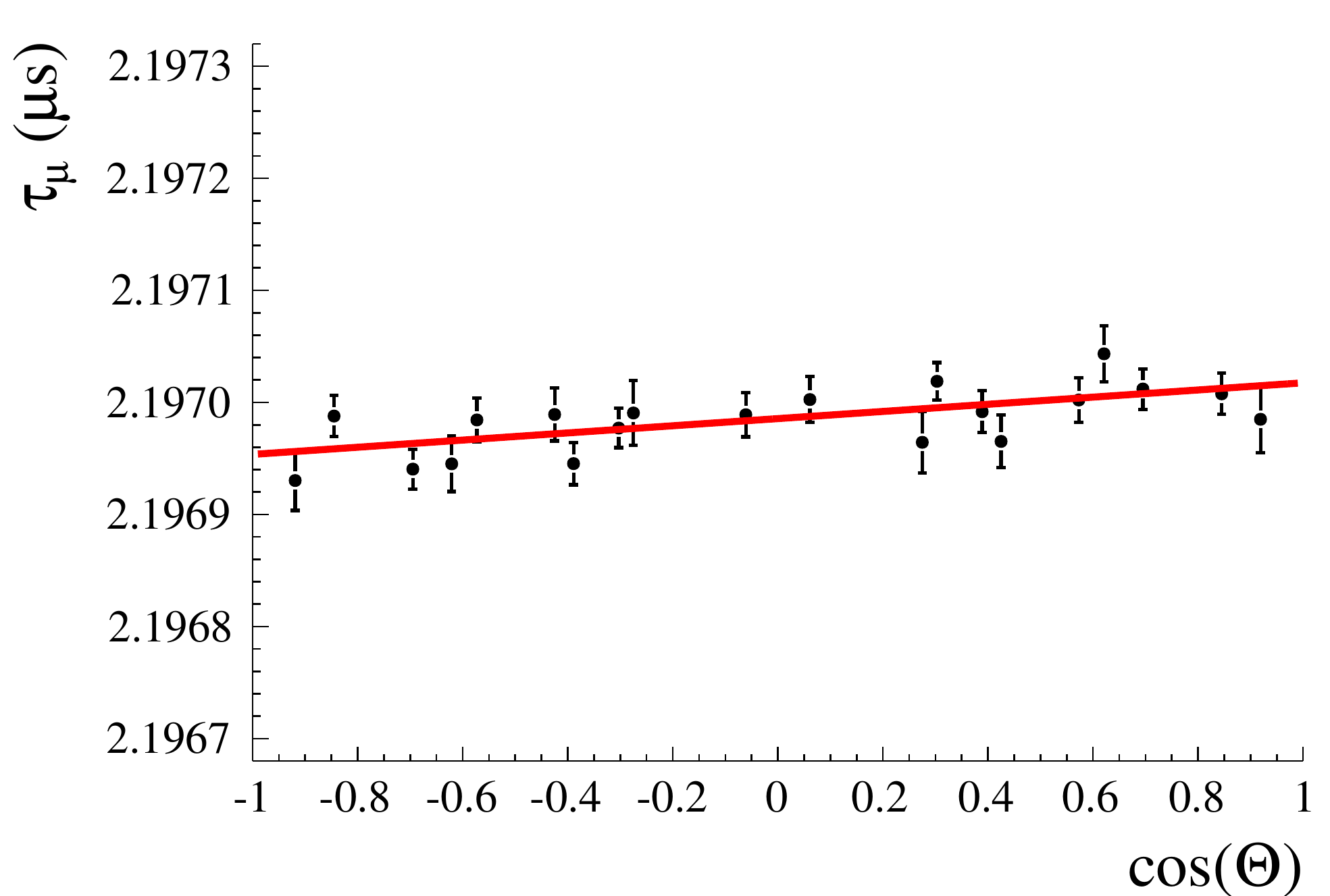}
\includegraphics[width=0.9\linewidth]{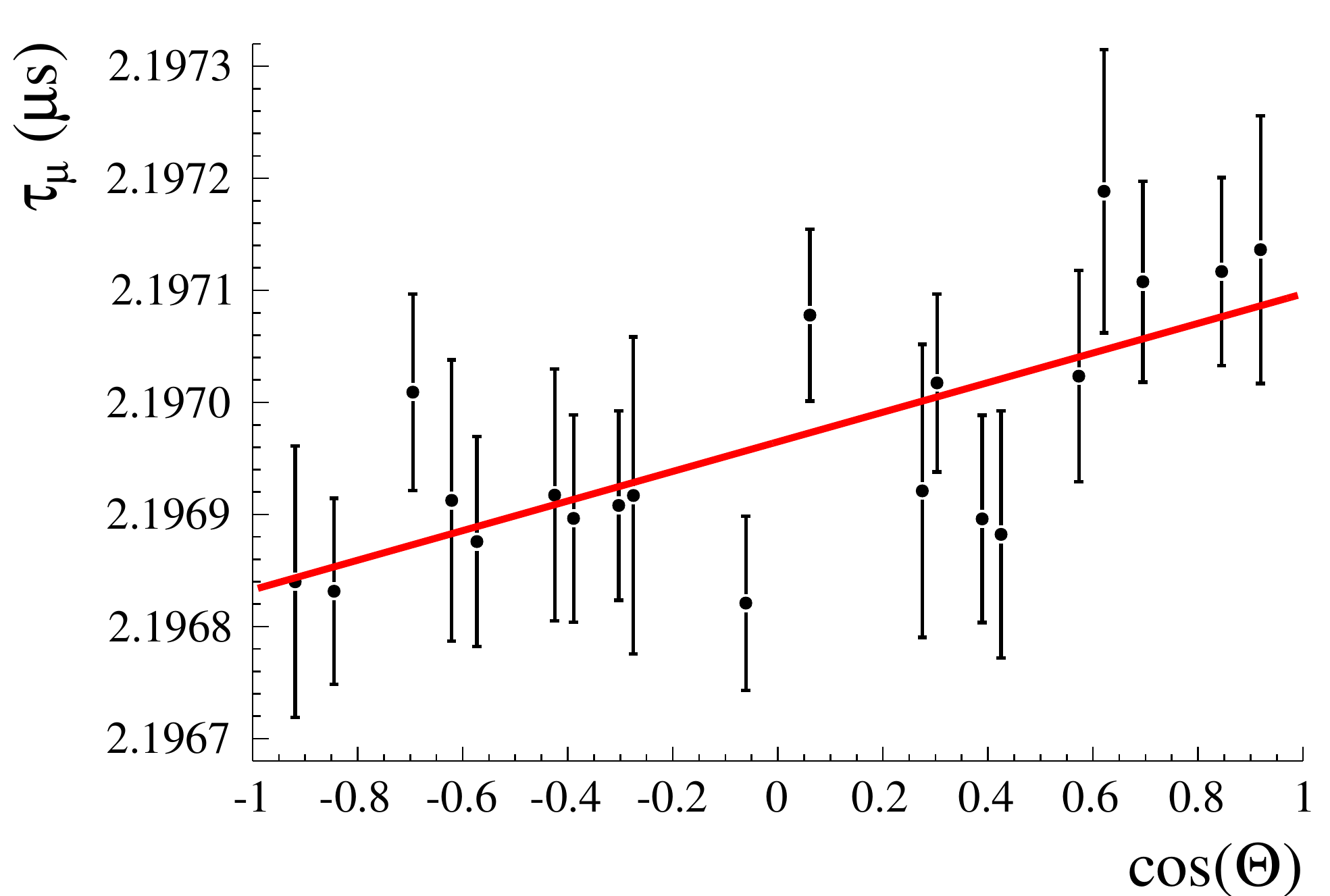}
\caption{Plots of the fitted lifetimes from 
the individual time histograms grouped by the detector angle $\theta$
relative to the beam axis. The upper panel represents
the AK-3 production data with the magnetization axis at 90 degrees
to the beam axis.  The lower panel represents
the AK-3 supplemental data with the magnetization axis at 45 degrees
to the beam axis. The four-fold increase in the $\theta$-dependence of
the fitted lifetime in the lower plot reflects the increase 
in the longitudinal polarization for the 45 degree dataset. Note the
vertical scales on the two plots are the same.}
\label{fig:AK3_LFmuSR}
\end{figure}

As shown in Fig.\ \ref{fig:AK3_LFmuSR}, a tiny variation  
of the fitted lifetime versus the detector angle $\theta$
relative to the beam axis is observed. This effect is attributed
to the slow relaxation of the longitudinal polarization $P_1$.

The longitudinal relaxation axis is approximately parallel to the beam axis
while the AK-3 magnetization axis is oriented perpendicular to the beam axis.
We believe this originates  from the partial alignment 
of the AK-3 magnetic domains along the AK-3 magnetization axis.
Consequently, while $P_1$
accrues only small contributions from stops 
in domains oriented nearly perpendicular to the $\mu$-spin axis,
it accrues much larger contributions from stops 
in domains  oriented more parallel to the $\mu$-spin axis. 
The result is an ensemble-averaged longitudinal polarization
that is roughly parallel to the beam axis.

The LF $\mu$SR effect on the fitted lifetime versus  
detector coordinate $\theta$ is roughly 10~ppm.
Supplemental data was acquired
that enhanced $P_1$.
The circular-cross section target, 
oriented at 90$^{\circ}$ to the beam axis, 
was replaced by an elliptical-cross section 
target, oriented at $45^{\circ}$ 
to the beam axis.
This increased the component of the AK-3 magnetization 
along the $\mu$-spin axis.
The elliptical-target data shows a four-fold increase in lifetime variation 
with detector coordinate $\theta$.
This is qualitatively consistent with the expected increase 
in the polarization $P_1$.\footnote{A quantitative prediction of the 
longitudinal polarization increase for the elliptical target 
requires a detailed knowledge of the domain orientations. 
However, the observed four-fold increase 
in the $\theta$-variation of the fitted lifetime 
is consistent with a simple model of populations of 
fully aligned domains and randomly aligned domains that reproduces the ratio 
between the known AK-3 remnant and saturation fields \cite{AK3}.} 
It supports our argument that the $\theta$-dependence of the fitted lifetime
originates from a longitudinal relaxation in the AK-3 target, 
and therefore will cancel in sums of opposite tile pairs.

\subsubsection{Transverse field $\mu$SR studies in quartz}
\label{sec:TFQuartzmuSR}

\begin{figure}
\includegraphics[width=0.9\linewidth]{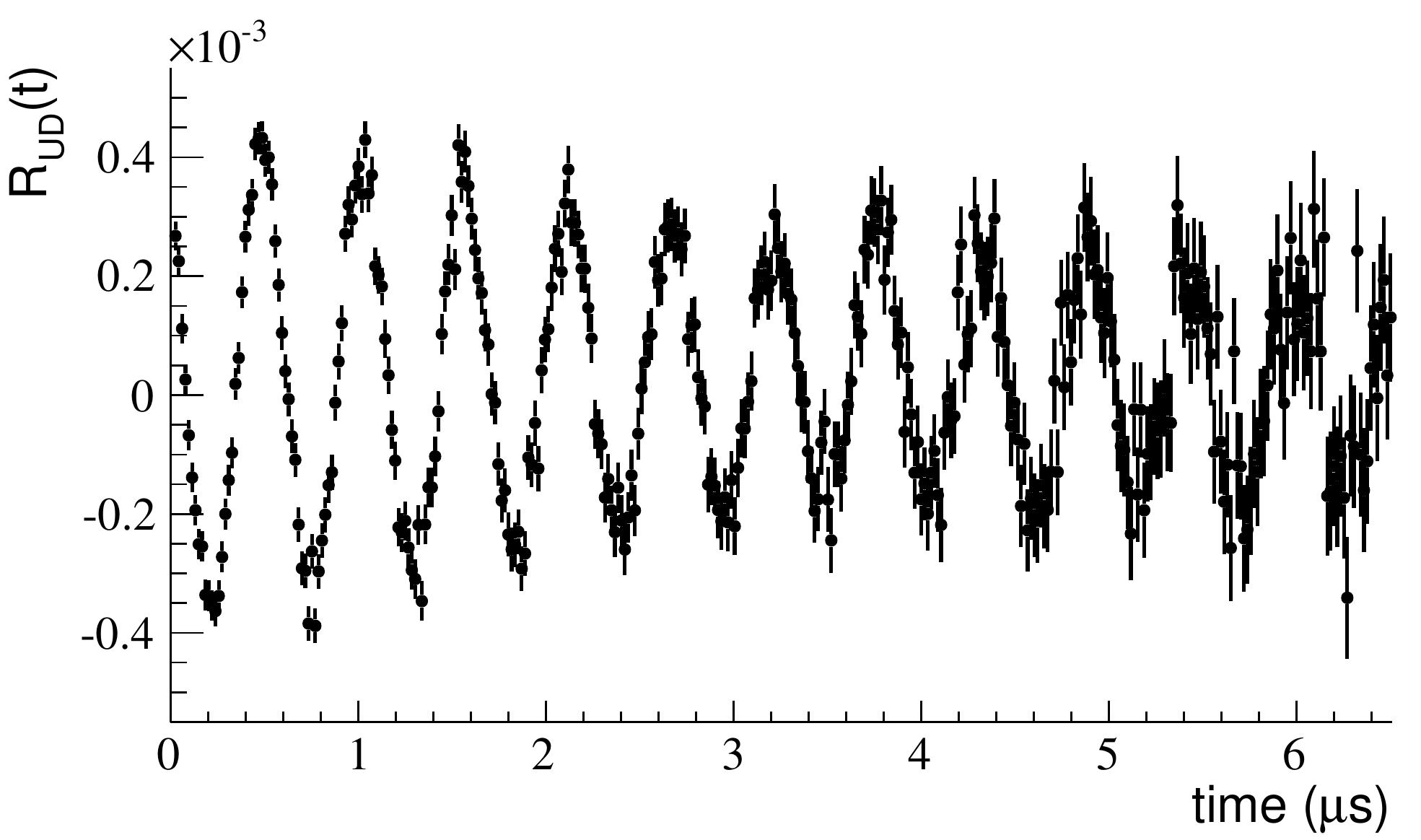}
\includegraphics[width=0.9\linewidth]{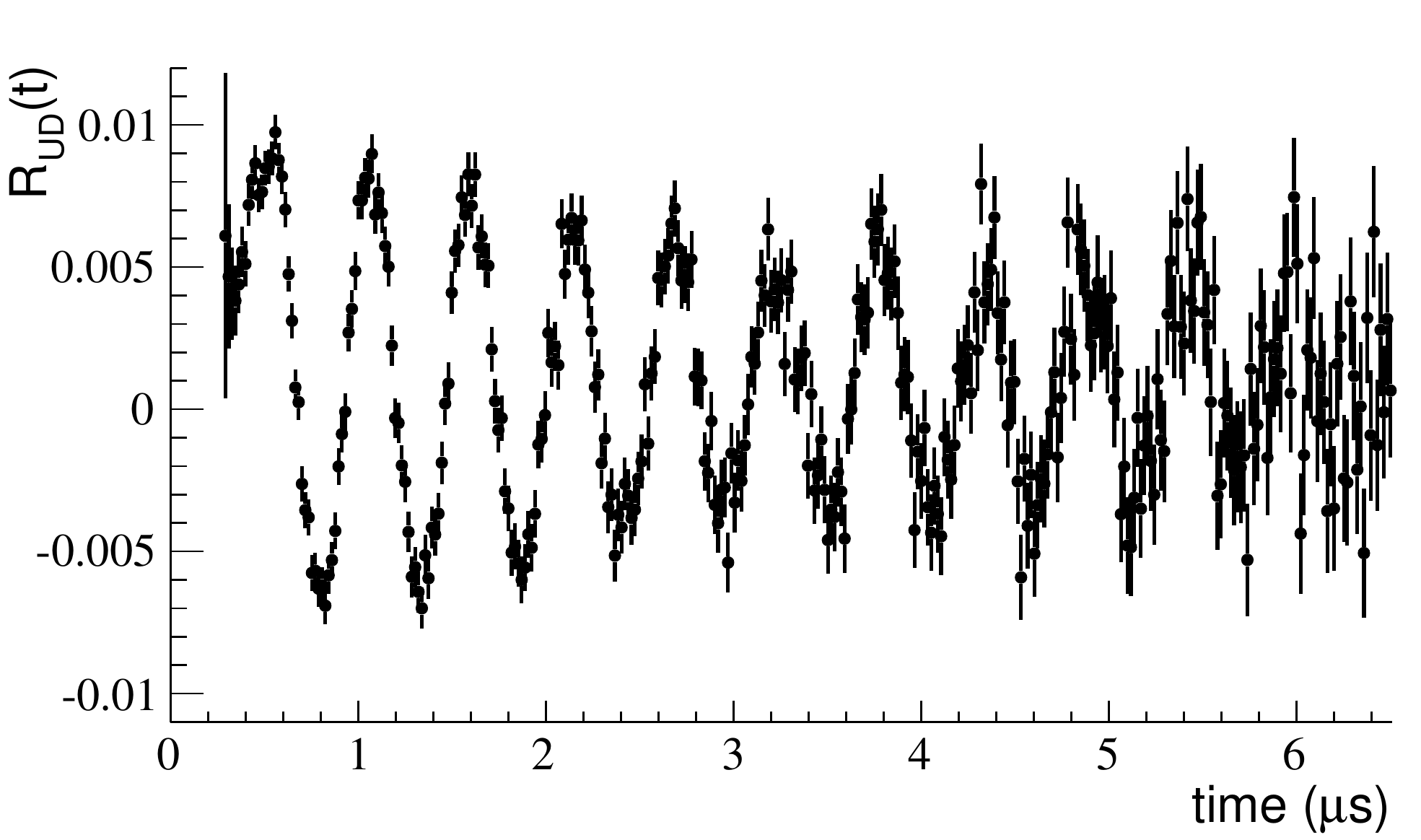}
\caption{Ratio $R_{UD} (t)$ of the normalized difference between the time distributions
in the upstream/downstream hemispheres of the positron detector. The upper
panel corresponds to production data with an accumulation period $T_A = 5.0$~$\mu$s
and the lower panel corresponds to supplemental data with a shortened accumulation period $T_A = 0.15$~$\mu$s
(note the different scales of the vertical axes).
The plots highlight the contribution of the TF $\mu$SR signal in the quartz data.}
\label{fig:R_UD}
\end{figure}

Figure \ref{fig:R_UD} (upper panel) shows the ratio 
\begin{equation}
R_{UD} (t) = {  N_U (t) - N_D (t)  \over  N_U (t) + N_D (t)  } 
\end{equation}
involving the normalized time distributions of the coincidence hits $N_{U/D} (t)$
in the upstream/downstream hemispheres of the positron detector.\footnote{
A subtraction of the time-independent background
and a normalization of the summed histogram contents,
is performed before deriving the ratio $R_{UD} (t)$ from 
the $N_{U/D} (t)$ spectra.}
This diagnostic is sensitive to time-dependent differences
between upstream/downstream hits and
clearly shows a TF precession signal  with a frequency $\omega \sim 11$~$\mu$s$^{-1}$ 
and a time constant $T_2 \sim 5$~$\mu$s.
The value of $\omega$ 
agrees with the expected value for a diamagnetic $\mu^+$ population 
in the 130~G Halbach magnet field.

The relative amplitude of the TF $\mu$SR signal in the production data
is only $\sim$$10^{-3}$.
To further study the TF $\mu$SR signal,
a supplemental dataset was acquired by shortening  
the accumulation period from $T_A = 5.0$ to $0.15$~$\mu$s
and thereby reducing roughly 20-fold the spin dephasing during muon accumulation.
As depicted in Fig.\ \ref{fig:R_UD} (lower panel), the ratio $R_{UD}(t)$ obtained with $T_A = 0.15$~$\mu$s
shows a large TF signal consistent with a 
twenty-fold enhancement of the production data signal.
Fits using Eqn.\ \ref{eq:Lambda:d} to the time histograms derived  
from the $T_A = 0.15$~$\mu$s dataset
indicate a quartz diamagnetic $\mu^+$ population
of roughly  6-7\%, in good agreement with published work \cite{Dawson:1994,Brewer2000425}.

\subsubsection{Longitudinal field $\mu$SR studies in quartz}
\label{sec:LFQuartzmuSR}

\begin{figure}
\includegraphics[width=0.9\linewidth]{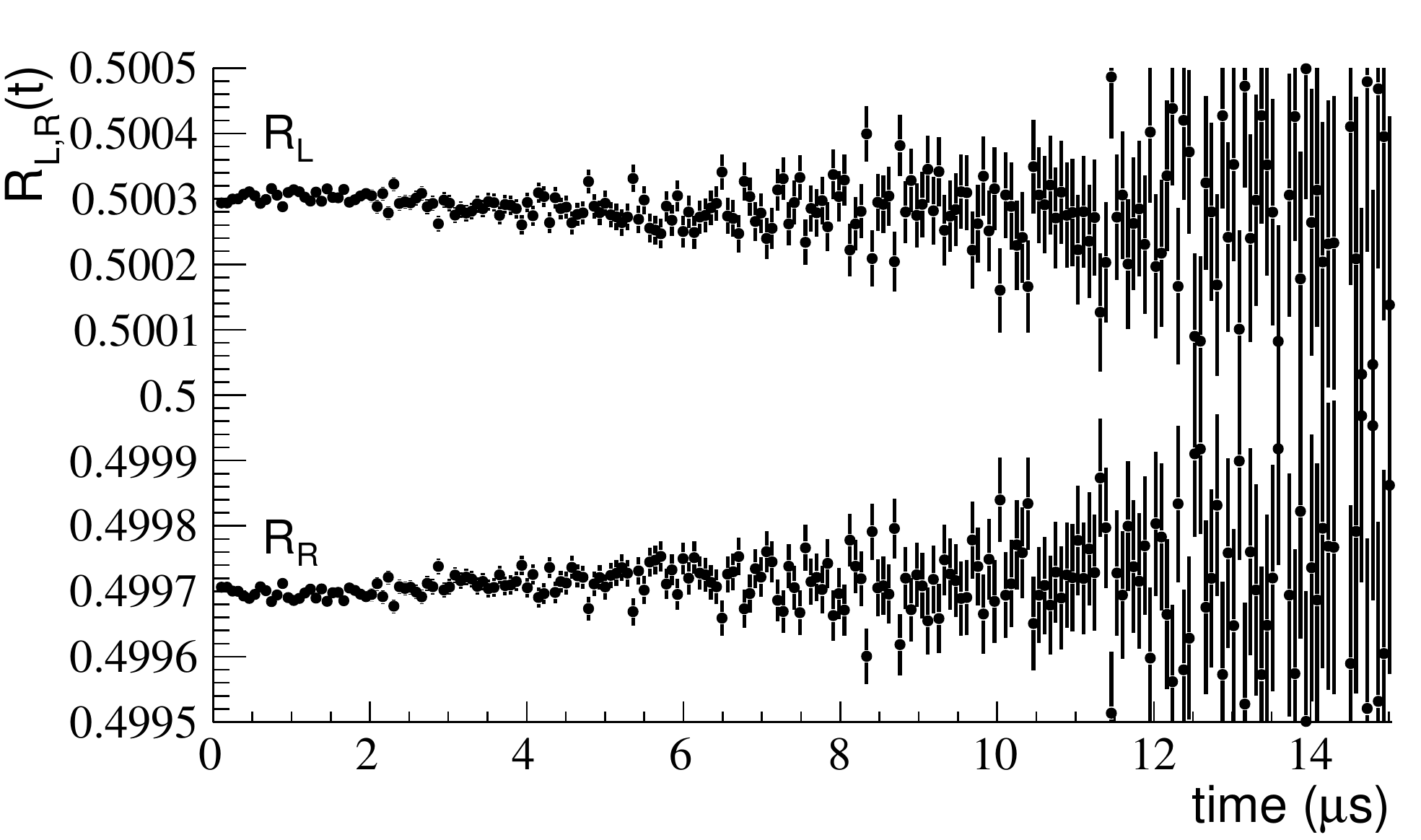}
\includegraphics[width=0.9\linewidth]{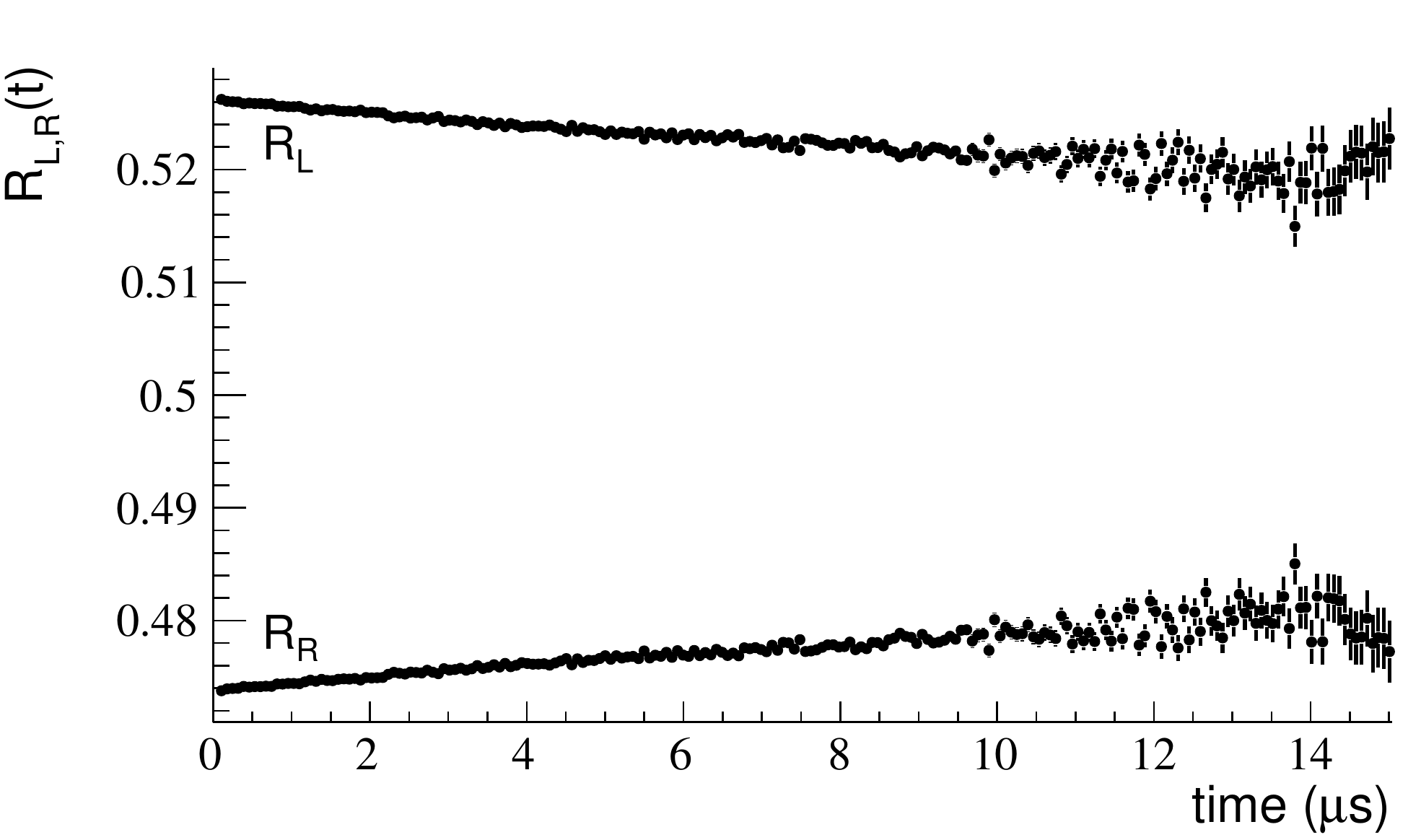}
\caption{Ratio of hit time distributions from the beam-left hemisphere ($R_\mathrm{L} (t)$) and
the beam-right hemisphere ($R_\mathrm{R} (t)$) to the
detector sum.  The upper
panel corresponds to production data with a longitudinal polarization $P_1 \sim 0.01$
and the lower panel corresponds to supplemental data with a longitudinal polarization $P_1 \sim 0.25$.
The gradual slopes in $R_\mathrm{L/R} (t)$ are the effect of the LF $\mu$SR in the quartz data.}
\label{fig:R_LR}
\end{figure}

The two ratios 
\begin{eqnarray}
R_L (t) & = &  N_L (t) / ( N_L (t) + N_R (t) ) \notag \\ 
R_R (t) & = &  N_R (t) / ( N_L (t) + N_R (t) ) \\ 
\notag 
\end{eqnarray}
are plotted in Fig.\ \ref{fig:R_LR}.
$N_{L}(t)$ and $N_{R}(t)$ correspond to the normalized time distributions of the coincidence hits
in the beam-left and beam-right hemispheres of the positron detector, respectively.\footnote{A 
(i) subtraction of the time-independent background 
and (ii) normalization of the summed histogram contents is performed
before deriving the ratios $R_{L/R} (t)$ from the $N_{L/R} (t)$ spectra.}
These rates are sensitive to time-dependent differences
between beam-left and right hits and suggest a gradual decrease (increase) 
in $R_L (t)$ ($R_R (t)$) consistent with a long-timescale relaxation 
of the muon longitudinal polarization along the $B$-field axis.
The time constant $T_1$ of the relaxation 
is much longer than the $\mu^+$ lifetime.

The relative amplitude of the LF $\mu$SR signal in the
production data is only $10^{-4}$.
Supplemental datasets were acquired with
enhanced LF effects by re-orientating the Halbach magnet
from a perpendicular alignment to the $\mu^+$ spin axis ($P_1 \leq 0.01$)
to non-perpendicular alignments ($P_1 \sim 0.05 - 0.25$).

Fig.\ \ref{fig:R_LR} shows the ratios $R_{L/R} (t)$ for
the $P_1 \sim 0.25$ dataset and clearly indicates
a gradual decrease (increase) in $R_L(t)$ ($R_R(t)$) 
consistent with a roughly twenty-fold enhancement 
of the LF $\mu$SR signal.
Fits of Eqn.\ \ref{eq:fitfunction:sophisticated} to the time histogram from the 
$P_1 = 0.05$ to $0.25$ datasets give a 
relaxation constant $T_1 = 28 \pm 8$~$\mu$s.

\subsection{Consistency checks}

A number of checks are
performed on the overall consistency of the lifetime results. 
They include comparisons of fit results for
different time ranges, run groups, detector positions 
and $B$-field orientations. 

\begin{figure}[ht] 
  \centering
\includegraphics[width=0.90\linewidth]{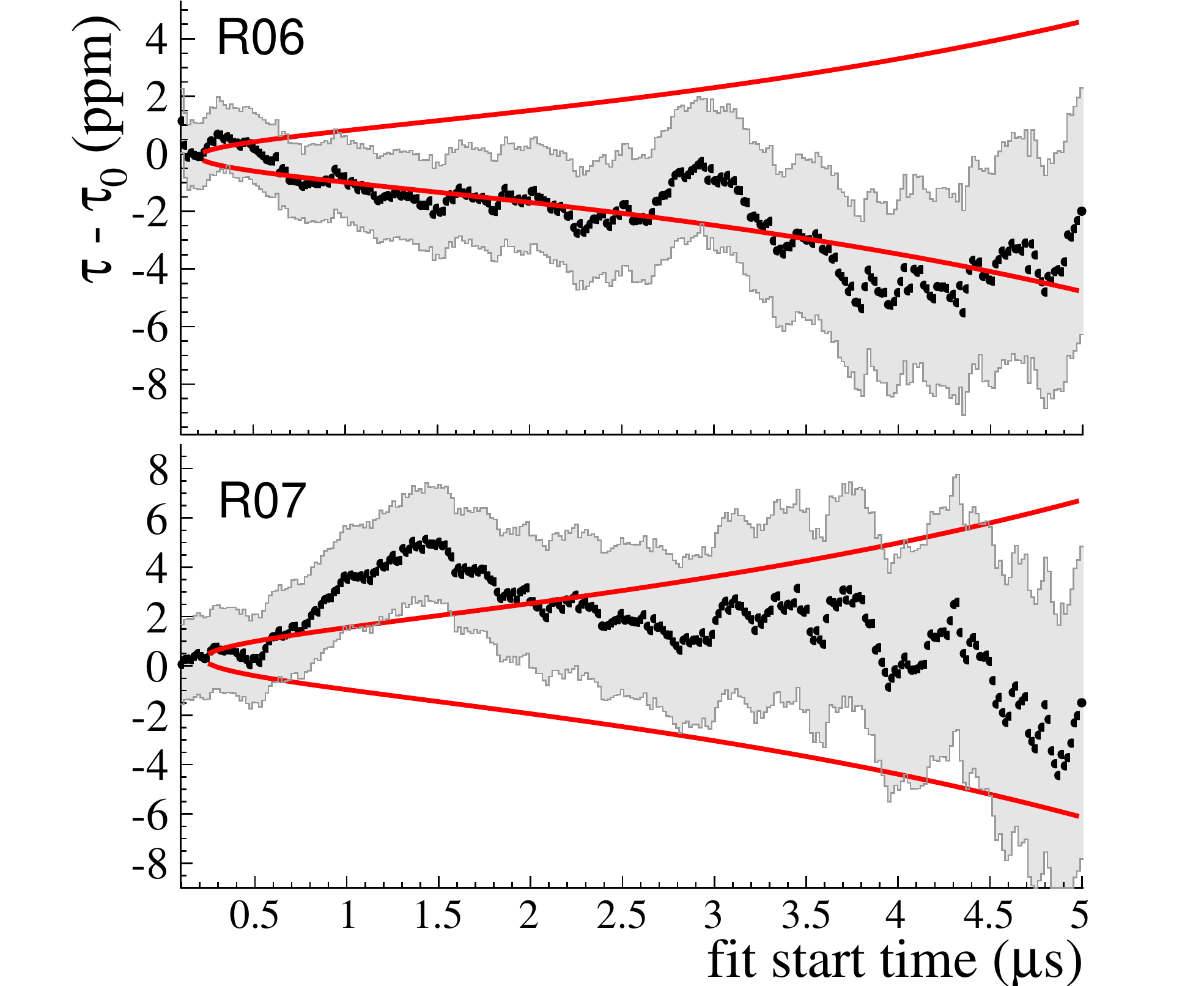}
\caption{Muon lifetime versus fit start-time for the AK-3
data (top) and the quartz data (bottom).
The data points and gray band indicate the lifetime results 
and corresponding uncertainties from individual fits.
The solid envelopes indicate 
the permissible statistical deviations (1$\sigma$) 
from the benchmark fits with a start-time 
$t_{start} = 0.2$~$\mu$s. The $1\sigma$ deviations 
are calculated with appropriate accounting
for dataset correlations between different fit ranges.}
  \label{fig:R_vs_starttime}
\end{figure}

Fig.\ \ref{fig:R_vs_starttime} shows the fitted lifetime 
versus the fit start time from $t_{start} = 0.1$ to $5.0$~$\mu$s
for the AK-3 and quartz data.
The points indicate the lifetime results 
from individual fits and the solid curves 
their permissible deviations (1$\sigma$) 
from the benchmark fit with a start time 
$t_{start} = 0.2$~$\mu$s. 
The deviations account for the
correlations between the individual fits
and the benchmark fit.
No evidence is seen for either a start-time
dependence or a stop-time dependence of
the lifetime.
This builds confidence in the handling 
of known time-dependent effects including
$\mu$SR signals, positron pileup
and gain changes, as well as the absence of any 
unidentified time-dependent distortions.

\begin{figure}
\begin{minipage}[t]{0.485\linewidth}
\centering
\includegraphics[width=\linewidth]{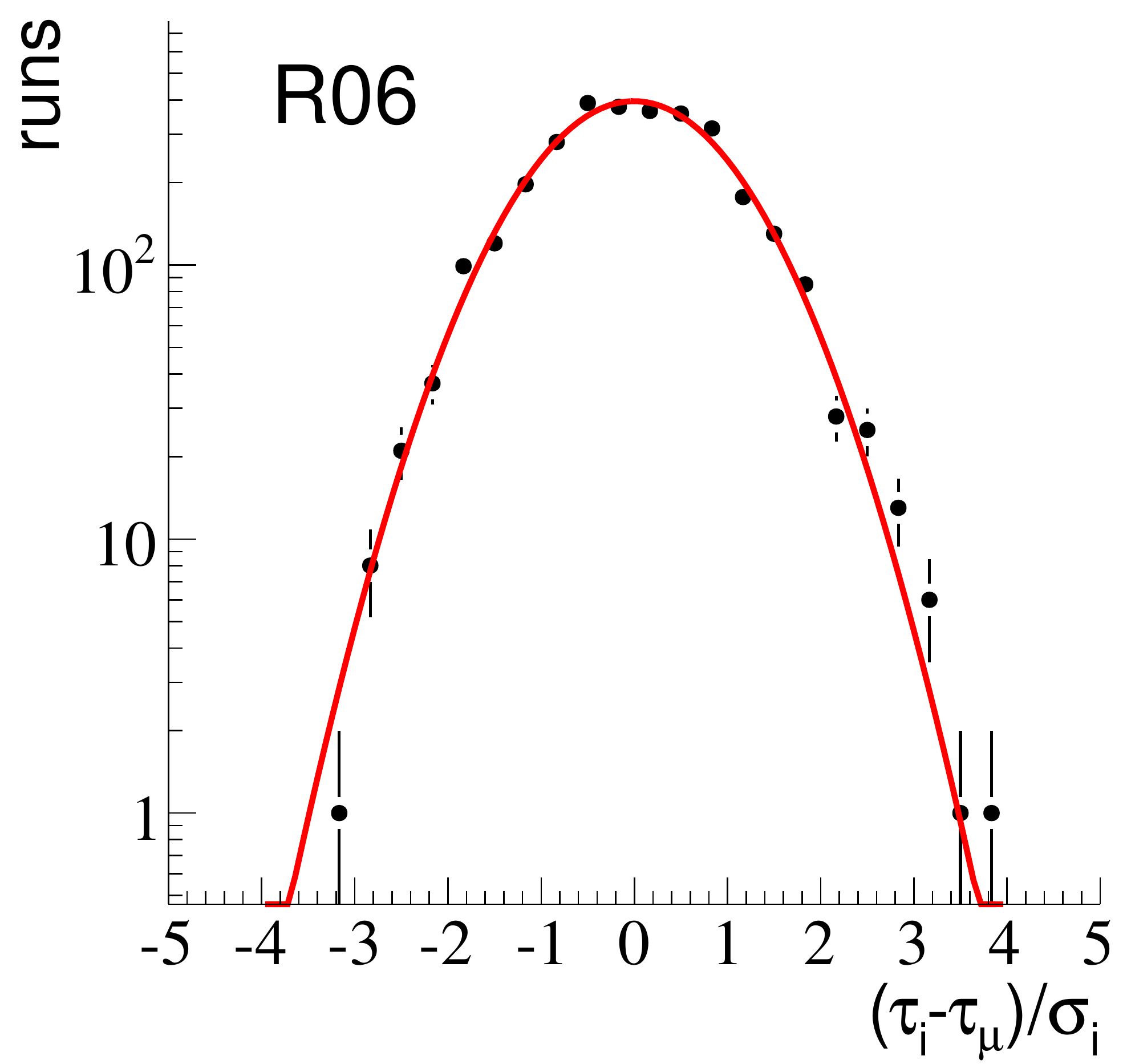}
\end{minipage}
\hspace{0.001\linewidth}
\begin{minipage}[t]{0.485\linewidth}
\centering
\includegraphics[width=\linewidth]{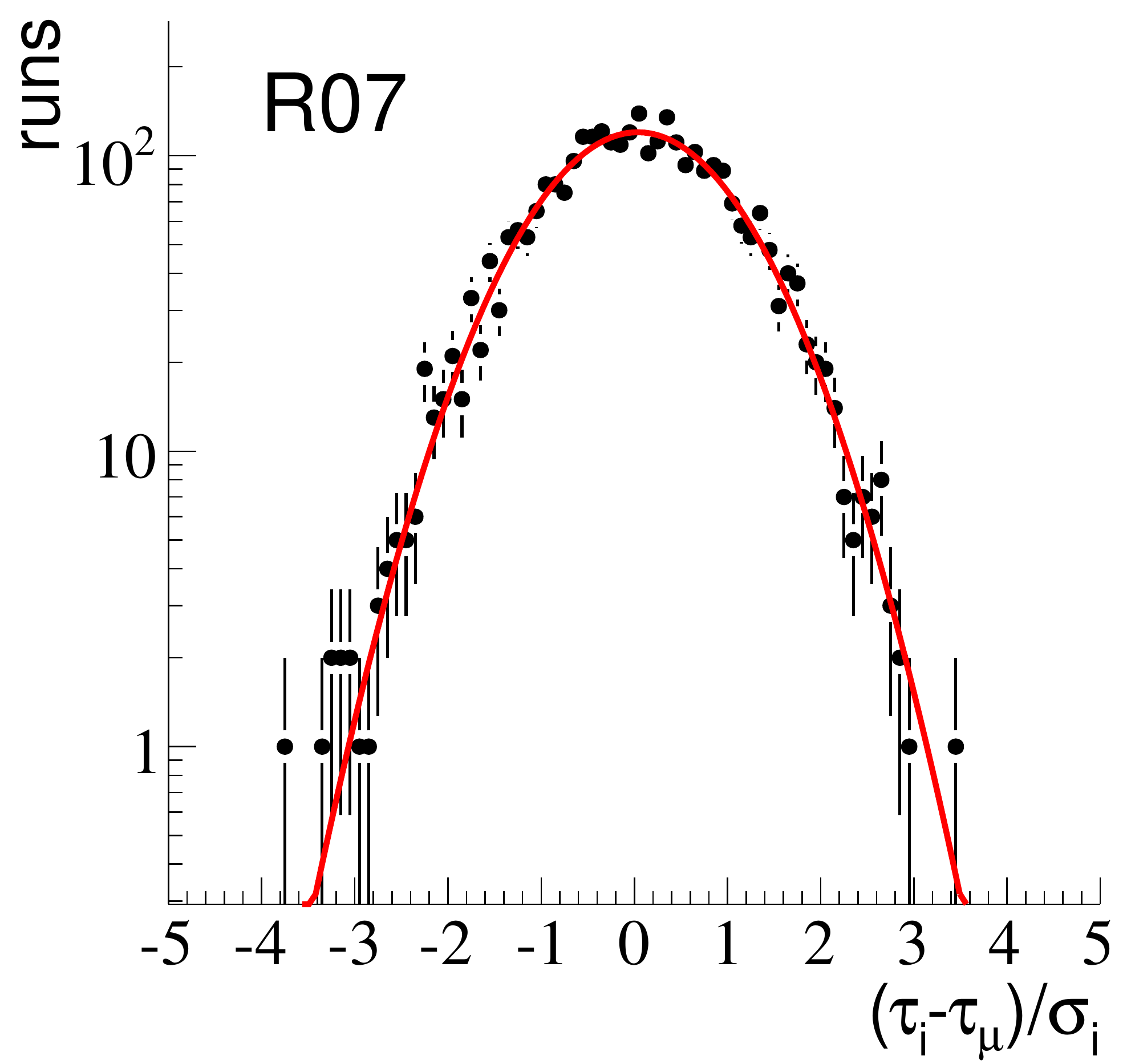}
\end{minipage}
\caption{The distribution $( \tau_i - \tau_{\mu} ) / \sigma_i$ of the normalized
deviations of the fitted lifetimes $\tau_i$ from the individual 
runs from the final result $\tau_{\mu}$. The data points are the run-by-run
results and the solid curve is a least squares fit to a Gaussian distribution. 
The left panel is the AK-3 dataset and the 
right panel is the quartz dataset.}
\label{fig:R_Ri_sigma}
\end{figure}

Fig.\ \ref{fig:R_Ri_sigma} shows $ ( \tau_i - \tau_{\mu} ) / \sigma_i $,
the normalized  deviations of 
lifetime values $\tau_i$ for the individual runs
from the benchmark result $\tau_{\mu}$.
The figure indicates the normalized deviations form 
a Gaussian distribution of mean zero and standard deviation 1.0,
{\it i.e.}, consistent with purely statistical variations.
These results for individual run variations,
and similar results for run group variations,
demonstrate the absence of effects on $\tau_{\mu}$ 
from long-timescale changes in gains, pedestals and thresholds.

Recall in AK-3 (quartz) tiny variations
of the fitted lifetime with the coordinate $\theta$ ($\theta_B$) are observed. 
These effects are attributed
to the slow relaxation of the longitudinal polarization and
are expected to cancel out in opposite-pair sums. 
However, such $\mu$SR cancellation grows increasingly imperfect
for a source that is displaced from the center of the detector.

\begin{figure}[ht] 
  \centering
\includegraphics[width=0.9\linewidth]{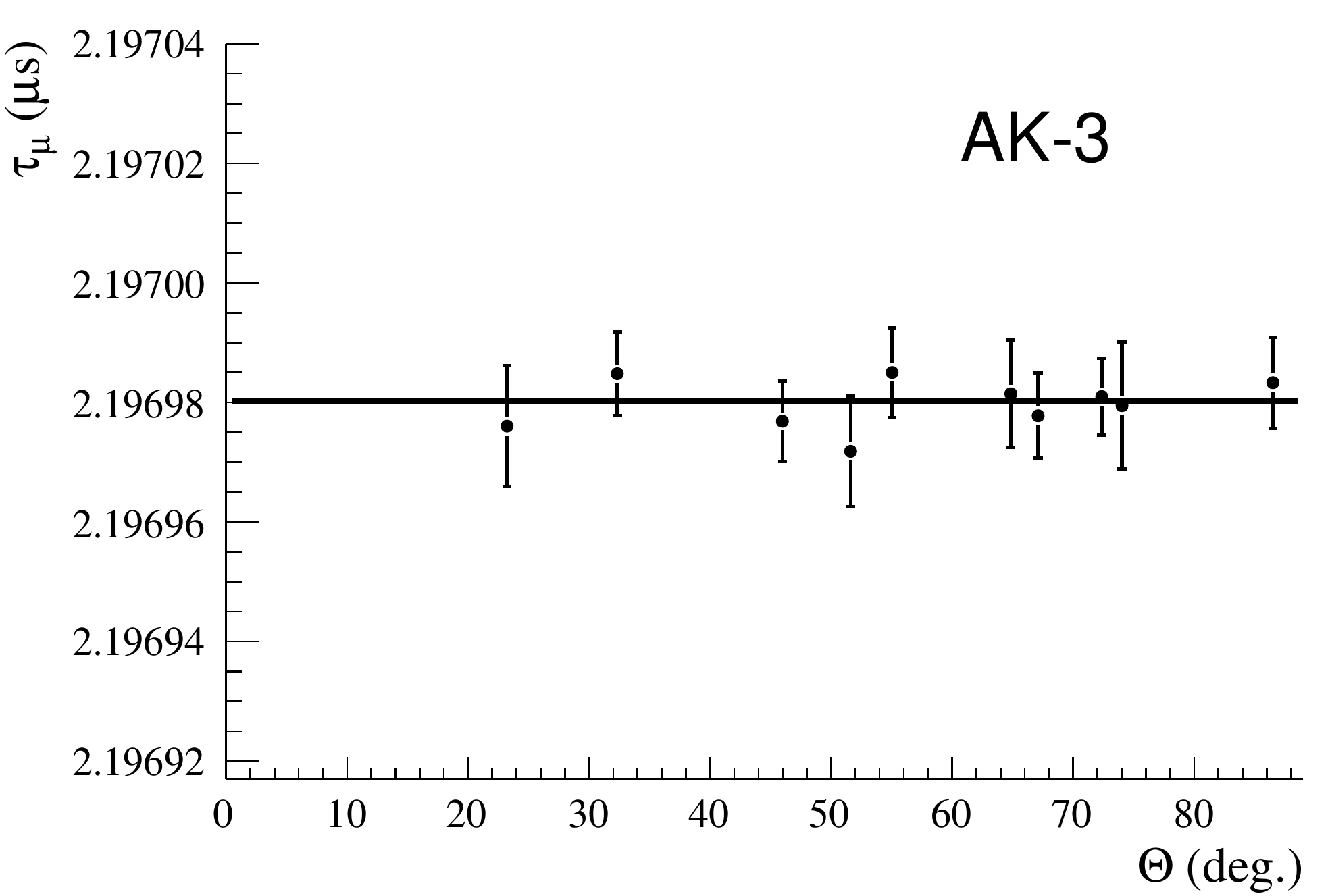}
\includegraphics[width=0.9\linewidth]{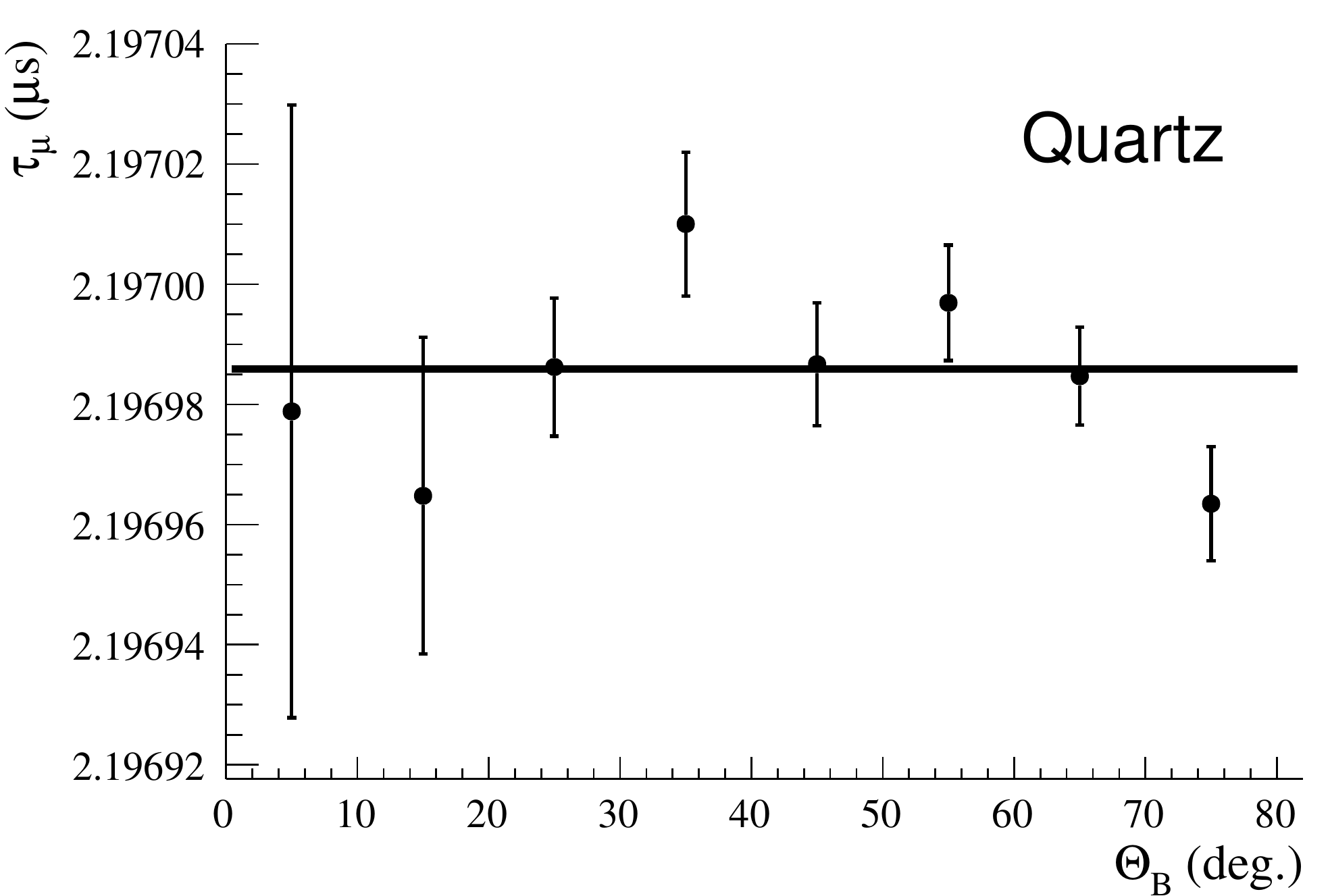}
  \caption{The upper panel shows the lifetime results from
opposite tile-pair sums grouped by detector coordinate $\theta$ 
for the entire AK-3 dataset. The lower panel shows the lifetime results from
opposite tile-pair sums grouped by detector coordinate $\theta_B$ 
for the entire quartz dataset. The plots demonstrate the cancellation  
of $\mu$SR distortions from target stops in opposite tile pairs.}
  \label{fig:R-det_theta}
\end{figure}

The $\theta$($\theta_B$)-distribution 
of the lifetime values obtained by fitting
summed time histograms of geometrically-opposite tile-pairs
in AK-3 (quartz) are plotted in Fig.\ \ref{fig:R-det_theta}. Neither show any evidence for a systematic 
$\theta$($\theta_B$)-dependence of the fitted lifetimes (as could originate 
from imperfect cancellation of $\mu$SR distortions from upstream stops).

A number of datasets were accumulated with different
magnetic field orientations and different positron detector offsets $( \delta x, \delta y, \delta z)$.
The agreement between lifetime results for different magnet and detector configurations
is an important requirement for the overall handling
of the $\mu$SR distortions.

We conducted the most exhaustive studies on the 
larger $\mu$SR signals from the quartz target. 
The quartz studies spanned 
a wide range of precession axes (from horizontal to vertical) 
and longitudinal polarizations (from $\sim 0.001$
to $\sim 0.25$). The specific datasets were
\begin{enumerate}

\item[(i)]
Production quartz data accumulated with the Halbach magnet
$B$ field oriented to the beam-left and the beam-right.
This change reverses the precession direction.

\item[(ii)]
Supplemental quartz data accumulated with
the Halbach magnet $B$-field tilted at pitch angles 
up to 40$^{\circ}$ from the vertical.
This changes both the axis of the TF precession 
and the magnitude of the LF relaxation.

\item[(iii)]
Both quartz production and supplemental data 
accumulated with different offsets $( \delta x, \delta y, \delta z)$ between 
the detector center and the 
stopping distribution (the changes in offset were $\leq$2~cm).
This changes the magnitude and direction of the Halbach $B$-field 
at the stopping location and therefore 
the frequency of the TF precession and 
the amplitude of the LF relaxation.

\end{enumerate}

\begin{figure}
\begin{center}
\includegraphics[width=\linewidth]{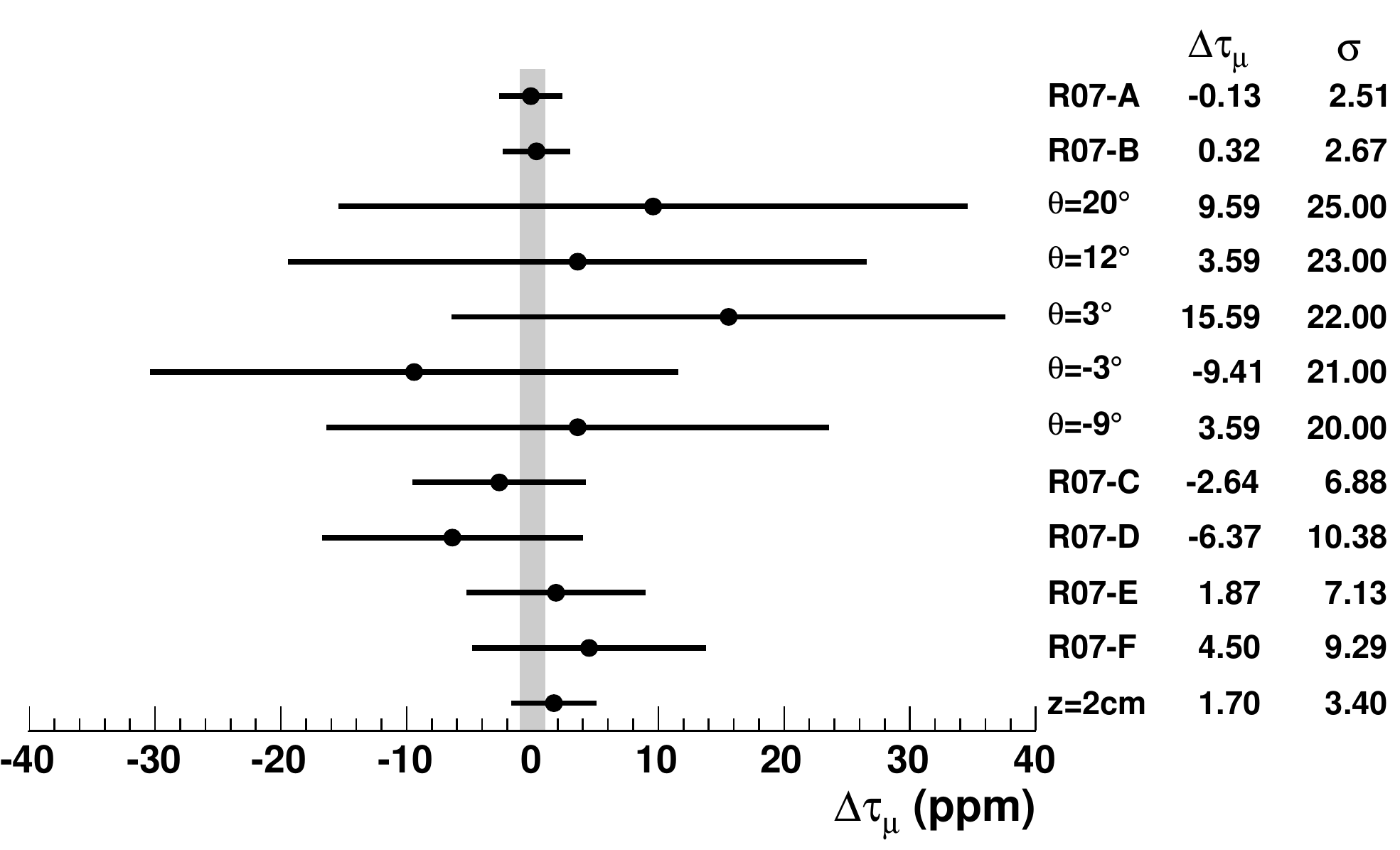}
\end{center}
\caption{Fitted lifetime for
quartz datasets with different magnetic field orientations
and positron detector offsets.
Entries R07-A and R07-B correspond to the quartz production data 
with the $B$-field orientation to the beam-left and beam-right, respectively.
Entries R07-C through R07-F correspond to quartz production data
with different detector positions and entries
$\theta = -9.0$$^{\circ}$ through $\theta = 20.0$$^{\circ}$ 
correspond to quartz systematics data with different magnet tilt angles.
The different detector positions and magnet angles
change the magnitudes and axes of the LF/TF polarizations 
and their corresponding $\mu$SR signals.}
\label{fig:R07_tau_summary}
\end{figure}

Fig.\ \ref{fig:R07_tau_summary} summarizes the lifetime values derived 
from the left/right $B$-field orientations, vertical/tilted $B$-field 
orientations and different target-detector offsets. 
The good agreement between lifetime values obtained
with different TF/LF $\mu$SR signals verifies 
that muon spin rotation in quartz was handled correctly.

Finally, we remark that the AK-3 production datasets were also accumulated
with the target magnetization oriented to the beam-left and 
the beam-right.

\subsection{Systematics studies}

This section addresses four classes of systematic uncertainties.
One class of effect is associated with corrections for pulse pileup
and gain changes in the histogram construction and 
another class is associated with $\mu$SR effects in the histogram fits.
A third class is related 
to possible time-dependent backgrounds
such as muon stops in the upstream beampipe
and a time-varying extinction of the muon beam.
A final class is the overall stability of the time measurement.
 
\subsubsection{Artificial deadtime correction}
\label{sec:pileup systematic}

A correction is applied for the lost hits
during the artificial deadtime (see Sec.\ \ref{sec:pileupprocedure}).
The pileup correction is performed for artificial deadtimes (ADTs) 
of 5 to 68~c.t.\ (11 to 151~ns). 
By increasing the ADT, one decreases the trigger-hit histogram contents
and increases the shadow-hit histogram contents.
Assuming the procedure correctly accounts for digital pileup,
the lifetime derived by summing trigger hit
and shadow-hit histograms is independent of the applied ADT.

\begin{figure}
\begin{center}
\includegraphics[width=\linewidth]{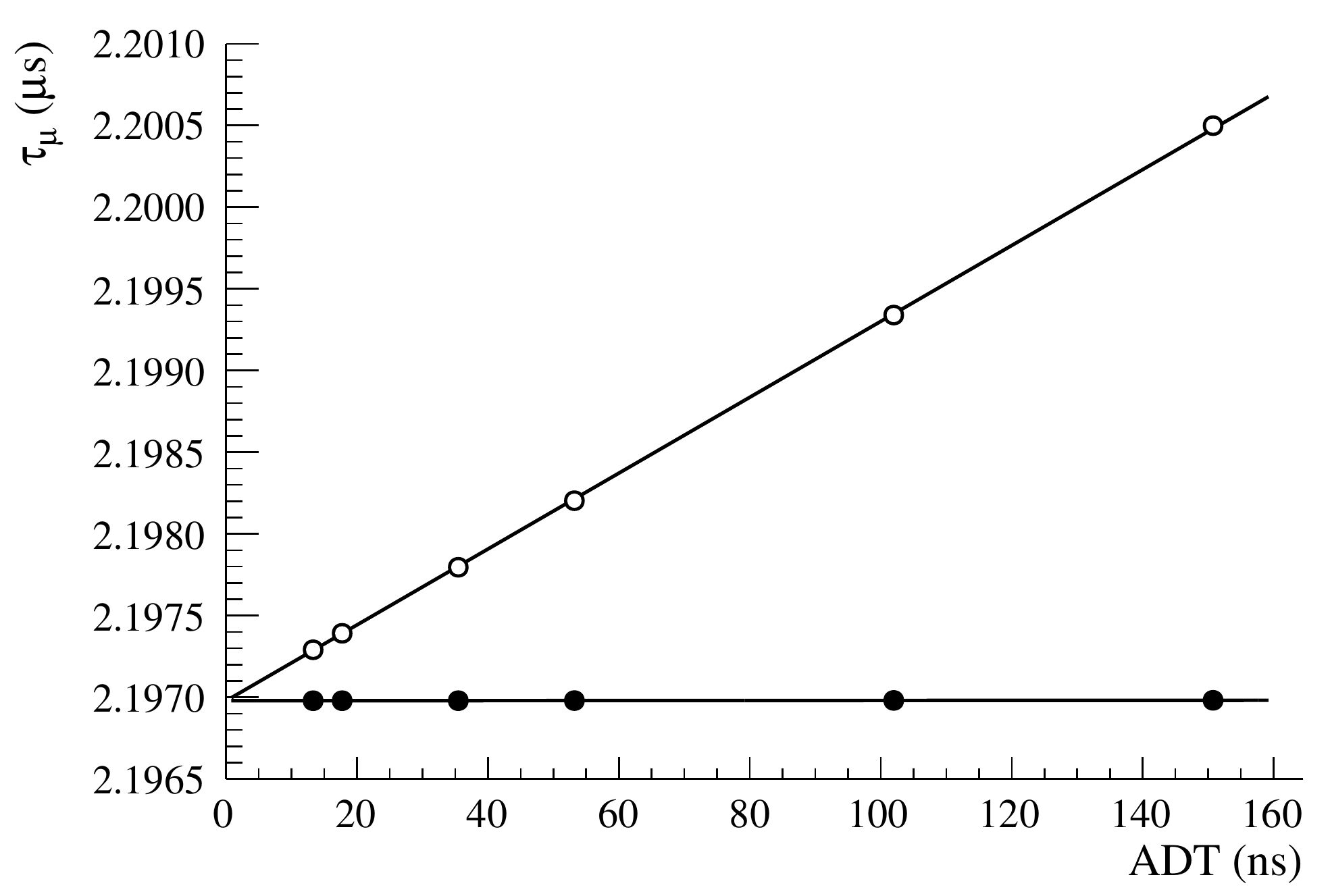}
\end{center}
\caption{Plot of $\tau_{\mu}$ versus ADT
before the pileup correction (open circles) 
and after the pileup correction (filled circles).
The statistical uncertainties associated with 
the pileup correction are substantially smaller than the statistical
uncertainty on the muon lifetime and not visible on this scale.
The solid lines through data points 
are only to guide the eye.}
\label{fig:tauVersusADT}
\end{figure}

Fig.\ \ref{fig:tauVersusADT} shows $\tau_{\mu}$ versus ADT
both before and after the pileup correction for the R07 dataset. 
Before the correction, the lifetime increases by 
10.5~ppm/ns (R06) and 10.6~ppm/ns (R07), while after the correction, 
the lifetime increases by 0.008~ppm/ns (R06) and 0.007~ppm/ns (R07),
a roughly 1000-fold reduction in the 
ADT-dependence of the lifetime $\tau_{\mu}$.

A ``perfect'' correction of pileup would yield an
ADT-independent lifetime. The tiny residual slope in
$\tau_{\mu}$ versus ADT after correcting for pileup 
thereby implies the procedure is slightly under-estimating 
the artificial deadtime effects at the level of  
about $10^{-3}$. One possible explanation is either the 
omission of a small, high-order, pileup term 
or the mis-accounting for a larger, known, pileup term.
However, as discussed in Sec.\ \ref{sec:pileupprocedure}, 
the procedure for accounting for digital pileup was verified
by Monte Carlo simulation and indicated no evidence 
of any omitted terms or mis-accounted terms 
at the $10^{-3}$ level.

Another possible explanation 
is a non-statistical variation of the muon stops in the 
accumulation periods.
Non-statistical variations lead to a pileup under-correction for 
``high rate'' fills and a pileup over-correction for ``low rate'' fills
that does not cancel out exactly.\footnote{The effects do not 
cancel as the number of under-corrected hits in high-rate fills exceeds 
the number of over-corrected hits in low-rate fills.}
Although sources of non-statistical variations are known---{\it e.g.}\ ion source fluctuations and production target 
rotation---their effects on muon stops are insufficient to explain 
the slope in $\tau_{\mu}$ versus ADT.

Although no conclusive explanation for the ADT-dependence 
of the fitted lifetime is identified, 
the observed variation of $\tau_{\mu}$ versus ADT is clearly linear.
Therefore the lifetimes derived 
from $\mathrm{ADT} = 6$~c.t.\ (13.3~ns) fits 
are corrected by linear extrapolations to zero deadtime. 
The extrapolation uses the slope obtained 
from a straight-line fit to $\tau_{\mu}$ versus ADT
from 5 to $68$~c.t.\ (11-151~ns). The extrapolation
yields a decrease in $\tau_{\mu}$ of 0.11~ppm 
for the AK-3 data and 0.10~ppm for the quartz data. 
A conservative systematic uncertainty of $\pm$0.2~ppm is assigned 
to the digital pileup correction procedure.

\subsubsection{Below-threshold pulses}
\label{sec:pile-on}

In the absence of pileup, below-threshold pulses cannot generate hits.
Two pileup effects involving below-threshold pulses are important:

\begin{enumerate}

\item[(i)]
``Pile-on'' involves the pileup of two below-threshold pulses.
If the pulses overlap, the below-threshold pulses 
can produce a single above-threshold pulse. 
Pile-on is a rate dependent increase
in hits that distorts the time distributions.
Because unaccounted pile-on adds hits at high rates (early times), it decreases 
the fitted $\tau_{\mu}$.

\item[(ii)]
``Pile-down'' involves the pileup of one below-threshold pulse
and one above-threshold pulse. 
If the pulses do not overlap, the below-threshold pulse
can raise the pedestal and lower the amplitude of 
the fitted above-threshold pulse.
Pile-down is a rate dependent decrease
in hits that distorts the time distributions.
Because unaccounted pile-down removes hits at high rates (early times), it increases 
the fitted $\tau_{\mu}$.

\end{enumerate}

The effects of pile-on
in promoting below-threshold pulses, 
and pile-down in demoting above-threshold pulses,
are not ADT-dependent  
and are not corrected 
by the shadow window technique.
The two effects cause opposite  shifts in $\tau_{\mu}$. 
For a pulse fitting procedure that is linear in the amplitude $A$ and the pedestal $P$
and equally weights all ADC samples, the effects 
exactly cancel. However, the exact cancellation 
is defeated if below-threshold pulses 
are first identified in the pulse fitting and then dismissed
by the amplitude cut.

A complete evaluation of pile-on and pile-down 
requires the measurement of the below-threshold amplitude distribution.
To avoid biases from the hardware threshold
of the waveform digitizers, the distribution was measured by 
acquiring data using an extended, 64-sample data block; 
the region of samples from 32 to 64 being used to obtain 
an unbiased amplitude distribution.
A relatively large number of pulses having amplitudes $<$25 ADC counts is found that carry no 
time dependence. A relatively small number of
pulses having amplitudes 25-35 ADC counts is found that do carry
the muon lifetime.

Using the measured amplitude distributions, the combined 
effects of pile-on and pile-down on $\tau_{\mu}$ 
are estimated by Monte Carlo simulation to be less than 0.2~ppm.
Importantly, the
combined effects of pile-on and pile-down 
are manifest in the time dependence
of the amplitude spectrum during the measurement period,
and therefore are compensated
by the application of the gain correction to the time histograms
(see Sec.\ \ref{sec:gaincorrection}). 

\subsubsection{Time pick-off stability}

Pileup can result in the loss of valid hits 
or the addition of extra hits.
Pileup can additionally ``pull'' the 
time determination of tile pulses
through effects including the absorption of 
a small, overlapping pulse into 
the trigger pulse fit or the interference of a 
large, preceding pulse with the trigger pulse fit.
Such distortions of the time ``pick-off''
are rate-dependent and therefore can 
distort $\tau_{\mu}$.
  
The determination of the time pick-off stability
is conducted using the 24 tiles instrumented
with the optical fibers from the
pulsed-laser system.
Approximately 400 laser runs
were accumulated in each production run.
The time difference $\delta t(t)$
between laser pulses in the scintillator tiles 
and the reference PMT  is sensitive to any
rate-dependent shifts in the time pick-off 
from the tile pulses.\footnote{The scintillator tile rate falls 
during the measurement period while the reference PMT rate
is constant during the measurement period.} 

To determine the time difference $\delta t(t)$ versus measurement
time $t$, a series of time difference histograms are
accumulated corresponding to sequential 2.5~$\mu$s-wide time windows 
on the laser pulse time during the measurement period.
These histograms are then fit to a Gaussian distribution
with the centroids determining $\delta t(t)$.
The fitted slopes of $\delta t(t)$ versus $t$ for 
all tiles are consistent with zero.
By combining results from all tiles, 
the average time pick-off was found to change by less 
than 0.25~ps over the measurement period. 
The limit implies a negligible distortion of $\tau_{\mu}$.

\subsubsection{Gain variations}

For the normal amplitude threshold $A_{thr}$, the lifetime shifts resulting
from gain variations are $0.50 \pm 0.25$~ppm (R06) and $0.53 \pm 0.25$~ppm (R07).
The uncertainties of $0.25$~ppm are conservative and based on variations between  
alternative procedures for making the gain corrections (see Sec.\ \ref{sec:gaincorrection}). 
The separate shifts for pairs instrumented with the Electron Tube PMTs (22 tile pairs) and
Photonis PMTs (148 tile pairs) are $+5.0$ and $-0.2$~ppm for the R06 dataset and $+5.3$ and $-0.1$~ppm 
for the R07 dataset. The separate values of $\tau_{\mu}$ obtained from Electron Tube and Photonis PMTs---after the 
appropriate correction for the gain variation---are in reasonable agreement for the two datasets.

To test the procedure for correcting for gain variations,
time spectra having amplitude thresholds $A_{thr}+20$, $+30$ and $+40$ ADC units were obtained.
Increasing the threshold increases the $\tau_{\mu}$ sensitivity
to gain changes due to a higher count fraction in the threshold amplitude bin (see Eqn.\ \ref{eqn:gainchange}).
The $A_{thr}+20$, $+30$ and $+40$ settings increase
the lifetime distortions from gain variations by factors of roughly 10, 30 and 50, respectively. 
A small error in the gain correction procedure at the normal amplitude
threshold thus implies a large error in the gain correction procedure 
at the higher amplitude thresholds.
Lifetimes obtained from a 10\% fraction of the R06/R07 datasets with the 
$A_{thr}+20$, $+30$ and $+40$ settings---with 50-fold differences in gain corrections---are 
in reasonable agreement.

\subsubsection{AK-3 $\mu$SR}

The application of the fitting method to the AK-3 data 
relies on geometrical cancellation of $\mu$SR effects,
in particular the $\sim$10~ppm variation
of the fitted lifetime with the tile-pair angle $\theta$ about the beam axis.
The cancellation is not exact owing to differences in the detector acceptances 
(for example, variations in tile efficiencies, 
electron absorption in intervening materials, and the mis-centering 
of the stopping distribution). An acceptance variation
between opposite tile pairs of several percent is estimated from
the measured variations of the positron rates in the scintillator tiles.
By combining the observed $\theta$-dependence of the lifetime $\tau_{\mu}$
with the estimated variation of the  opposite tile-pair acceptances,
a 0.1~ppm systematic uncertainty from AK-3 $\mu$SR effects is derived.

\subsubsection{Quartz $\mu$SR}

As described in Sec.\ ~\ref{sec:quartzfit}, the handling of
$\mu$SR effects in quartz requires specifying the beam-detector offset, 
$B$-field axis and several $\mu$SR parameters.

\begin{figure}
\includegraphics[width=0.9\linewidth]{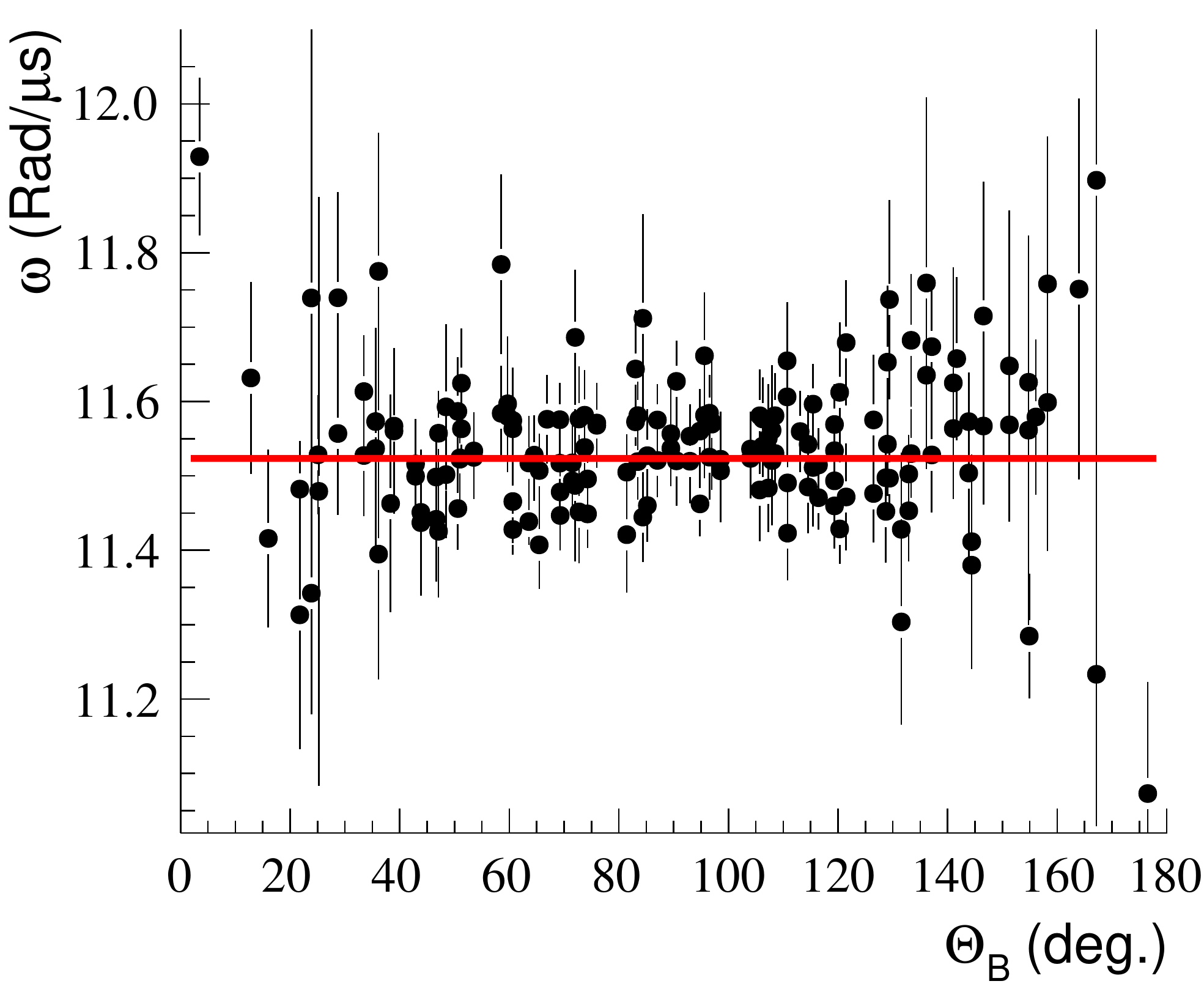}
\includegraphics[width=0.9\linewidth]{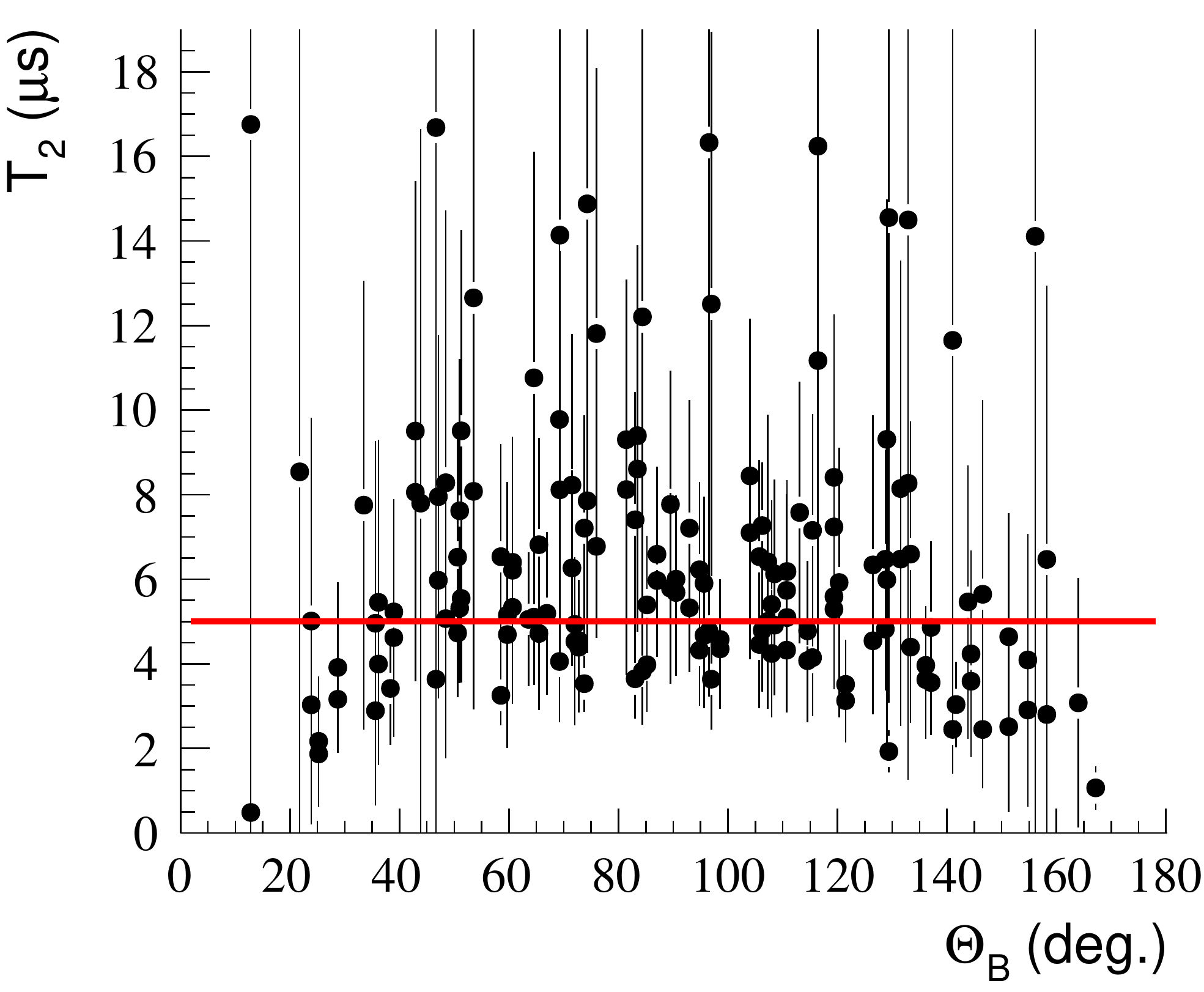}
\caption{Examples of precession frequency $\omega$ (upper panel) and relaxation constant $T_2$ 
(lower panel) versus tile coordinate $\theta_B$ obtained from 7-parameter fits 
using Eqn.\ \ref{eq:Lambda:d}. The solid horizontal lines indicate the fixed ``best fit'' values 
of $\omega$ and $T_2$ that are used in the fits of the tile-dependent effective lifetimes.}
\label{fig:omega:T2}
\end{figure}

In fitting the time histograms of tile-pairs,
the precession frequency $\omega$ and relaxation constant $T_2$ are fixed
to their ``best fit'' values (see Fig.\ \ref{fig:omega:T2}).
The change in $\tau_{\mu}$ 
when $\omega$ and $T_2$ are varied 
by one standard deviation from 
their ``best-fit'' values 
is adopted as the associated systematic uncertainties.
The procedure yields uncertainties in $\tau_{\mu}$ 
of 0.003~ppm from fixing $\omega$ and $0.02$~ppm from fixing $T_2$.

The fit to the angular distribution of the effective lifetimes
requires the determination of the tile coordinates $( \theta_B , \phi_B )$
relative to the $B$-field axis. 
A $\pm 2^\circ$  uncertainty is assigned to 
the knowledge of the $B$-field axis based on the uncertainties 
in the mechanical alignment of the Halbach magnet 
and the $B$-field non-uniformities at the target location.\footnote{The 
estimate of a $\pm$2$^{\circ}$ uncertainty 
in the $B$-field axis is consistent with the
observed longitudinal polarizations
in the quartz production data.}
The uncertainty in $\tau_{\mu}$ 
is obtained by correspondingly varying the $B$-field axis 
in the fit procedure for the effective lifetimes (Sec.\ \ref{sec:quartzfit}).
The approach yields an associated uncertainty 
of $0.004$~ppm in $\tau_{\mu}$.

Also required in the fitting procedure to the effective lifetimes
are the offsets $( \delta x, \delta y, \delta z )$ between the centers of the positron
detector and the stopping distribution.
These offsets are determined from measured differences 
between positron rates in opposite tile-pairs.
A $\pm2$~mm uncertainty is assigned to the determination of the 
offsets $( \delta x, \delta y, \delta z )$, based on the estimated variations in the 
tile-pair detection efficiencies.
The corresponding uncertainty in $\tau_{\mu}$ is obtained
by varying each offset by $\pm$2~mm  
in the fit procedure to the effective lifetimes.
The approach yields an associated uncertainty 
of $0.19$~ppm in $\tau_{\mu}$. The dominant contribution
originates from the offset $\delta x$ along the direction
of the longitudinal relaxation.

\begin{table}[htbp]
  \centering
  \caption{Compilation of the contributions 
to the systematic uncertainty  in the lifetime $\tau_{\mu}$
from the fitting procedure for the $\mu$SR effects in the 
quartz data.}
  \label{tbl:syst:muSR}
  \begin{tabular}{lc}
    \hline
    \hline
      parameter\hspace{1cm} &    $\delta\tau_\mu$ (ppm)    \\
    \hline
        $\omega$    &    0.003                     \\
        $T_2$       &    0.021                     \\
        $( \theta_B , \phi_B )$ &    0.004                     \\
        $\delta x$  &    0.19                      \\
        $\delta y$  &    0.001                    \\
        $\delta z$  &    0.001                    \\
        $A$         &    0.05                      \\
    \hline
        Total uncertainty   &    0.20                      \\
    \hline
    \hline
  \end{tabular} 
\end{table}

Finally, an uncertainty in $\tau_{\mu}$ is associated 
with the positron asymmetries $A( \theta_B , \phi_B )$ required
in the effective lifetime fit procedure.
This uncertainty is based on the difference between 
$\tau_{\mu}$ obtained with the tile-dependent, GEANT-calculated asymmetries 
and the tile-independent, energy-integrated theoretical asymmetry $A = 1/3$. 
The approach yields an associated
uncertainty of 0.05~ppm in $\tau_\mu$.

Table~\ref{tbl:syst:muSR} summarizes the various contributions 
to the systematic uncertainty in $\tau_\mu$ associated with the fitting procedure for the  
quartz $\mu$SR effects. 
The total uncertainty in $\tau_\mu$ from quartz $\mu$SR effects is 0.2~ppm.

\subsubsection{Upstream muon stops}\label{sec:upstreamin_stops}
\label{sec:errant_stops}

A small fraction of muons stop on the inner walls
of the beampipe upstream of the target disk.
These stops include both incoming muons in the beam halo
and back-scattered muons from the stopping target. 
Both target disks were sufficiently thick 
to eliminate downstream stops.

The concern with upstream stops 
is potential shifts of $\tau_{\mu}$
arising from $\mu$SR. Unlike target stops, 
where $\mu$SR signals cancel in opposite tile-pair sums,
for upstream stops, the $\mu$SR signals do not cancel out.
Therefore the presence of $\mu$SR signals from upstream stops 
could distort $\tau_{\mu}$.

Recall that following the downstream beamline elements,
the vacuum-pipe diameter was reduced from 35.5~cm to 20.2~cm 
at a location 52~cm upstream of the stopping target.
To minimize any $\mu$SR effects,
a 67~cm long section of the pipe immediately upstream 
of the stopping target was lined with 0.3~mm-thick AK-3 foil.
To maximize dephasing, the AK-3 magnetization 
was oriented perpendicular to the beam axis.

Several measurements were conducted
to estimate the effects of upstream stops.
Two regions of beampipe are particularly worrisome;
(i) the location of the beampipe constriction 
and (ii) the section of the reduced-diameter beampipe.

To estimate the stopping fraction in the annular face of the pipe constriction,
a 2$\times$3~cm$^2$ cross section, 3-mm thick, plastic scintillator was mounted
on a vacuum feed-through in the beampipe region that was immediately upstream 
of the pipe constriction. The setup determined the radial dependence of the muon rate
along horizontal, vertical and 45$^{\circ}$-angled axes perpendicular to the beam direction. 
The measurements yielded 
an estimate of ($0.5 \pm 0.1$)\% for the stopping 
fraction in the pipe constriction.

To estimate the stopping fraction in the 20.2~cm diameter beam pipe,
a plastic scintillator telescope  was used to detect the outgoing positrons
from beampipe stops. The telescope consists of
two pairs of separated scintillator tiles 
that viewed a $\sim$10~cm diameter region of beampipe.
The measurement yielded a stopping fraction of (0.10$\pm$0.05)\%  
in the 20.2~cm diameter beampipe.

The combined results of measurements (i) and (ii) indicate
less than a 1.0\% upstream stopping fraction.
To determine the effects of upstream stops,  
a dedicated dataset of upstream stops was collected.
The dataset comprised $\sim$2$\times$10$^{10}$ stops
in magnetized AK-3 at twelve upstream locations 
of 5 to 80~cm from the detector center.
The lifetimes derived from this upstream stop data
and the production data are in good agreement.
When the upstream stop data and the production data
are added in proportions of 1:100
the lifetime decreases by 0.1~ppm.

\subsubsection{Beam extinction stability}
\label{sec:beam stability}

A time-dependent beam extinction ($\epsilon \sim 900$) will cause a time-dependent positron background.
A variation of extinction can arise 
from a systematic change in the kicker voltage 
during the measurement period.

A limit on lifetime distortions  
from voltage instabilities is derived by
combining: (i) a  measurement of the kicker voltage $V(t)$ over the measurement period,
and (ii) a measurement of the beam extinction $\epsilon (V)$ versus the kicker voltage,
with (iii) a simulation of the effects of a time-dependent background on the lifetime determination.

The kicker voltage was measured with a high voltage probe connected to the kicker deflector plates. 
The probe was read out using a digital oscilloscope 
and the average voltage $V(t)$ from many kicker cycles was fit to determine a limit
on the voltage drift during the measurement period. Data were obtained 
on several occasions during both production runs
and yield limits of $\Delta V < 300$~mV (R06) and $\Delta V < 150$~mV (R07).\footnote{The ``worst-case'' voltage drift
is obtained by examining a variety of fits ({\it e.g.}\ linear, exponential)
to the kicker voltage during the measurement period. The values quoted 
for $\Delta V$ are obtained from a linear fit.} 

The variation in beam extinction with kicker 
voltage ${\Delta\varepsilon}/{\Delta V}$ was measured 
during both production runs using the beam monitor. 
Over the kicker voltage range 23.5-25.0~kV the extinction 
increased linearly with increased voltage. 
Combining the limits on $\Delta V$ 
with the determinations of ${\Delta\varepsilon}/{\Delta V}$,
the limits on changes in beam extinction 
from voltage instabilities of $\Delta \epsilon < 0.20$ (R06)  
and $\Delta \epsilon < 0.06$ (R07) are obtained.

A simulation is used to estimate the effect of an extinction variation $\Delta \epsilon$ on the
fitted lifetime $\tau_{\mu}$.
First, a time distribution is generated, consisting
of an exponential decay curve with a time-dependent background
due to extinction variation.
Then, the resulting distribution is fit to an exponential decay curve
with a time-independent background.
The difference $\Delta \tau_{\mu}$ between the simulated lifetime
and the fitted lifetime thereby determines the distortion 
of the lifetime originating from an unaccounted variation 
of the beam extinction.
The procedure yields upper limits of $0.20$~ppm (R06) and 
$0.07$~ppm  (R07) on the lifetime distortions from the kicker voltage drifts.

These limits are 20-40 times smaller  than the limit quoted 
in our 2004 commissioning measurement.
The improvements in limits were achieved through better suppression of
voltage drifts, better measurements of voltage stabilities, and increased 
beam extinction.

\subsubsection{Master clock stability}

Finally, the master clock output was compared at several frequencies 
to a calibrated rubidium frequency standard. The comparisons  
were made at the beginning and the end of each running period.
The observed deviations were less than 0.03~ppm
over the duration of the experiment.  A systematic uncertainty 
of 0.03~ppm on $\tau_{\mu}$ is assigned
to the time calibration of the master clock.
\section{Results}

\subsection{Summary of uncertainties}

Table~\ref{tbl:ErrorTable} summarizes
the statistical and systematic uncertainties that contribute 
to the muon lifetime measurements
in the R06 and R07 running periods.
The uncertainties associated
with $\mu$SR effects and kicker instabilities
are considered to be independent contributions to the R06 and R07
systematics. The 
larger $\mu$SR uncertainty in R07 data-taking
reflects the larger $\mu$SR distortions in the quartz data
and the smaller kicker uncertainty
in R07 data-taking reflects the improved 
limits on voltage instabilities.

By comparison, the uncertainties associated
with pulse pileup, gain variations, upstream stops,
time pick-off stability and clock calibration 
are considered to be common contributions to the
R06/R07 running periods. In each case
the same methods are used to obtain
estimates of systematic uncertainties.

\begin{table}[h!]
  \centering
\caption{Sources of systematic uncertainties on the
muon lifetime measurements in 
the R06/R07 running periods. The uncertainties listed
in single-column format are common 
uncertainties and those listed 
in two-column format are uncorrelated uncertainties. 
The last two rows are the combined systematic
uncertainties and the overall statistical uncertainties for the R06/R07 datasets.}
  \label{tbl:ErrorTable}
  \begin{tabular}{lcc}
    \hline
    \hline
    Uncertainty      & ~~R06~~       &  ~~R07~~ \\
 & ~~(ppm)~~ & ~~(ppm)~~ \\
    \hline
    Kicker stability                & ~~0.20~~  & ~~0.07~~ \\
    $\mu$SR distortions   & ~~0.10~~  & ~~0.20~~ \\
    Pulse pileup                         & \multicolumn{2}{c}{0.20} \\
    Gain variations                   & \multicolumn{2}{c}{0.25} \\
    Upstream stops             & \multicolumn{2}{c}{0.10} \\
    Timing pick-off stability               & \multicolumn{2}{c}{0.12} \\
    Master clock calibration               & \multicolumn{2}{c}{0.03} \\
    \hline
    Combined systematic uncertainty                & 0.42 & 0.42 \\
    Statistical uncertainty         & 1.14 & 1.68  \\
    \hline
    \hline
  \end{tabular}
\end{table}

\subsection{Results for muon lifetime, $\tau_{\mu}$}

The individual results for the muon lifetime 
from the fits to the R06 (AK-3) dataset and the R07 (quartz) dataset,
after the $-0.11$~ppm (R06) and $-0.10$~ppm (R07) adjustments for the 
$\mathrm{ADT} = 0$ extrapolation, are 
\begin{equation}
\tau_{\mu}({\rm R06}) =  2~196~979.9 \pm 2.5 (stat) \pm 0.9 (syst) {\rm ~ps}
\end{equation}
and
\begin{equation}
\tau_{\mu}({\rm R07}) = 2~196~981.2 \pm 3.7 (stat) \pm 0.9 (syst) {\rm ~ps}.
\end{equation}
The combined result 
\begin{equation}
\tau_{\mu}({\rm MuLan}) = 2~196~980.3 \pm  2.1 (stat) \pm 0.7 (syst) {\rm ~ps}
\label{finalresult}
\end{equation}
is obtained from the weighted average of the individual R06/R07 values with
the appropriate accounting for the correlated uncertainties.
The result corresponds to an overall uncertainty in the muon lifetime 
of 2.2~ps or 1.0~ppm.

\begin{figure}
\begin{center}
\includegraphics[width=0.99\linewidth]{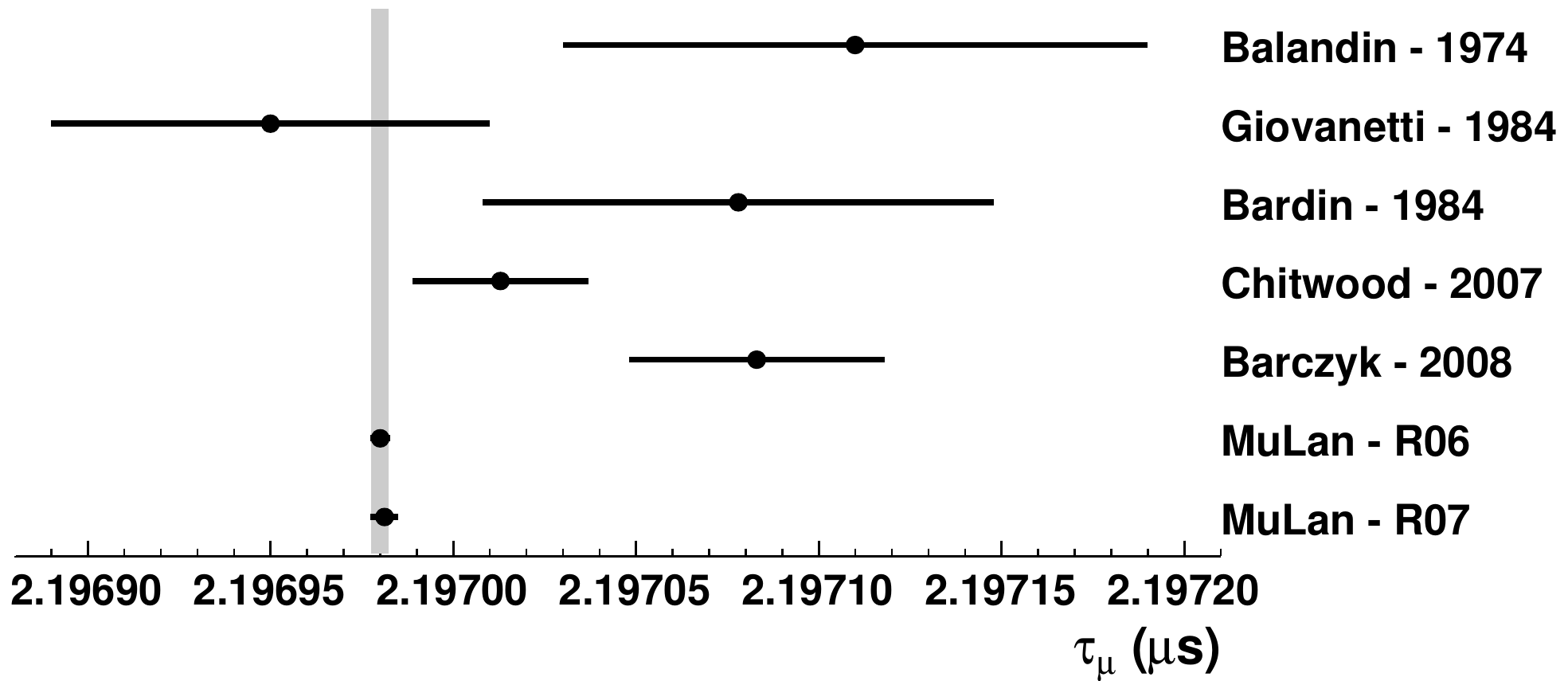}
\end{center}
\caption{Plot of post-1970 measurements
of the positive muon lifetime. The data points represent 
the individual measurements and the vertical band corresponds to
the combined R06 and R07 result.
The result of Duclos {\it et al.} is not shown in the figure.}
\label{fig:lifetimeplot}
\end{figure}

Figure \ref{fig:lifetimeplot} and Table \ref{tab:mulife} summarizes  
the measurements of the positive muon lifetime over the
past forty years. They include
the MuLan R06 and R07 results,
the MuLan commissioning measurement (Chitwood) \cite{Chitwood:2007pa}, 
the FAST commissioning measurement (Barczyk) \cite{Barczyk:2007hp}, 
and the earlier work of Duclos {\it et al.}\ \cite{Duclos:1974gg}, 
Balandin {\it et al.}\ \cite{Balandin:1974mq,Balandin:1975fe}, 
Giovanetti {\it et al.}\ \cite{Giovanetti:1984yw}
and Bardin {\it et al.}\  \cite{Bardin:1984ie}.
The final result from the combined R06 and R07 datasets 
is a factor of 10(15) improvement over
the MuLan(FAST) commissioning measurements 
and a factor of 30-150 improvement over the earlier experiments.

\begin{table}[htb]
\centering \caption {Compilation of post-1970 measurements
of the positive muon lifetime. The first three entries
are our final results from the combined R06+R07 datasets
and the individual R06/R07 datasets. Barczyk {\it et al.}\ is the
result of the FAST commissioning measurement and 
Chitwood {\it et al.}\ is the result of the MuLan commissioning
measurement. The remaining entries are previous generations
of lifetime experiments.
In column one the uncertainty represents the 
combined statistical and systematic errors.}
\vspace{0.3cm}
\begin{tabular}{ccc} 
\hline
\hline
  Measured   & Reference & Publication \\ 
lifetime ($\mu$s) & & year \\
\hline
 $2.196~9803 \pm 0.000~0022$ & R06+R07 &   \\
 $2.196~9799 \pm 0.000~0027$ & R06 &  \\
 $2.196~9812 \pm 0.000~0038$ & R07 &  \\
\hline
 $2.197~083 \pm 0.000~035$ & Barczyk~\cite{Barczyk:2007hp} & 2008 \\
 $2.197~013 \pm 0.000~024$ & Chitwood~\cite{Chitwood:2007pa} & 2007 \\
\hline
 $2.197~078 \pm 0.000~073$  & Bardin~\cite{Bardin:1984ie} & 1984 \\  
 $2.196~95 \pm 0.000~06$ & Giovanetti~\cite{Giovanetti:1984yw} & 1984\\  
  $2.197~11 \pm 0.000~08$ & Balandin~\cite{Balandin:1974mq,Balandin:1975fe} & 1974\\ 
  $2.197~3 \pm 0.000~3$ &  Duclos~\cite{Duclos:1974gg} &
  1973 \\ 
\hline 
\hline 
\end{tabular}
\label{tab:mulife}
\end{table}

The final result for $\tau_{\mu}$ is in reasonable agreement 
with the earlier work of Duclos {\it et al.}, Balandin {\it et al.},
Giovanetti {\it et al.} and Bardin {\it et al.} 
While it is in reasonable
agreement with the MuLan commissioning measurement 
(a difference of 1.3~$\sigma$) it is in 
marginal disagreement with the FAST commissioning measurement
(a difference of 2.9~$\sigma$). 
The weighted average of all results in Table \ref{tab:mulife}
gives a lifetime $\tau_{\mu} = 2\, 196\, 981.1 \pm 2.2$~ps with a chi-squared
$\chi^2/{\rm dof} = 2.7$. The weighted average is dominated 
by our final R06$+$R07 result with the poor chi-squared 
originating from the disagreement between our final result 
and the FAST commissioning result. 

\subsection{Result for Fermi constant, \GF}

We use the relation obtained by van Ritbergen and Stuart (vRS) \cite{vanRitbergen:1999fi,vanRitbergen:1998yd,vanRitbergen:1998hn}
for the extraction of the Fermi constant $G_F$ from a
measurement of the muon lifetime $\tau_{\mu}$. 
The vRS relation is derived using the
$V$-$A$ current-current Fermi interaction with 
QED corrections evaluated to 2-loop order. It yields
\begin{gather}
\GF = \sqrt{\frac{192\pi^3}{\tau_\mu m_\mu^5} \frac{1}{1 + \Delta q^{(0)} + \Delta q^{(1)} +
\Delta q^{(2)}}}\
\label{eq:gf_inversion}
\end{gather}
where $\tau_{\mu}$ is the measured muon lifetime, 
$m_{\mu}$ is the measured muon mass, and $\Delta q^{(0)}$,
$\Delta q^{(1)}$ and $\Delta q^{(2)}$ are theoretical corrections
with  $\Delta q^{(0)}$ accounting for the effects of the non-zero 
electron mass on the $\mu$-decay phase space and $\Delta q^{(1/2)}$ accounting
for the contributions of the 1-/2-loop radiative corrections
to the $\mu$-decay amplitude. The effects of non-zero neutrino 
mass are completely negligible.

For the muon lifetime, the weighted average 
of the R06$+$R07 results, $\tau_{\mu} = 2\, 196\, 980.3 \pm 2.2$~ps, is used.
The determination of \GF in  units of GeV$^{-2}$ 
from the measurement of $\tau_{\mu}$ in units of ps
requires a unit conversion 
via Planck's constant, $\hbar$. 
For Planck's constant, the value recommended  
by the 2010 CODATA committee \cite{Mohr:2012tt} of 
$\hbar = \unit[6.582\, 119\, 28(15) \times 10^{-25}]{GeV \cdot s}$ is used. 

For the muon mass, the recommended value  
of $m_\mu = \unit[105.658\, 371~5(35)]{MeV}$ \cite{Mohr:2012tt} is used.
This value is derived from the combination 
of the measurements of the electron mass and 
the electron-to-muon mass ratio. 

Computing the theoretical corrections requires both
the electron-to-muon mass ratio ${m_e} / {m_\mu}$ and 
the fine structure constant $\alpha( m_{\mu})$ 
at the momentum transfer $q = m_{\mu}$ of the $\mu$-decay process.
The recommended value  of ${m_e} / {m_\mu} = 4.836\, 331\, 66(12) \times 10^{-3}$
\cite{Mohr:2012tt} is used.
For the fine structure constant $\alpha( m_{\mu})$
the value $\alpha( m_{\mu})  = 1.0/135.902\, 660\, 087(44)$ is used.
The value is obtained from Eqn.\ 4.13 of van Ritbergen and Stuart \cite{vanRitbergen:1999fi}
using the CODATA value of the fine structure constant 
$\alpha (0) = 1.0/137.035\, 999\, 074(44)$ \cite{Mohr:2012tt} at
zero-momentum transfer.

The value for the phase space term $\Delta q^{(0)} = -187.1$~ppm
is obtained from Eqn.\  2.7 in Ref.\ \cite{vanRitbergen:1999fi}
and the value for the 1-loop QED correction $\Delta q^{(1)} = -4233.7$~ppm
is obtained from Eqn.\ 2.8 in Ref.\  \cite{vanRitbergen:1999fi}.
Note that Eqn.\ 2.8 of  Ref.\  \cite{vanRitbergen:1999fi} incorporates the effects of
the non-zero electron mass on the 1-loop QED correction.\footnote{The 1-loop 
QED correction including electron mass effects 
was originally derived by Nir \cite{Nir:1989rm}.}

The value for the 2-loop QED correction $\Delta q^{(2)} = +36.3$~ppm
is obtained by summing the individual contributions from
purely-photonic loops (Eqn.\ 9 in Ref.\ \cite{vanRitbergen:1998yd}),
electron loops (Eqn.\ 10 in Ref.\ \cite{vanRitbergen:1998yd}),
muon loops (Eqn.\ 19 in Ref.\ \cite{vanRitbergen:1998hn}),
tau loops (Eqn.\ 20 in Ref.\ \cite{vanRitbergen:1998hn})
and hadronic loops (Eqn.\ 16 in Ref.\ \cite{vanRitbergen:1998hn}).
Additionally, a correction of $-0.4$~ppm, first evaluated by Pak and 
Czarnecki \cite{Pak:2008qt}, is included to account
for the effects of the non-zero electron mass 
on the 2-loop QED correction.

Using Eqn.\ \ref{eq:gf_inversion} and the aforementioned values for
the $\tau_{\mu}$, $m_{\mu}$,
and the theoretical corrections $\Delta q^{(0)}$, $\Delta q^{(1)}$ and $\Delta q^{(2)}$, 
we obtain
\begin{gather}
\GF ({\rm MuLan}) = 
\unit[1.166\, 378\, 7(6)\times 10^{-5}]{GeV^{-2}}\ (0.5~\mathrm{ppm}).
\label{eq:gf_mulan_final}
\end{gather}
This result represents a thirty-fold improvement 
over the 1999 PDG value obtained before the vRS theoretical work
and the lifetime measurements pre-dating MuLan.

The error in $G_F$ of \unit[0.5]{ppm} 
is dominated by the 1.0~ppm uncertainty of the muon lifetime
measurement with additional contributions of 0.08~ppm from 
the muon mass measurement and 0.14~ppm from the theoretical corrections
in Eqn.\ \ref{eq:gf_inversion}.
Note the theoretical uncertainty of 0.3~ppm originally quoted
by van Ritbergen and Stuart \cite{vanRitbergen:1999fi,vanRitbergen:1998yd,vanRitbergen:1998hn}
arose from estimates of 
electron mass terms in the 2-loop QED corrections (0.17~ppm), 
hadronic contributions in the 2-loop QED corrections (0.02~ppm),
and 3-loop QED corrections (0.14~ppm).
Subsequently, Pak and Czarnecki \cite{Pak:2008qt} 
extended the work of vRS and calculated
the 2-loop order, electron mass terms.
Consequently, the overwhelming theoretical uncertainty 
is now the 0.14~ppm contribution associated with 3-loop QED corrections.
The uncertainties associated with $\hbar$, $m_e / m_{\mu}$ and
$\alpha( m_{\mu} )$ are all negligible.

The determination of $G_F$ from $\tau_{\mu}$
has additionally been discussed by Erler and Langacker 
in the PDG Reviews of Particle Physics \cite{Beringer:2012}.
Their articles first appeared in 1992 with later
updates as measurements improved and theory advanced.  
The authors maintained a formulation that 
differs slightly from vRS with their placement of the phase space correction $F( \rho ) $
as a multiplicative factor for the radiative corrections, {\it e.g.}
\begin{equation}
\frac{1}{\tau_\mu} = \frac{\GF^2 m_\mu^5}{192\pi^3} F( \rho ) \left[ 1 
+ H_1 ( \rho ) { \alpha ( m_{\mu} ) \over \pi } + H_2 ( \rho ) { \alpha^2 ( m_{\mu} ) \over \pi^2 } \right] 
\label{eq:el_lifetime}
\end{equation}
where $\rho = m_e^2 / m_{\mu}^2$ is the electron-muon mass ratio squared,
$\alpha ( m_{\mu} )$ is the fine structure constant evaluated at the relevant momentum transfer,
and $ H_1 ( \rho )$ and  $H_2 ( \rho )$ represent the 
1- and 2-loop radiative corrections.
Although the phase space correction $F ( \rho )$ is identical to $1 + \Delta q^{(0)}$,
and the 1-loop radiative correction $F ( \rho ) H_1 ( \rho ) {\alpha ( m_{\mu} ) / \pi }$
is identical to $\Delta q^{(1)}$ (see Ref.\ \cite{Nir:1989rm}), the 2-loop radiative corrections 
$F ( \rho ) H_2 ( \rho ) { \alpha^2 ( m_{\mu} ) / \pi^2 }$ and $\Delta q^{(2)}$
differ by $F ( \rho )$---thus introducing a small inconsistency 
between Eqn.\ \ref{eq:gf_inversion} of vRS and Eqn.\ \ref{eq:el_lifetime}
of Erler and Langacker. However, as $F ( \rho ) = 0.99981295$ is almost unity, 
the evaluation of $G_F$ using Eqn.\ \ref{eq:gf_inversion} 
and Eqn.\ \ref{eq:el_lifetime} gives the same value to a precision 0.1~ppm.\footnote{In addition, the expressions used 
by vRS and Erler and Langacker to evaluate the running of $\alpha ( m_{\mu} )$ 
are slightly different. However, these
differences are numerically unimportant for the Fermi constant
at the part-per-million level.}

The above extraction of $G_F$  from $\tau_{\mu}$ is based on the assumption 
of the $V-A$ chiral structure of the charged-current interaction.
If the assumption of the $V-A$ character
is relinquished, the relation between $\tau_{\mu}$ and $G_F$ is modified \cite{PhysRevLett.94.021802}.

\subsection{Discussion of results}

Below we compare the results 
from the AK-3 (R06) dataset and the quartz (R07) dataset and 
comment on the significance of the 1.0~ppm measurement of the
muon lifetime and the 0.5~ppm
determination of the Fermi constant.
 
\subsubsection{Comparison between R06/R07 datsets}

The two values of $\tau_{\mu}$ extracted from the R06/R07 datasets 
are in good agreement. Because the two datasets involved different targets 
having greatly varying magnitudes, time dependencies and angular distributions
of $\mu$SR signals, the natural interpretation
is that the spin rotation aspects have been properly handled. 

However, a second concern for a precision lifetime measurement 
is any difference between the positive muon lifetime in free space and the
positive muon lifetime in matter.
For example, a muon can reside in matter as either a bound muonium atom or an unbound positive muon. 
The relative populations of bound muonium and unbound $\mu^+$'s
can vary with material and ranges from small  muonium fractions ({\it e.g.}\ AK-3) 
to large muonium fractions ({\it e.g.}\ quartz).
A significant difference in lifetimes between muonium and $\mu^+$'s
could jeopardize the interpretation 
of lifetime measurements using stopped muons.

We stress: the two targets involve different populations
of $\mu^+$'s (the dominant species in AK-3) 
and muonium atoms (the dominant species in quartz).
The possibility of a lifetime difference between the 
free muon and the muonium atom 
was addressed by Czarnecki, 
LePage and Marciano \cite{Czarnecki:1999yj}. 
They found that the leading-order contributions from the two largest 
corrections---the $\mu^+$$e^-$ binding energy and $e^+$$e^-$ 
final-state interaction---cancel out and the resulting difference between 
their lifetimes to be only $\sim$0.6~parts-per-billion.\footnote{This 
cancellation between final-state interaction effects and binding energy effects 
is familiar from work by Huff \cite{Huff:1961} and Uberall {\it et al.}\ \cite{Uberall:1960zz} for muonic atoms. 
So-called Huff factors, that represent the difference between the $\mu^-$ decay rate in free space 
and the $\mu^-$ decay rate in muonic atoms, have been calculated across the periodic table.
The calculated effects for bound negative muons are supported by experimental data 
from muonic atoms.}. This difference is negligible compared to the part-per-million precision of 
the R06/R07 lifetime measurements.  

In addition, a lifetime difference between free muons and
muonium atoms could arise from $\mu^+$$e^-$ annihilation in muonium atoms.
For example, the rate of $\mu^+ e^- \rightarrow \bar{\nu_{\mu}} \nu_e$ annihilation
was considered by Czarnecki {\it et al.}\ \cite{Czarnecki:1999yj} and earlier by Li {\it et al.}\ \cite{Li:1988xb}.
However, the SM prediction for the annihilation rate is only $10^{-13}$ of the muon decay
rate and therefore negligible compared to the ppm precision of 
the R06/R07 lifetime measurements.

\subsubsection{Relevance to precision muon capture studies}

A precise determination of $\tau_{\mu}$ is important 
to current work on nuclear muon capture. The MuCap experiment \cite{Andreev:2007wg}
has recently measured the $\mu^-$p singlet capture rate $\Lambda_s$ 
and the MuSun experiment \cite{Andreev:2010wd} is presently measuring 
the $\mu^-$d  doublet capture rate $\Lambda_d$.
Both experiments derive the muon capture rates from the
tiny difference $( \Lambda_o - \Lambda )$ between the positive muon decay rate ($\Lambda_o = 1/\tau_{\mu^+}$) and 
the muonic atom disappearance rates ($\Lambda = 1/\tau_{\mu^-Z}$).
Because the capture rates are very small,
a precise determination of
the muonic atom disappearance rates 
and the free muon decay rate is necessary.

Prior to this work the muon decay rate $\Lambda_o$
was known to 9~ppm or 5~s$^{-1}$, an uncertainty
that limited the capture rate determinations
from disappearance rate experiments.
The new result for the muon lifetime---yielding 
a precision of 1~ppm or 0.5~$s^{-1}$ in 
$\Lambda_o$---has eliminated this source of uncertainty.

\subsubsection{Relevance to fundamental constant determinations}
 
In 1999 the uncertainty in $G_F$ was 17~ppm,  a combination of a
9~ppm experimental uncertainty and a 15~ppm theoretical uncertainty. 
The combined outcome of the vRS calculations \cite{vanRitbergen:1999fi,vanRitbergen:1998yd} 
and the MuLan experiment is a thirty fold reduction
in the total uncertainty in the Fermi constant---to a precision 0.5~ppm.

The electroweak sector of the standard model involves three parameters, 
the two gauge coupling constants $g$, $g\prime$
and the Higgs energy density $v$.
Their values are fixed by measurements 
of the fine structure constant, $\alpha$, 
Fermi coupling constant, $G_F$,
and Z boson mass, $M_Z$. The thirty-fold improvement 
in the determination of $G_F$, together with recent improvements 
in the determinations of $\alpha$ and $M_Z$,
allows for improved tests 
of the Standard Model. 

\section{Conclusion}

In summary, we report a measurement of the positive muon lifetime $\tau_{\mu}$
to 1.0~ppm  and a determination of the Fermi constant $G_F$ to 0.5~ppm.
The experiment used a novel, time-structured muon beam
with a segmented, fast-timing, positron detector instrumented
with 450~MHz waveform digitizer readout.
The results for $\tau_{\mu}$ and $G_F$ are substantial improvements 
over earlier determinations from previous generations of lifetime experiments. 
Improved knowledge of $\tau_{\mu}$ is important to other experimental programs
and improved knowledge of $G_F$ is important for numerous electroweak processes.

In the determination of the Fermi constant,
we are fortunate that the uncertainties on many
quantities--{\it e.g.} the muon mass and  Planck's
constant---are significantly below the ppm-level.
To improve the determination of \GF by another
order of magnitude will require, in addition to 
another order of magnitude improvement in $\tau_{\mu}$,
a significant experimental improvement in the muon mass determination
and a significant improvement in the various theoretical corrections.

\section{Acknowledgments}

We thank E.~Morenzoni, R.~Scheuermann and A.~Stoykov
for their assistance with $\mu$SR studies in AK-3 and quartz. 
We thank the PSI staff, especially D.~Renker, K.~Deiters, 
and M.~Hildebrandt. We thank M.~ Barnes and G.~Wait from
TRIUMF for the development of the kicker, and thank
N.~Bondar, T.~Ferguson and R.~Prieels for their contributions
to the electronics readout for the beam monitor. We also thank the National
Center for Supercomputing Applications (NCSA) for supporting 
the data analysis, and the U.S. National 
Science Foundation for their financial support.

\bibliography{mulan}

\end{document}